\documentclass{cimento}

\usepackage{amsmath}
\usepackage{graphicx}% Include figure files
\usepackage{dcolumn}% Align table columns on decimal point
\usepackage{bm}% bold math
\usepackage{color}
\usepackage[normalem]{ulem}

\title{Virtual social science}
\author{Stefan Thurner %\from{} \from{ins:y} \from{ins:z} \from{ins:zz} 
}
\instlist{
\inst{}Section for the Science of Complex Systems, Medical University of Vienna - Vienna, Austria
\inst{} Santa Fe Institute - Santa Fe, USA
\inst{} IIASA - Laxenburg, Austria
\inst{} Complexity Science Hub Vienna - Vienna, Austria
}

\begin{document}

\maketitle

\begin{abstract}
Can we describe social systems quantitatively and predictively, when we know all the actions, 
interactions, and states of individuals?
We interpret human societies as co-evolutionary systems of individuals and their interactions. 
Based on unique data of a society of computer game players, where all actions and interactions between all players are 
known, we show that this might indeed be possible. Within this framework we address a number 
of sociological classics, including formation of social networks, strength of relations, group formation, 
hierarchical organization, aggression management, gender differences, mobility, and wealth-inequality. 
We discover behavioral and organizational patterns of the homo sapiens and its society that were not visible with
traditional methodology from the social sciences. 
\end{abstract}

%%%%%%%%%%%%%%%%%%%%%%%%%%%%%%%%%%%%%%%%%%%%%%%%%%%%%%%
\section{Introduction}

In the first years of the nineteenth century Auguste Comte   
suggested to copy what had been done in physics at this point in time: 
to establish a natural and experimental science of social systems. He suggested to call this 
hopeless endeavor {\em sociophysics}. No-one followed him. He died poor and alone,  
without any acknowledgement of this vision of his. Some remember him today as the 
inventor of the term {\em sociology}. 
Of course, his vision was bound to fail. There was no way of understanding what the rules of interactions 
between the components of societies were. Even had he known the interpersonal interaction laws, he would have failed. 
He would not yet have had a way to aggregate information of many interactors to levels that could
have been interpreted in any meaningful way. Statistical mechanics was not invented yet, 
nor was the computer; Gauss was barely born.  
Had he overcome all these difficulties, he would have most likely failed because he did not have a way to know 
what non-linearity can do to the predictability of dynamical systems. 
He had to fail. 

%\begin{figure}[t]
%\begin{center}
%\includegraphics[width=0.25\textwidth]{plots/auguste.pdf}
%\end{center}
%\caption{Auguste Comte, luckless visionary and inventor of the words {\em sociophysics} and 
%{\em sociology}. 
%}
%\label{gene}
%\end{figure}

Now, more than 200 years later, we try it again: to transform the social science as we know it today, 
into an experimental physical science: fully quantitative,  predictive, and experimentally testable.
A science that is based on the elementary interactions between humans, and between them and 
their environment. Immediately criticism from the traditional representatives of the social 
sciences and humanities arises: 
(i) humans have a free will, their actions and interactions can not be predicted; 
(ii) societies have way too complicated interactions, which will never be quantifiable, and 
(iii) data on actions and interactions on a society-wide level does not exist. 

In the 21st century we do not have to accept these arguments. 
None of them states a fundamental reason, why a new attempt would have to fail. 
In response to the free will argument (i), we can argue that atoms have even more ``free will'' than humans do, 
and that on a 
quantum mechanics level the situation is even worse: not only could an atom decay spontaneously, 
not even its position and momentum can be known simultaneously. How should we ever be able to predict 
properties of matter that is composed of such volatile things? But we can.
A counter argument to statement (ii), that human interactions are too complicated to be quantified, is 
found in the very existence and availability of exactly this kind of data, together with  the computer power to process it. 
We might be able to record a large fraction of all interactions soon, 
given that we are not stopping the current trends in digitation. 
The same holds true for argument (iii), that there is a lack of data on social actions and interactions. 
The world has changed.
%, traditional sociologists might have a hard time with these counter arguments soon. 

This does not mean that those who try to realize the vision of Comte again will be successful. 
In this lecture I will show that in fact, we are getting to the point, where we will be able to record every single interaction in a 
social system, meaning that we have records about all interactions, at all times, and between all individuals. 
I will show a special example where we have exactly this situation: a dataset of a human society 
of players ``living'' in  a massive multiplayer online game, 
where all information about actions, interactions and properties of all avatars is available. 
The game is called \verb|Pardus|, and was developed, programmed, and 
maintained by my former students and collaborators Michael Szell and Werner Payer. 
In this game, which attracted almost half a million players since 2004, avatars act out an open-ended 
``second life'', over large periods of time---often over years. 
All actions and interactions, all properties and characteristics of all players are recorded, 
at every point in time.

We have studied this human society in the past years. In this lecture notes, I summarize some of the 
work that was done together with Michael Szell, Peter Klimek, Benedikt Fuchs, Renaud Lambiotte, 
Vito Latora, Roberta Sinatra, Didier Sornette, Olesya Mryglod, Yurko Holovatch, Bernat Corominas-Murtra, 
Maximilian Sadilek, and others. The original works are found in  
\cite{Szell2010msd,Szell2010,Thurner2012,Szell2012,Szell2012B,Szell2012C,Klimek2013,CorominasMurtra2014,CorominasMurtra2014B,Thurner2014,Fuchs2014,Fuchs2014B,Mryglod2015,Klimek2016,Sadilek2018}.

%%%%%%%%%%%%%%%%%%%%%%%%%%%%%%%%%%%%%%%%%%%%%%%%%%%%%%%
\subsection{What is social science?}
Social science is the science of social interactions and their implications for society.
Traditionally, social science has neither been very quantitative or predictive, nor does it produce experimentally
testable predictions. It is largely qualitative and descriptive. 
Until recently, there has been a tremendous shortage of data that are both, time-resolved
(longitudinal) and multidimensional. As stated, the situation is changing fast with the tendency of 
homo sapiens to leave electronic fingerprints in practically all dimensions of life. The century-old
data problem of the social sciences is rapidly disappearing. Another fundamental problem in the
social sciences is the lack of reproducibility and repeatability. On many occasions, an event takes place once 
in history, and no repeats are possible.

As in biology, social processes are hard to understand mathematically because they are evolutionary,
path-dependent, out-of-equilibrium, and context-dependent. They are high-dimensional and
involve interactions on multiple levels and scales. The methodological tools used by traditional social
scientists, which rarely extend much beyond linear regression models, and basic Gaussian statistics, 
are not powerful enough to address these issues appropriately. 
However, two important innovations have been developed within the social sciences 
that play a crucial role in the theory of complex systems:

{\em Multilayer interaction networks.} 
In social systems, interactions  between individuals and institutions, 
happen simultaneously on more or less the same strength scale on a 
multitude of superimposed interaction networks. Social scientists, in particular sociologists, have recognized 
the importance of social networks, 
starting in the 1970s \cite{Wasserman1994,Granovetter1973}. 
	 
{\em Game theory.} Another contribution is game theory, a concept that allows us to
determine the outcome of rational interactions between agents trying to optimize their
payoff or utility \cite{vonNeumann1944}. Each agent is aware of the fact that the other agent is 
rational and that he/she also knows that the other agent is rational.  
Before computers arrived on the scene, game theory was one of the 
few methods of dealing with complex systems (in equilibrium). Game theory can easily be transferred to dynamical
situations, and it was believed for a long time that iterative game-theoretic interactions were
a way of understanding the origin of cooperation in societies. This view is now severely
challenged by the discovery of  so-called zero-determinant games \cite{Press2012}. Game theory
was first developed and used in economics, but later penetrated other fields of the social, 
behavioural, and life sciences.

\subsubsection{Social systems are continuously restructuring networks}

Social systems can be thought of as time-varying multilayer networks. 
Nodes are individuals or institutions; links are interactions of different type. Interactions change over
time. The types of link can be friendship, family ties, processes of good exchange, payments, trust,
communication, enmity, and so on. 
Every type of link is represented by a separate network layer, see e.g. Fig. \ref{MPN}. Individuals
interact through a superposition of these different interaction types (multilayer network), which
happen simultaneously, and are often of the same order of magnitude in ``strength''. Often, networks
at one level interact with networks at other levels. Networks that characterize social systems
show a rich spectrum of growth patterns  and a high level of plasticity. This plasticity of networks arises from
restructuring processes through link creation, re-linking, and link removal. 
Understanding and describing the underlying restructuring dynamics can be challenging. 
However, there are a few typical and recurring dynamical patterns that allow us to make scientific progress.

Individuals are represented by ``states'', a set of characteristics that describe their wealth, gender, education level, political
opinion, age, and so on. Some of these states change dynamically over time. States typically have an
influence on the linking dynamics of their corresponding node. If that is the case, a
tight connection exists between network structure and node states. The joint dynamics of network
re-structuring and changes of states by individuals is a classic example of {\em co-evolution}\index{co-evolutionary dynamics}.

%%%%%%%%%%%%%%%%%%%%%%%%%%%%%%%%%%%%%%%%%%%%%%%%%%%%%
\subsection{Social systems are complex systems}

If social systems are complex systems---what are complex systems? 
We follow the definition of complex systems given in \cite{Thurner2018book}: 
{\em Complex systems are co-evolving multilayer networks.}		
This statement summarizes ten facts about complex systems:

\begin{figure}[t]
 \begin{minipage}[c]{0.4\textwidth}
\includegraphics[width=\textwidth]{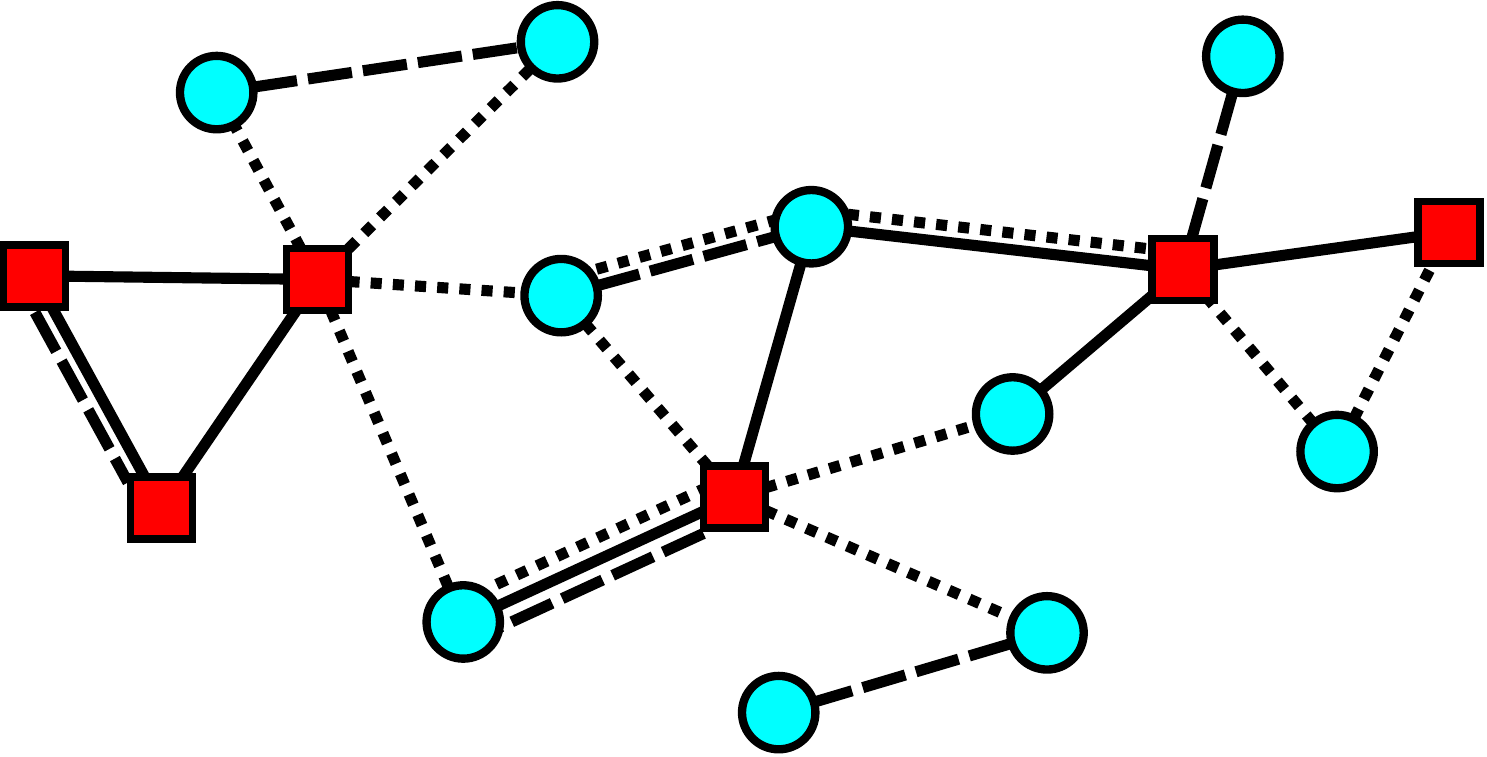}
  \end{minipage}\hfill
  \begin{minipage}[c]{0.55\textwidth}
    \caption{Schematic representation of a multilayer network. Here nodes are characterized by a
two-dimensional state vector. The first component is given by colors (red, blue), 
the second by shapes (circles, squares). Nodes interact through three types of interaction 
that are represented by (full, broken, and dotted) lines. 
The system is complex if states simultaneously change as a function of
the interaction network, and at the same time, if interactions change as a function of the states,  see Eqs. (\ref{coev_intro}).
%For example, the multilayer network can represent a network of banks at a given moment, where shapes represent 
%the wealth of the bank and the links represent financial assets that connect banks with each other.
%A full line could mean a credit relation; a broken line could represent derivatives trading, and dotted
%lines indicate if one bank owns shares of another bank. Depending on that network, the banks will
%make a profit or loss in the next time step. See text for a different example. 
It is a co-evolving system, where states and interactions 
update each other, similar to what an algorithm is doing.
           } 
\label{MPN}
\end{minipage}
\end{figure}

\begin{enumerate}	
\item Complex systems are composed of multiple elements, labelled by latin indices, $i$.

\item These elements interact with each other through one or more interaction type, labelled by greek indices, $\alpha$. 
Interactions between elements are specific, not everyone interacts with all the others. 
To keep track of which elements interact we use networks. Interactions are represented as links, 
the interacting elements are  nodes.
 Every interaction type can be seen as one network layer in a multilayer network, see Fig. \ref{MPN}. 
 A multilayer network is a collection of networks linking the same set of nodes. If these networks evolve independently, 
 multilayer networks are superpositions of networks. However, there are often interactions between interaction layers. 

\item Interactions change over time. 
 We use the following notation to keep track of interactions in the system. The strength of an interaction 
of type $\alpha$ between two elements $i$ and $j$ at time $t$ is denoted by,   
\[M_{ij}^{\alpha} (t) \quad {\rm  interaction} \ {\rm strength} \quad . \]
Interactions can be physical, financial, emotional, economical, hostile, or symbolic, to just mention a few. 
Most interactions are mediated by an exchange process of some sort between nodes. 
In that sense, interaction strength is
often related to the quantity of ``things'' or units exchanged (money for financial interactions,
love letters for emotional interactions, bullets for hostile interactions, bottles of wine
for positive social interactions, and so on). Interactions can be deterministic or stochastic.
 
\item Elements are characterized by states. States can be scalar; if an element has various 
independent states, it will be described by a state vector, or a state tensor. 
States also evolve over time. We denote the state vectors by,   
\[\sigma_i(t) \quad {\rm  \ state \ vector}  \quad .  \] 
States can be the education level, gender, political believe, wealth, aggression level of a person, or the 
capitalization, and risk aversion levels of a bank. 
State changes can be deterministic or stochastic. 
Changes can be the result of the endogenous dynamics in the multilayer network, or of external driving. 

\item States and interactions are often not independent 
but evolve together by mutually influencing each other; states and interactions {\em co-evolve}.  
The way in which states and interactions are coupled can be deterministic or stochastic. 

\item The dynamics of co-evolving multilayer networks is usually non-linear. 

\item Complex systems are context-dependent. Multilayer networks provide that context. 
To be more precise, for any dynamic process  
happening on a given network layer, the other layers represent the ``context'' in the sense that they provide the 
only other ways in which elements in the initial layer can be influenced. Multilayer networks sometimes 
allow complex systems to be interpreted as ``closed systems''. They can be externally driven. Then they are  
dissipative and non-Hamiltonian. 
  
\item Complex systems are {\em algorithmic}, they behave rather like an algorithm than a system that is described by 
a set of differential equations.  
The algorithmic nature is a direct consequence of the discrete 
interactions between interaction networks and states. 

\item Complex systems are path-dependent, and therefore often non-ergodic. Given that the network 
dynamics (the dynamics of links) is sufficiently slow, the networks in the various layers can be seen as 
a ``memory'' that stores the recent past of the system.

\item Complex systems often have memory. Information about the past can be stored in nodes if they have explicit 
memory, or in the network structure in the various layers. 
\end{enumerate}
In the following, we assume that a co-evolving multilayer network structure 
is the fundamental dynamical backbone of social systems. 
A snapshot of a co-evolving multilayer network is shown in Fig. \ref{MPN}. Nodes are given 
by a state vector with two components, colour (blue, red) and shape (circles and boxes). Nodes
interact through three types of interaction  (full, broken, and dotted lines). The system is
a complex system if states change as a function (deterministic or stochastic) of the interaction
network and, simultaneously, interactions (the networks) change as a function of the states. 
The shown multilayer network could represent a social network of 
individuals with certain time-dependent properties. Think of wealth represented by shape (rich=circle, poor=square), 
and gender (blue=male, red=female), and the different 
different link-types represent communication, trade, and friendship.
While the state of wealth can change as a function of the trading links, gender will not change because of trading links, 
but might because of friendship links. Changes in wealth will have a positive influence on the creation of future trading links, 
and maybe a negative effect on friendship links. 

\subsubsection{What is co-evolution?}

Interactions can change the states of elements. 
%In physics, gravitational interaction changes the momentum of massive objects;  electromagnetic interactions lead to spin flips; chemical interactions may change the binding state of proteins; economic interactions change the portfolios of traders; and social interactions (exchanging gifts) may change sympathy levels.
The interaction partners of a node in a (multilayer) network  can be seen as the
local ``environment'' of that node. The environment often determines the future state of the node. In
social systems, interactions do change over time. For example, people establish new friendships
or economic relations, countries terminate diplomatic relations. The state of nodes determines
(fully or in part) the future state of the links, whether it exists in the future or not, and if it exists,
the strength and the direction that it will have.
The essence of co-evolution is expressed in the statement: 
{\em The state  of the network (topology and weights) determines the future states of the nodes. 
The state of the nodes determines the future states of the links of the network.}\\

Formally, co-evolving multilayer networks can be written as 
\begin{eqnarray}
\frac{d}{dt} \sigma_i (t) &\sim&  F \left( M_{ij}^{\alpha} (t), \sigma_j (t) \right) \nonumber  \\
&{\rm and}& \nonumber  \\
\frac{d}{dt} M_{ij}^{\alpha} (t) &\sim& G\left(   M_{ij}^{\beta} (t), \sigma_j (t)   \right) \quad . 
\label{coev_intro}
\end{eqnarray}
Here, the derivatives indicate ``change within the next time step'', and are not continuous derivatives. 
The first equation means that the states of element $i$ change as a function, $F$, that depends on the present states 
of $\sigma_i$, and the present multilayer network states, $M_{ij}^{\alpha} (t)$.
The function (or functional), $F$, can be deterministic or stochastic and contains all summations over 
all greek indices and over $j$. 
The first equation depicts the analytical nature of physics that characterized the past 300 years of science. 
Once one specifies $F$, and the initial conditions, say, $\sigma_i(t=0)$, the solution of the equation 
provides us with the trajectories of the elements of the system. 
In physics the interaction matrix, $M_{ij}^{\alpha} (t)$, could represent the four forces. 
Usually it only involves a single interaction type $\alpha$, that is static, fully connected, 
and interaction strength only depends on the relative distance between $i$ and $j$. 
Typically, systems that can be described with the first equation alone are not complex systems, 
however complicated they may be.

The second equation specifies how the interactions evolve over time as a function $G$ 
that depends on the same inputs, states and interaction networks. 
$G$ can be a deterministic or stochastic function or functional.
Now, interactions evolve in time. In physics this is very rarely the case. 
The combination of both equations makes the system co-evolving and complex. 
Co-evolving systems of this type are, in general, no longer analytically solvable.
%\footnote{Except for simple examples or situations, where the timescale of the dynamics of the states is clearly different from the dynamics of the interaction networks.} 
One cannot solve these systems using the rationale of physics because the 
environment---or the boundary conditions---specified by $M$, change as the system evolves. 
Equations (\ref{coev_intro}) are not useful until the functions $G$ and $F$ are well specified. 
The science of complex systems often tries to identify these functions for a concrete system at hand. 
Often this is done in an algorithmic way, meaning that $F$ and $G$ can be given as ``update rules''. 

More and more data sets containing full information about an entire system are becoming available, 
meaning that all state changes and all interactions between the elements are recorded. It is becoming
technically and computationally possible to monitor cell-phone communications  on
a national scale \cite{Onnela2007PNAS}, to track all airplanes in motion, or to track all legal financial
transactions on the planet. Longitudinal data about states and interactions
can be used to visualize Eqs. (\ref{coev_intro}); 
all the necessary components are in the data at any point in time: 
the interaction networks, $M_{ij}^{\alpha}$, the states of the elements, $\sigma_i$, and all the changes 
$\frac{d}{dt} \sigma_i$ and $\frac{d}{dt} M_{ij}^{\alpha} $. 
Even though Eqs. (\ref{coev_intro}) might not be analytically solvable, it is becoming possible for more
and more situations to ``watch'' them. 
%One can now design agent-based models in the form of Eqs. \ref{coev_intro}, which allows us to make quantitative and testable predictions. 

The structure of Eqs. (\ref{coev_intro}) is not the most general. 
One generalization is to endow multilayer networks with a second greek index, $M_{ij}^{\alpha \beta}$, 
that captures cross-layer interactions between elements \cite{Kivela2014}. 
It is conceivable that elements and interactions are embedded in space and time; 
indices labelling the elements and interactions could carry such additional information, 
$i(x,t,...)$ or $\{ij \} ^{\alpha \beta}(x,t,...)$.
One can introduce memory to the elements and interactions. 
We will make use of these generalizations in the following, 
where we use complete information on all the actions and interactions 
of ten thousands of players engaged in a massive multiplayer online game, 
with their millions of interactions and state-changes. 
 
%%%%%%%%%%%%%%%%%%%%%%%%%%%%%%%%%%%%%%%%%%%%%%%%%%%%%%%
\section{A virtual society}

%%%%%%%%%%%%%%%%%%%%%%%%%%%%%%%%%%%%%%%%%%%%%%%%%%%%%%%
\subsection{The universe: the Pardus game}

Our habitat is \verb|Pardus| (http://www.pardus.at), a 
browser-based MMOG in a science-fiction setting. 
There are about  480,000 registered \verb|Pardus| players, about 16,000 active players on a given day, 
the game is online since Sep 2004, and is free of charge.
MMOGs are characterized by a substantial number of users playing together in the 
same virtual environment connected by an internet browser \cite{Bartle2004,Castronova2005}.
In \verb|Pardus| every player owns one account with one single {\em character} or avatar. 
This avatar is a pilot owning a spacecraft with a certain cargo capacity, 
moving around in a virtual universe to produce and trade commodities and products, socialize, engage in social activities, 
wage wars, engage in administration, and much more. 
An important component of \verb|Pardus| is that the actions and interactions of players are strongly 
driven by social factors such as friendship, cooperation, and competition. 
There is no explicit goal in \verb|Pardus|, nor are there forbidden moves 
(with a few exceptions concerning decent behavior). 
\verb|Pardus| is a {\em virtual world}  with a gameplay based on socializing and role-playing, 
with massive interactions between players' avatars, and with interactions with a non-player environment.
%\begin{figure}[t]
%    \begin{center}
%        \includegraphics{szell2009pardussoc1images/activecharacters-eps-converted-to.pdf}    
%    \end{center}
%    \caption{Evolution of number of active characters in the game universes. The large increase of players in Orion between days $\approx\,800$ and 1,000 is due to ad campaigns after  October 2006. At day 1,000 (dotted line) Artemis and Pegasus were opened. Some thousand players abandoned their Orion characters focusing on their new Artemis characters. This  explains the mirrored development in these two universes after day 1,000.}
%    \label{fig:activecharacters}
%\end{figure}
%
Players simultaneously engage in three forms of ``life'':

{\em Economic life.} They produce, distribute and trade goods and services, and 
make profit from economic activity. They spend money on goods and services, 
such as new space ships, equipment, and consumables. Status symbols play a big role 
in the purchase and consumption of goods.
{\em Social life.} Players communicate and share information to organize in social structures and collective actions, 
be it social, political, legal, hostile, or economical. Often actions are driven by the wish to accumulate social status through 
one of many forms of earning recognition. 
{\em Exploratory life.} Players explore their universe. They produced maps of the universe and it resources, 
classified its game-specific fauna.  Some even have engaged in research on the ``physical nature''  of the game.
    
While playing, players form groups of different sizes and structures. These include political parties, 
special interest groups, criminal gangs, cartels, banks, courts, self-defence groups, or armies.  
They also organize in clubs that are called {\em alliances}.

Daily database backups are recorded, and 
are available for more than a decade, starting from Sept 2005. 
These backups contain longitudinal information of the actions and properties (states) of the avatars, 
as well as all the interactions between them, and between them and the environment. 

\subsubsection{The census of avatars}

{\em Age and nationality.}
In a poll taken 
in 2005, 5\% reported an age below 15 years, 18\% between 15--19, 34\% between  20--24,  
23\% between  25--29, and  20\% are older than 29 years. The distribution of player 
nationalities was estimated as follows: US 40\%, UK 14\%, 
Canada 5\%, Austria 4\%, Germany 4\%, Australia 4\%, other 29\%.

{\em Lifetimes of characters.}
Next to automatic deletion after 120 inactive days, every player can delete her account at any time. 
Of all characters, about $8\%$ have a lifetime of 0 days, i.e. they delete themselves on the same day they signed up. 
About $13.4\%$ of all characters become inactive after their first day. 
The probability that players play more than 120 days is about $0.7$.
More than $31\%$ of all deletions are self-induced. 

{\em Gender composition.}
When signing up for the first time, players have to choose between a male and female character. 
The decision is irreversible.
Depending on gender, a male or female avatar is displayed in various places and occasions in the game. 
About $90\%$ of all characters are male.  

%\begin{figure}[t]
%\begin{center}
%\includegraphics[width=0.35\textwidth]{plots/pardusspacefarm.jpg}
%\end{center}
%\caption{What a Pardus player might think of his home in the universe.
%}
%\label{gene}
%\end{figure}

\subsubsection{The structure of the universe}
\begin{figure}[t]
 \begin{minipage}[c]{0.5\textwidth}
       \includegraphics[width=0.52\textwidth]{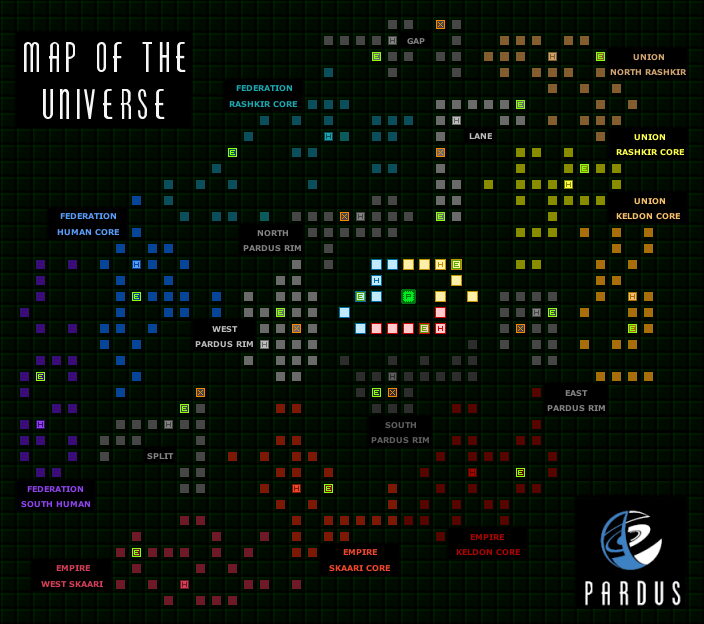}   
         \includegraphics[width=0.46\textwidth]{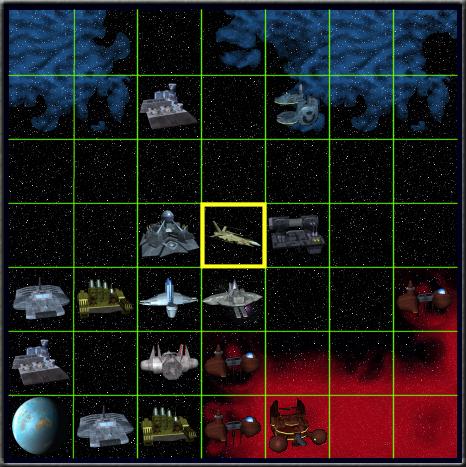}    
  \end{minipage}\hfill
  \begin{minipage}[c]{0.45\textwidth}
 %   \begin{center}
 %    \end{center}
    \caption{(left) Map of a universe. Squares are sectors consisting of 
    15$\times$15 fields. Colors indicate clusters. Sectors are connected by wormholes  (streets, not shown).
    (right) Space chart.  7$\times$7 field segment of the universe visible on the players' navigation screen; 
    the current position is in the center (yellow box). 
    Clicking on another field moves the ship to that location.
    From \cite{Szell2010msd}.
    }
    \label{fig:pardusuniverse}
\end{minipage}  
\end{figure}

Space in \verb|Pardus| is two-dimensional. The universe is divided into 400 {\em sectors}, 
Fig. \ref{fig:pardusuniverse} (top), each sector consisting of about 15$\times$15 {\em fields},  
the smallest units of space. 
They form a square grid on which ship movement is possible by clicking on the 
desired destination field within the {\em space chart}. This chart is a 7$\times$7 field 
segment of the universe, visible to the player with their current position located in the central field, 
Fig.~\ref{fig:pardusuniverse} (bottom). 
Moving between nearby sectors (think of them as a village or city) is possible by tunnelling 
through {\em wormholes}, which play the role of roads. 
A collection of about nearby 20 sectors is called a {\em cluster}. 
Typical spatial movements and the activities of avatars are usually 
confined to one cluster for several weeks or longer.

{\em Action Points -- the unit of time.}
Many game actions carried out by a  player (trade, travel etc.) cost a certain amount of so-called \emph{Action Points} (APs). 
These points can never exceed a maximum of 6,100 APs per avatar. 
For avatars owning less APs than their maximum, every six minutes 24 APs are automatically regenerated, 
i.e. 5,760 APs per day. Once a player's character is out of APs, she has to wait to be able to play on. 
As a result, the typical player logs in once per day to spend all APs on several activities within a few minutes. 
This makes APs the game's unit of time; it is the most valuable factor. 
Players that use APs most efficiently can experience the fastest progress or earn the highest profits. 
Social activities, such as communication, do not reduce APs. 
Highly involved players spend a lot of real time on socializing and on planning their future moves.

\subsubsection{Trade and economy}
The \verb|Pardus| currency unit is the so-called \emph{credit}. 
It is not convertible to real currencies. Every player starts life with 5,000 credits. 
Since most objects, such as ships, ship equipment, buildings, are traded in credits, 
it is of fundamental interest to earn money. 
There exist a number of possibilities to do this, usually through participation in the economy.
The richest players own hundreds of millions of credits. 
There exist over 30 commodity types. Some of these are renewable and exist in the environment. 
These ``raw materials'' can be mined, for example, gas from nebulas or ore from asteroids. 
Most commodities, however, are processed from more basic ones in player-owned firms. 
For example, a brewery manufactures expensive liquor out of cheap energy, water, food, and chemical supplies. 
Every player has the possibility to found a small number of such firms. 
Production chains follow a fixed production tree, and coordination of several players is needed to establish 
a sucessful industry. Most end-products at the top of the production tree are usable commodities. 
For example, manufactured drugs may be consumed to create APs, or 
droid modules can be installed for powerful building defences against hostile attacks. 
Needs generated by the society are the driving force behind the development of industries.
Besides player-owned firms there are game-owned sites that trade and consume commodities. 
Prices are exclusively determined by local supply and demand: when commodities are abundant, prices are low, 
if they are rare, prices rise.
Players face an economic life that is known from the real world. It involves collaboration, 
competition, cartellization, fraud, and so on.

\subsubsection{Communication}
There are three communication channels in the game: 
The {\em Chat:} players can simultaneously communicate with many others. 
    Chat entries scroll up and disappear; they are good for temporary talks.
The {\em Forum:} messages, called {\em posts}, consist of several lines and stay for a long time. 
    Posts are organized within {\em threads}, which bundle into topics. 
The {\em Private message} (PM): a system similar to email, where messages can be sent to any other player.
    The PM content is only seen by sender and receiver. 
    PMs always have exactly one recipient.  
    %A daily total of about $10,000$ PMs are exchanged. 
%
These communication channels can be used independently from game-mechanic states, 
such as ship's location, wealth, etc. In the following we focus on PMs, and 
call a PM between two players a ``communication event''.

\subsubsection{Friends and enemies}

\begin{figure}[t]
    \begin{center}
        \mbox{(a)}\hspace*{-0.5cm}\includegraphics[width=0.4\textwidth]{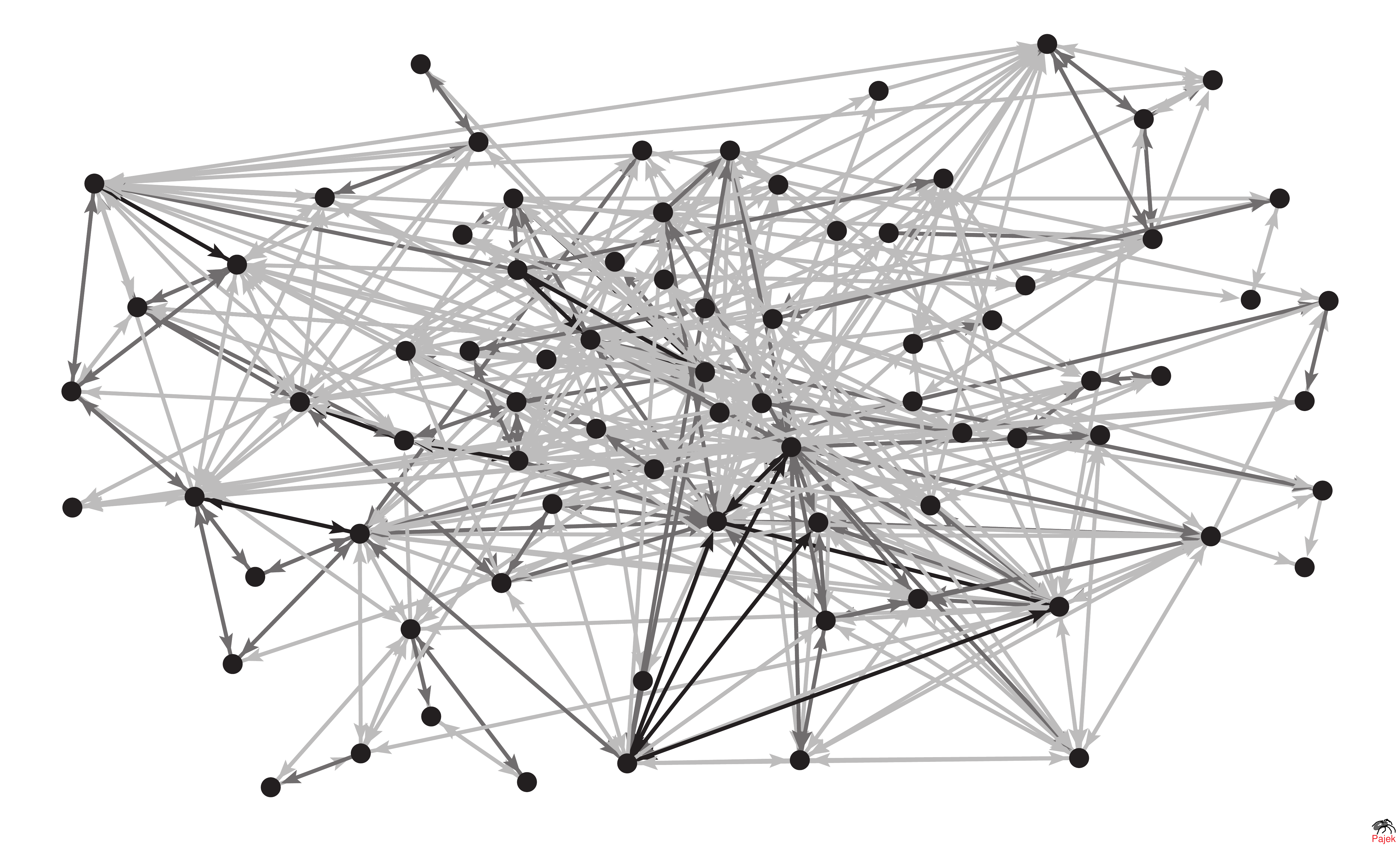}
        \mbox{(b)}\hspace*{-0.5cm}\includegraphics[width=0.4\textwidth]{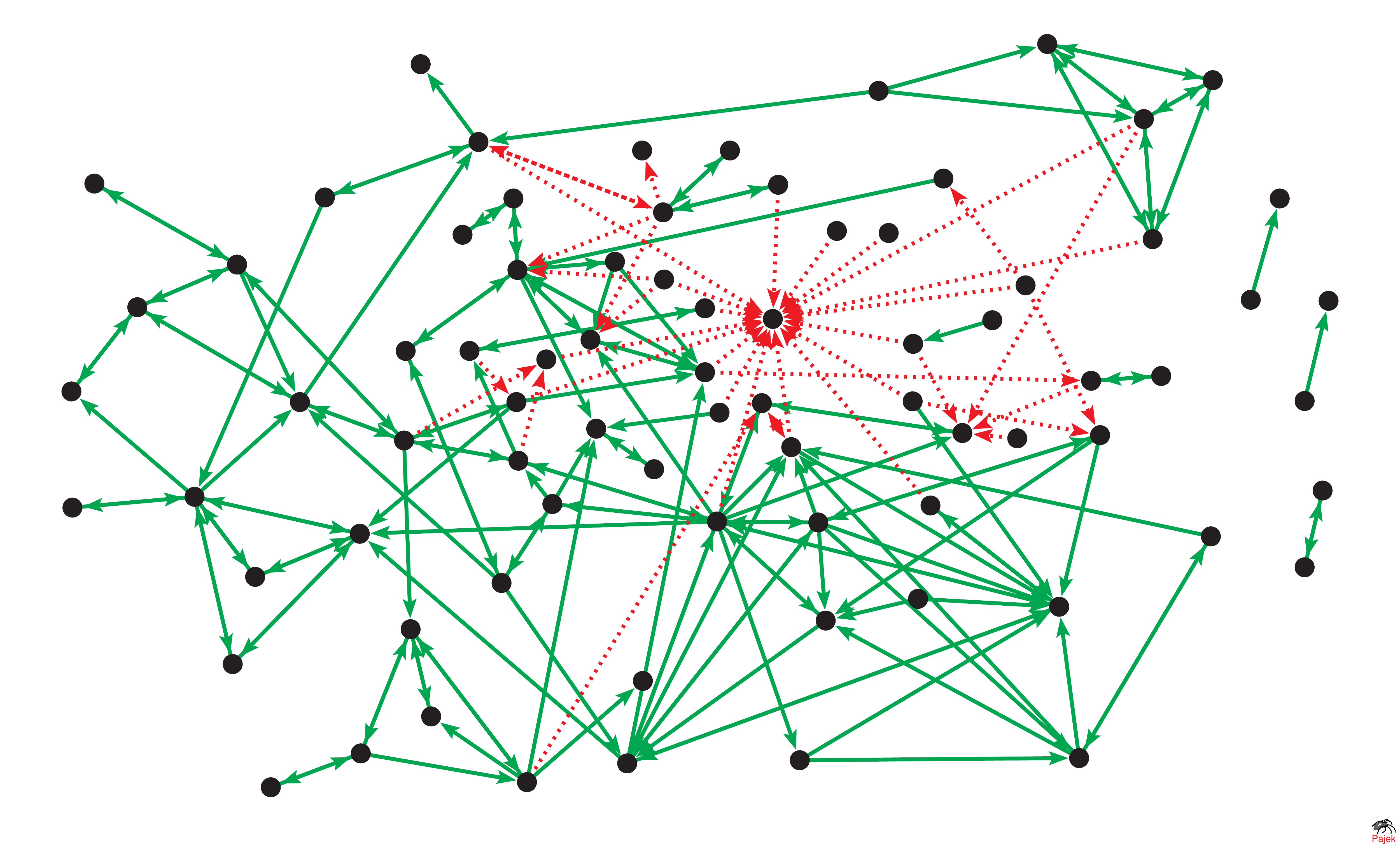}
    \end{center}
    \caption{
    	(a) Accumulated communications over 445 days between 78 randomly selected individuals in the early universe, $\sum_{t=\tau}^{\tau+445} M^{\rm comm}_{ij}(t)$. Light gray, gray, and black correspond to 1--10, 11--100, and 101--1000 PMs sent, respectively. 
        (b) Friend (green, solid), $M^{\rm friend}_{ij}(445)$, and enemy (red, dashed), $M^{\rm enemy}_{ij}(445)$, 
        relations on day 445 between the same individuals. One pretty hated guy is visible. From \cite{Szell2010msd}.
    }
    \label{fig:pajeknet}
\end{figure}

For a small amount of APs, players can mark others as their {\em friend} or {\em enemy}.
This can be done for any reason. Marked characters are added to the markers personal 
{\em friends} or {\em enemies lists}. Every player also has a personal {\em friend of} 
and {\em enemy of} list, displaying all players who have marked them as friend or enemy, respectively. 
When marked or unmarked as friend or enemy the player is informed. 
One can mark others as either friend or enemy, not both. 
Lists are private, meaning that no one except the marking and marked players have 
information about their ties. 
%It is not possible to see second degree neighbors (e.g. friends of friends) or the number of ties another player has.\footnote{On 2008-08-24 the {\em profile} feature of the game was extended, allowing players with Premium accounts to publicly display their numbers of friends or enemies. Since this feature was introduced at the very end of our last measured data (2008-09-01) and only a negligible proportion of players are making use of it, it is irrelevant here.} Note that this is in contrast to many online social networking services such as \verb|Facebook|, where usually second degree neighbors and number of friends are visible.
In this respect the \verb|Pardus| system does not introduce a bias toward accumulation of friends, 
and represents a more realistic social situation, where social ties are not immediately publicly accessible, 
but have to be found out by communication, or observation of others.
The friends and enemies lists also serve game-mechanic purposes: friends/enemies are 
included/excluded for certain actions. For example, enemies of building owners are not able to 
use services in the buildings. 
Friend and enemy markings need not reflect affective relations, 
they rather indicate a degree of cooperativeness.
Friend and enemy relations as well as communication events (PM) 
are temporal networks, Fig. \ref{fig:pajeknet}. We denote them by $M^{\rm friend}_{ij}(t)$, 
$M^{\rm enemy}_{ij}(t)$, and $M^{\rm comm}_{ij}(t)$, respectively. 

\subsubsection{Performance measures of players---``states''}

Players $i$ are characterized by a number of time dependent states, $\sigma_i(t)$, that may change as 
a result of interactions with others. 
These states can be achievement-factors that quantify various skills of players.
The efficiency in harvesting natural resources is quantified by the {\em farming skill}, $\sigma_i^{\rm farm}(t)$. 
Other performance measures are {\em combat skill}, $\sigma_i^{\rm comb}(t)$, that quantifies fighting skills, 
and the {\em experience points}, $\sigma_i^{\rm XP}(t)$, that keep a record of fighting and other activities.
Players may become members of political {\em factions} (parties), 
which sometimes engage in large-scale conflicts (wars).
{\em Faction rank}, $\sigma_i^{\rm fr}(t)$, is a measure of influence within a faction: 
above a certain threshold, the faction rank grants the privilege to take part in  
decisions on war or peace. 
Finally, players are characterized by a certain wealth level, $\sigma_i^{\rm wealth}(t)$, 
that depends strongly on their economic activities. 
Some players regard high combat skill, faction rank, wealth, or XP as their main goals in their \verb|Pardus| life.

\subsubsection{Alliances}

For various purposes players organize in social groups called {\em alliances}. 
Often players share the same interests, or cooperate in pirate groups, exploration teams, self-defense units, etc. 
Usually groups do not get larger than about 140 members. 
Note the proximity to the Dunbar number, 150 \cite{Dunbar1993}.
The game provides administration tools for officially declared alliances.
Alliances have a common cash pool, which they use for their goals, like defence or production.
Often alliances are used for economic purposes. 
There existed 161 alliances with an average size of 23 members at day 1200. 
Being a member of an alliance is a social commitment. 

%%%%%%%%%%%%%%%%%%%%%%%%%%%%%%%%%%%%%%%%%%%%%%%%%%%%%%%%%%%%%%%
\section{How do people interact?}

 \begin{figure}[t]
\begin{center}
\includegraphics[width=0.65\textwidth]{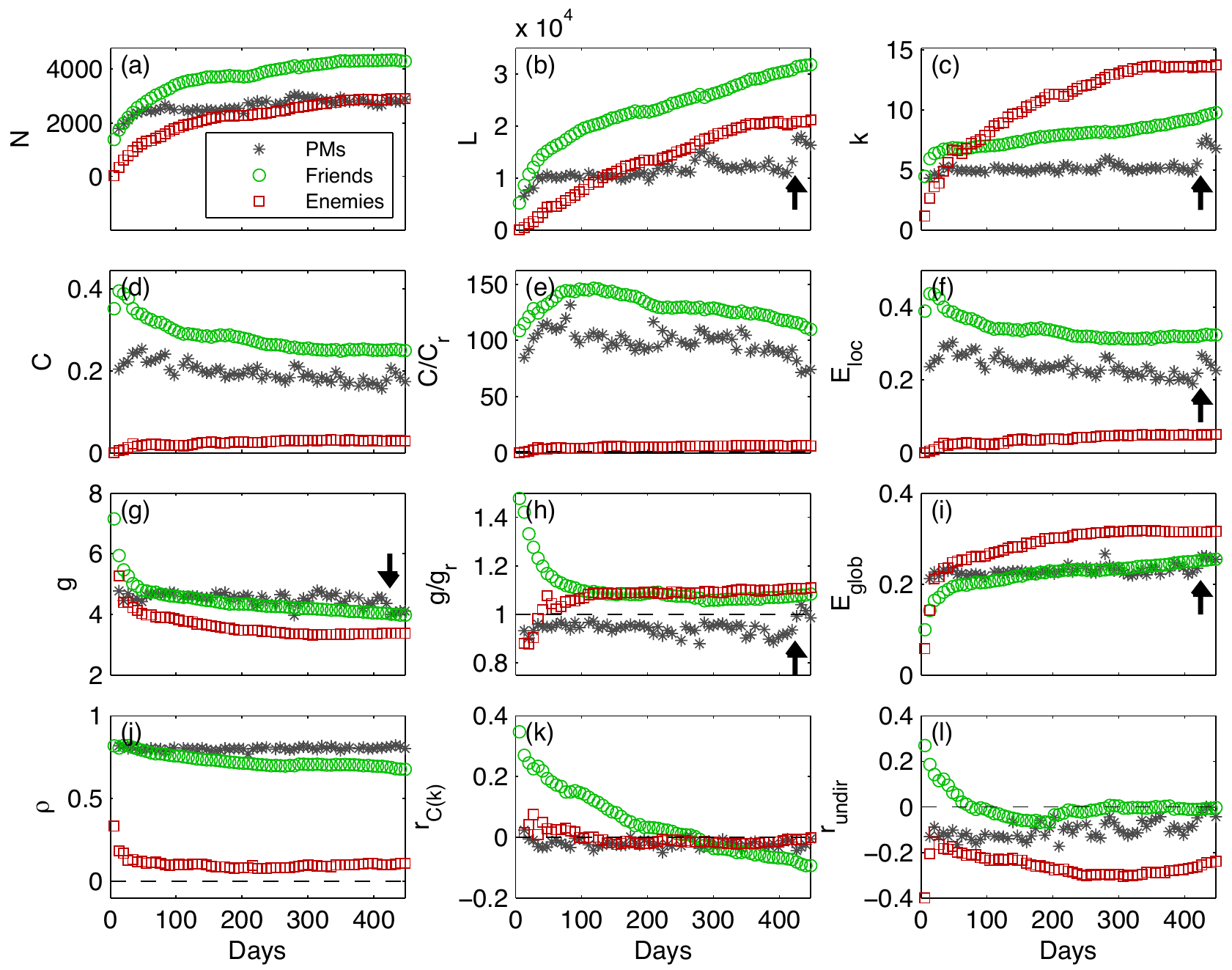}
\end{center}
\caption{Network properties over time: (a) number of nodes $N$, (b) number of (directed) links $L$, (c) average degree $\bar{k}$, (d) clustering coefficient $C$, (e) clustering coefficient $C$ divided by clustering coefficient of corresponding random graph $C_\mathrm{r}$, (f) local efficiency $E_{\mathrm{loc}}$, (g) average geodesic $\bar{g}$, (h) average geodesic $\bar{g}$ divided by average geodesic of corresponding random graph $\bar{g}_\mathrm{r}$, (i) global efficiency $E_{\mathrm{glob}}$, (j) reciprocity $\rho$, (k) assortative mixing coefficient $r_{C(k)}$, (l) assortative mixing coefficient  $r_\mathrm{undir}$. Arrows mark the beginning of a war at day 422.
 	From \cite{Szell2010msd}. 
}
\label{NWprop}
\end{figure}

One fascinating aspect of this game is that at all times, all social networks are available.
Nodes $i$ represent avatars. Links are individual social interactions of type $\alpha$ 
that happen from player $j$ to player $i$ at time $t$, $M^{\rm \alpha}_{ij}(t)$.  
This allows us to measure how people interact and organize. 
We focus on mainly six interaction types: 

{\em Communication networks.}
We consider all PM communications, usually on an aggregated (e.g. weekly) timescale. 
A weighted link pointing from node $i$ to node $j$ exists if avatar $i$ has sent at least one PM to  
$j$ within the aggregation period. The weight is the number of sent PMs. 
Figure \ref{fig:pajeknet} (a) illustrates a subgraph of 
the communication network, accumulated over 445 days between 78 randomly selected characters.

{\em Friends and enemies.}
A link is defined from $i$ to $j$ if character $i$ has marked character $j$ as friend/enemy. 
Friend/enemy markings exist until they are actively removed by the players. 
Friend and enemy networks are unweighted.
Since links of friend- and enemy links never coincide, we can see the union of friend- and enemy networks as 
{\em signed} networks. 
Figure \ref{fig:pajeknet} (b) shows a  signed friend--enemy network as observed on 
day 445. Note the cliquishness and reciprocity of friends, and a strong enemy in-hub. 

{\em Commercial (trading) networks.}
Trade networks are extracted by considering two kinds of trading between players: 
either players meet and exchange {\em credits} for commodities, 
or they visit commercial outlets of other players and buy/sell commodities or equipment there.

{\em Aggression networks.}
Bounties and Attacks are two forms of how aggression can be expressed in the game. 
 {\em Attack links} are defined as attacks carried out by one player on another (or on her commercial outlets). 
 {\em Bounty links} represent (weighted) bounties, which are amounts ({\em credits}) placed on other 
 players. Any player can collect a bounty by attacking the bountied player, or by harming his commercial outlet. 
% These networks are usually aggregated. 
 \\
 
 Other networks in the game are for example production and mobility networks. 
These do not directly constitute social interactions and will not be of interest here. 

\begin{figure}[t]
 \begin{minipage}[c]{0.5\textwidth}
    \includegraphics[width=0.49\textwidth]{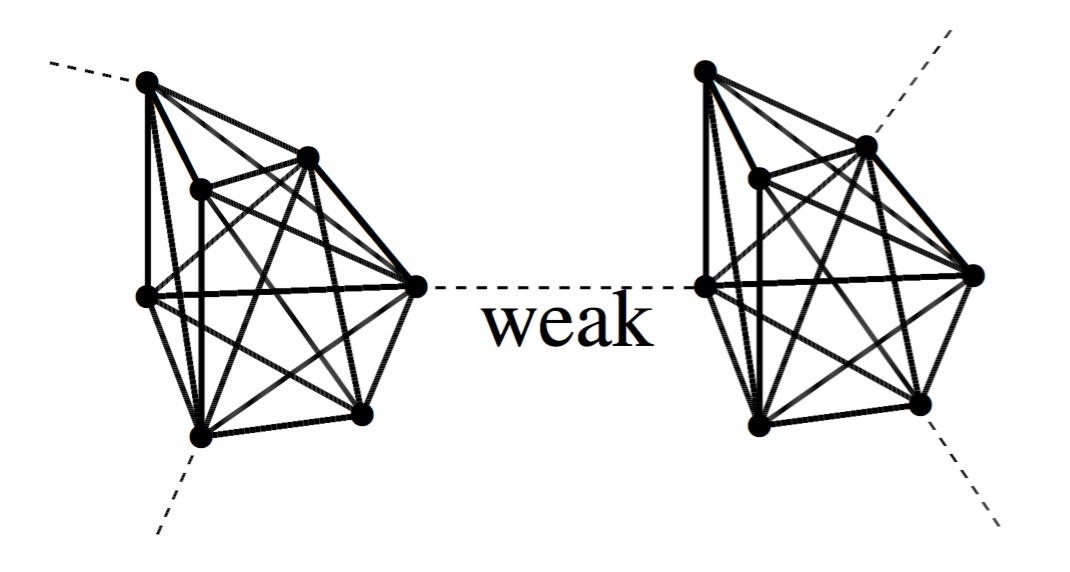}
    \includegraphics[width=0.49\textwidth]{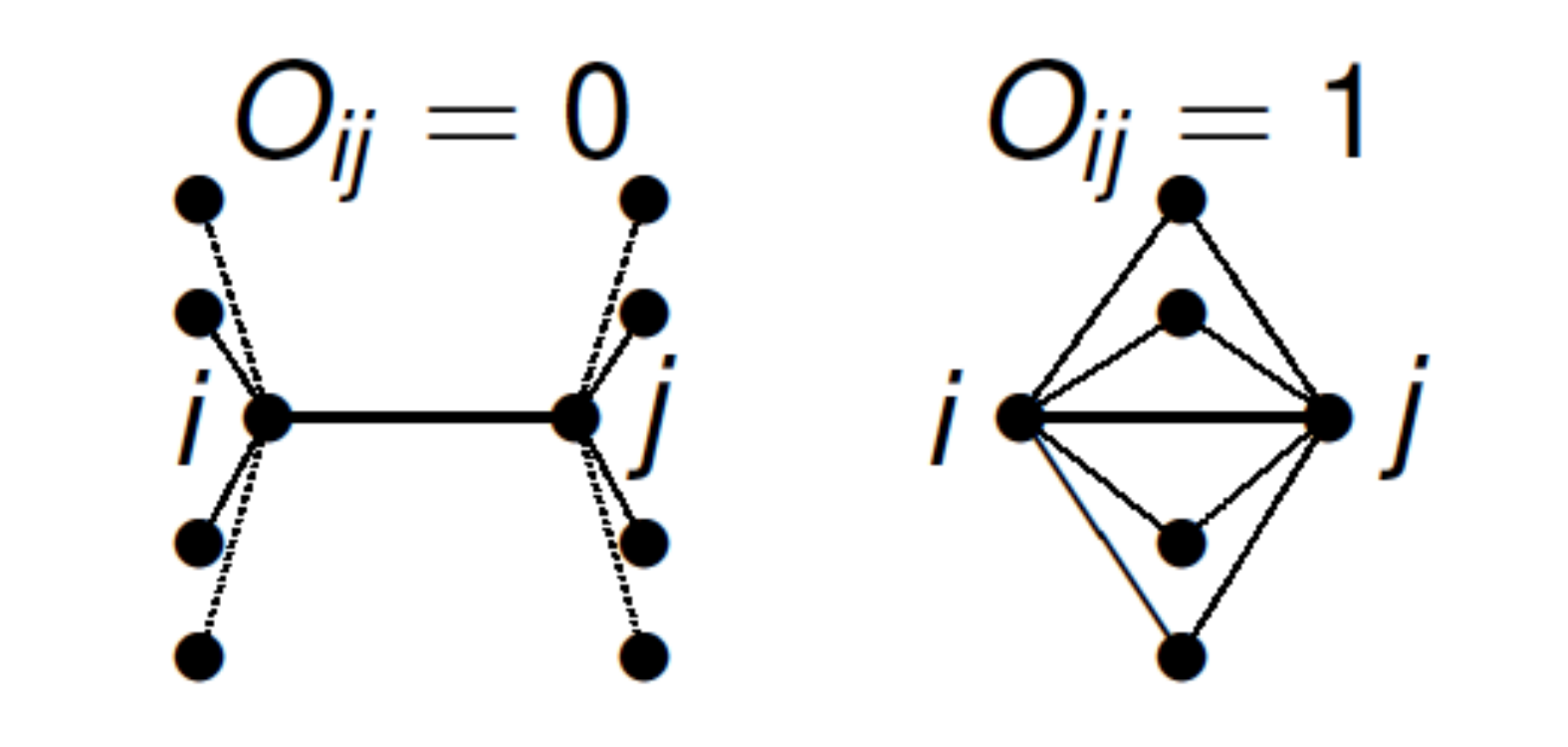}
  \end{minipage}\hfill
  \begin{minipage}[c]{0.47\textwidth}
%\end{center}
\caption{(left) Illustration of what Granovetter means by a weak tie. In network language, weak ties have high link betweenness. 
(right) Definition of overlap between two nodes $i$ and $j$.
}
\label{weaktie}
  \end{minipage}\hfill
\end{figure}

%%%%%%%%%%%%%%%%%%%%%%%%%%%%%%%%%%%%%%%%%%%%%%%%%%%%%%%%%%%%%%%%%%%%%%%%%%
%\subsubsection{Evolution of social network properties}
    
Interaction networks change over time. In Fig. \ref{NWprop} we show a number of network measures as they 
unfold for communication, friendship, and enmity during the first 422 days, clearly a period in transition. 
Networks are characterized by growing average degrees, and shrinking diameters;  
networks densify. After about 1,000 days, most measures become approximately stationary (not shown).
Real world communication \cite{Onnela2007PNAS} and friendship networks networks 
are similar to those observed in \verb|Pardus|. 
Not many real world studies exist on enmity networks---people seem to avoid to list their enemies. 
 
%%%%%%%%%%%%%%%%%%%%%%%%%%%%%%%%%%%%%%%%%%%%%%%%%%%%%%%%%%%%
\subsection{Testing a classic sociological hypothesis of social interaction: weak ties}

Given these networks, we can immediately test a classic hypothesis in sociology stated 
by Mark Granovetter in the 1970s. The so-called the {\em weak ties hypothesis} makes a statement 
about the importance of weak links, that connect communities, see Fig. \ref{weaktie}. 
It states that 
``the degree of overlap of two individual's friendship networks varies directly with the strength of their tie to 
one another'',  \cite{Granovetter1973}. 
Weak ties (for example casual acquaintances) are assumed to be important 
because they can link communities, which would otherwise be separated. 
While weak ties are {\em local bridges} between communities, strong ties (e.g., good friendships) 
are easily replaceable intra-community connections. In network language,  
weak ties are links with a high {\em link-betweenness}.
Link-betweenness of link $l_{ij}$ is 
\begin{equation} 
b_{ij} = \sum_{m \in \mathcal{N}} \sum_{n \in \mathcal{N}\setminus{\left\{m\right\}}} \frac{\rho_{mn}(l_{ij})}{\rho_{mn}} \quad, 
\end{equation}
where $\rho_{mn}$ is the number of all paths between $m$ and $n$, and $\rho_{mn}(l_{ij})$, is the number
of paths that contain the link between $i$ and $j$.
\begin{figure}[t]
%\begin{center}
 \begin{minipage}[c]{0.3\textwidth}
 \includegraphics[width=1.0\textwidth]{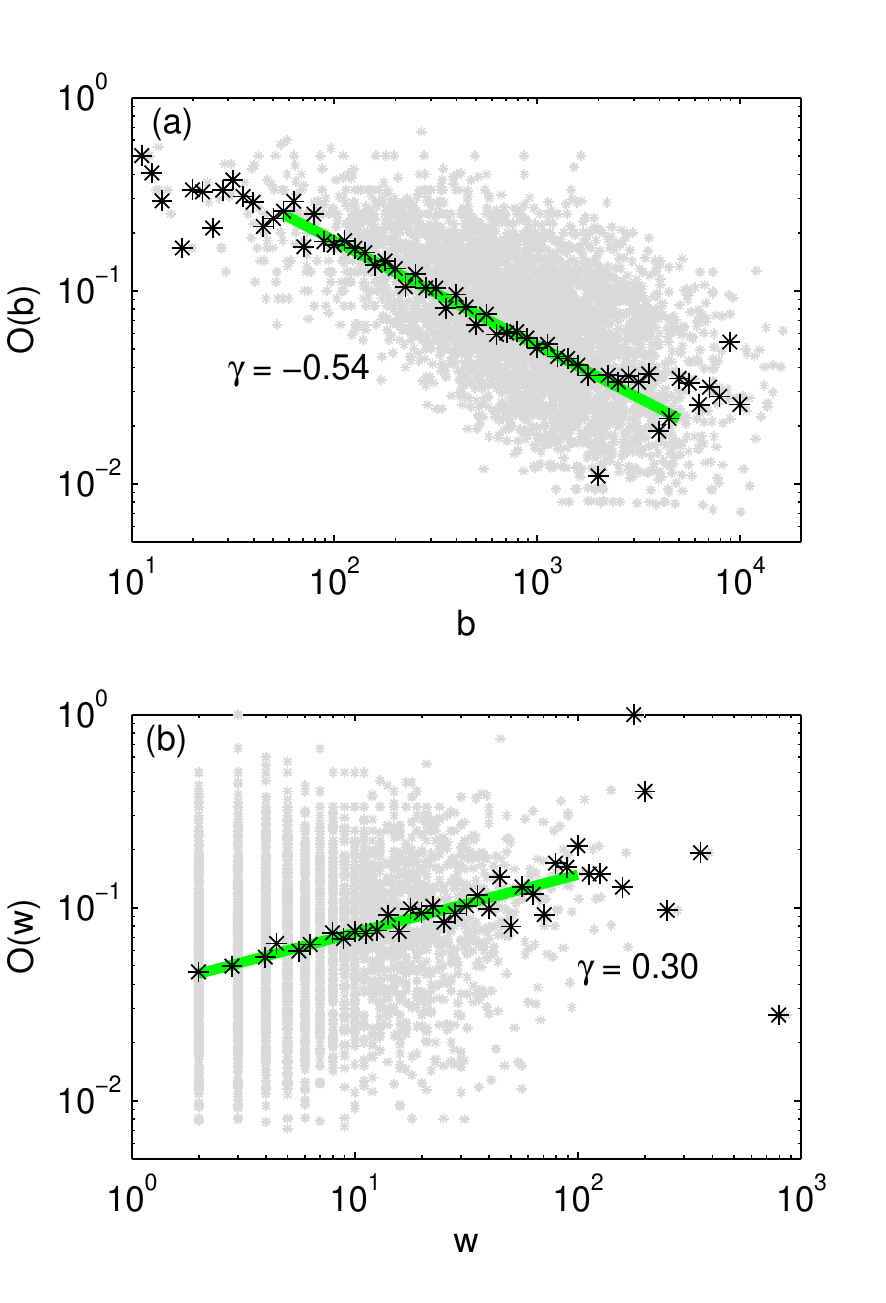}
  \end{minipage}\hfill
  \begin{minipage}[c]{0.67\textwidth}
%\end{center}
\caption{
Overlap versus (a) betweenness, and versus (b) weight in largest connected 
component  of the communication network at day 422. Grey markers show individual  overlap values of the links. 
%2,336 links (out of 6,502) have overlap $O = 0$. 
Black markers denote logarithmically binned averages, green lines are least squares fits.
From \cite{Szell2010msd}.
}
\label{fig:obow}
  \end{minipage}\hfill
\end{figure}

For testing the hypothesis, we have to clarify how to measure ``strength''. 
We define interaction strength between two individuals 
as the  number of exchanged messages $w=M^{\rm comm}_{ij}(t)$, over a given aggregation period.    
The hypothesis predicts an increasing function of overlap,  $O(w)$ , versus weight $w$. 
The overlap between two nodes $i$ and $j$ is 
\begin{equation}
    O_{ij} = \frac{n_{ij}}{(k_i-1)+(k_j-1)-n_{ij}} \quad , 
\end{equation}
where $n_{ij}$ is the number of common neighbors of the nodes. 
The expected relationship is clearly realized  for communication networks, see Fig. \ref{fig:obow} (b), 
where we find an approximate cube root law 
\begin{equation} 
O(w) = w^{0.30} \sim  \,  ^3\!\!\! \sqrt{w} \quad . 
\end{equation}
%Note that this method does not encounter sampling issues as in e.g. \cite{Onnela2007}, and this result is bias free.
%A direct way of testing the weak ties hypothesis is to examine the correlation between betweenness and overlap. 
By the weak ties hypothesis, the overlap $O(b)$, as a function of betweenness, should decrease. 
Figure \ref{fig:obow} (a) confirms this prediction, and suggests an inverse square root law 
\begin{equation} 
O(b) = b^{-0.54} \sim \frac{1}{ \sqrt{b}} \quad . 
\end{equation}
These results are in agreement with real world communication networks as obtained from 
mobile phone call data \cite{Onnela2007PNAS}, and are robust across game universes and 
various accumulation times. 
The weak ties hypothesis was been tested with real small-scale social networks \cite{Friedkin1980}.
 The weak ties hypothesis is fully confirmed in the \verb|Pardus| society. 

%%%%%%%%%%%%%%%%%%%%%%%%%%%%%%%%%%%%%%%%%%%%%%%%%%%%%%%%%%
\subsubsection{How strong do people interact?---Kepler's law}

We can eliminate the overlap from the equations obtained from Fig. \ref{fig:obow},  
$b = \frac{1}{O^2} $ and $w = O ^3 $, and get 
 \begin{equation}
 	w^2 =  (1/b)  ^{3} \quad , 
\end{equation}
which immediately reminds us at Kepler's third law of the motion of planets. 
This relation is interesting in the sense that it relates interaction strength, which is a local quantity 
between individuals, with the betweenness of a link, which is a global, society-wide quantity: 
the strength of a (positively connoted) individual relation seems to be related to the structure of the entire network. 
It remains to be verified if this relation is generally true for real societies. 

%%%%%%%%%%%%%%%%%%%%%%%%%%%%%%%%%%%%%%%%%%%%%%%%%%%%%%%%%%
\subsection{Forces between avatars---Newton's law for social interactions?}

\begin{figure}[t]
\begin{center}
\includegraphics[width=0.7\textwidth]{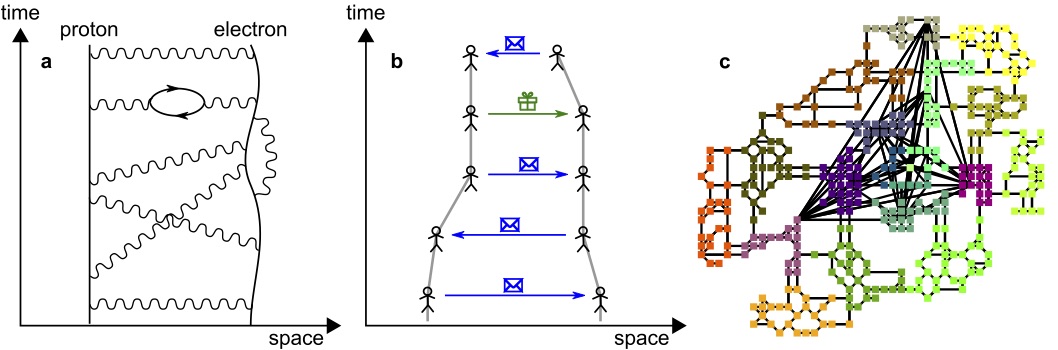}
\end{center}
\caption{Interactions mediated by exchange of particles.
	(a) Electromagnetic interaction between a proton and an electron 
	is mediated by the exchange of virtual photons \cite{Feynman1985}. 
	(b) Two players interact by exchanging messages, $M^{\rm comm}_{ij}(t)$, 
	and goods, $M^{\rm trade}_{ij}(t)$. 
	(c) Map of the universe, where nodes are sectors (cities), lines are 
	connections (wormholes or streets).
	Colors represent different regions (countries). From \cite{Thurner2014}.
}
\label{fig:Illustration}
\end{figure}

In physics there are four fundamental forces, 
the electromagnetic %force, that exists between electrically charged and magnetic objects, 
the weak force, %that acts on subatomic particles,  
the strong force, % that is responsible for the interactions between quarks, 
and  gravitation. %, that acts between objects that have a mass. 
The origin of the forces has been clarified in the 20th century. The current view is that they result  
from the exchange of virtual gauge bosons between interacting particles, see e.g. \cite{Feynman1985}. 
Electromagnetism results from the exchange of photons, see Fig. \ref{fig:Illustration} (a), 
the weak and strong force comes from the exchange of W- and Z-bosons, and gluons, respectively. 
%Gravitation is thought to be mediated by the exchange of (hypothetical) gravitons.
%By treating virtual exchange particles as excitations of a field the functional form of the interaction potential 
%can be derived \cite{Feynman1998,Wilson1974} from first principles.
%
In classical physics a force can be expressed as a negative gradient of a potential $V_{}(\mathbf{x})$
\begin{equation}
	m\, a = m\frac {d^2} {d t^2} \mathbf{x} = -\nabla V_{}(\mathbf{x})  \quad .
\label{eq:Newton}
\end{equation}
If a central force is present, meaning that only the distance $r$ between two bodies matters, 
the potential becomes a function of $r$,  $V_{}(\mathbf{x})=V_{}(r)$, and we get 
\begin{equation}\label{eq:NewtonRadial}
	ma = -\frac{d}{d r}\left[V_{}(r)+V^0(r) \right]  \quad , 
\end{equation}
where $V^0(r)$ is an effective-potential. 
%, which can arise for example by the presence of  angular momenta  $V^0(r) = \frac{L^2}{2mr^2}$ (in cylindrical coordinates).
Similar to physics, many human interactions are also based on exchange. 
Exchanged objects can be messages, goods, money, presents, promises, aggression, bullets, and so on. 
In Fig. \ref{fig:Illustration} (b) we schematically show the trajectories of two individuals, 
who exchange messages and a gift; as a result their relative distance reduces over time.
Up to now it was not possible to determine if exchange events generate effective attractive or repulsive 
forces that influence relative motion. This is due to the lack of simultaneous information on 
exchange events and the trajectories of the involved individuals.   
%The existence of potentials causing and influencing the relative motion of humans is not new and has been conjectured in \cite{Helbing1995}. New technologies in data acquisition and storage are about to change the experimental situation. 
This situation is about to change. Data from mobile phone networks, email networks, 
and online social networks show that the probability for interaction events 
decays with distance as an approximate power law, $P\sim r^{-\gamma}$ \cite{Lambiotte2008,Krings2009,Levy2014,Liben-Nowell2005,Adamic2005,Backstrom2010,Scellato2011,Cho2011,Grabowicz2014}, with exponents ranging from  
 $\gamma=0.83$ \cite{Cho2011} to $\gamma=2.0$ \cite{Lambiotte2008,Krings2009,Levy2014}.
%Few empirical studies go beyond the analysis of the relation between distance and social dynamics.  
%It was found  that humans mostly travel towards others with whom they share a (weak) tie \cite{Phithakkitnukoon2012}. 
%In \cite{Cho2011} human mobility is described as a combination of a periodic daily pattern (from ``home''  to  ``work'') 
%and long-distance travels which are influenced by social networks. This model was successfully applied to mobile phone data 
%and the social networks {\em Brightkite} and {\em Gowalla} \cite{Cho2011}. 
% In a slightly wider context, the role of the number of interaction partners in spacial iterated prisoner's dilemma games on regular lattices has been shown to have an influence on the size of collaborative clusters \cite{new1,new2}. There however the  players are static, and do not represent real individuals.

\begin{figure}[t]
%\begin{center}
 \begin{minipage}[c]{0.4\textwidth}
\includegraphics[width=0.9\textwidth]{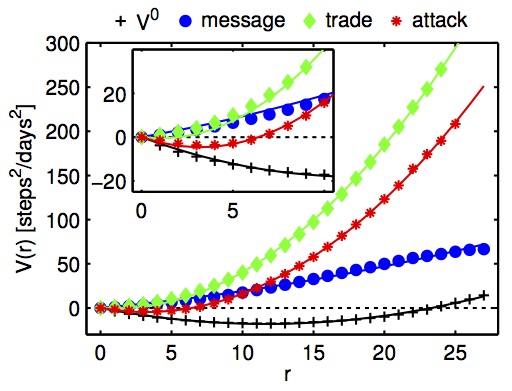}
  \end{minipage}\hfill
  \begin{minipage}[c]{0.6\textwidth}
%\end{center}
\caption{Interaction-specific potentials for messages, trade, and attacks.
	Solid lines are least-squares fits to a harmonic potential in Eq. (\ref{eq:HO}). 
	$V^0$ is a result of the background motion of non-interacting pairs of players. 
	The inset is a blow-up for small distances. The potential for attack shows a 
	minimum at $r^{\rm attack}\sim3$. From \cite{Thurner2014}.
	%For all fits (lines) the explained variance is $R^2>0.99$.
}
\label{fig:V_r}
  \end{minipage}\hfill
\end{figure}
The game is constrained to a 2-dimensional virtual universe that is partitioned into 400 sectors (cities)
that are connected by 1,064 local, and 77 long-range connections (roads), see Fig. \ref{fig:Illustration} (c). 
Movement is not for free, long-distance travel costs more than short moves. 
Travel can be fast but it takes time; traversing the entire universe needs about three days. 
We define the distance between two neighbouring sectors as one ``step'' (network- or Dijkstra distance 1). 
Given the Dijkstra metric we have relative distances, $r_{ij}(t)$, 
between players $i$ and $j$, their relative velocities  $v_{ij}(t)$, and accelerations $a_{ij}(t)$. 
At the same time, we have the exchange densities given by the multilayer network, $M^{\alpha}_{ij}(t)$. 
We can now address the question of how social interactions between humans influence their relative motion. 
We obtain the interaction potentials by integrating Eq. (\ref{eq:NewtonRadial}) for cases 
where particular interactions are predominant. We get
 \begin{equation}
	V^\alpha(r) = \kappa_\alpha r^2 - b_\alpha r \quad .
\label{eq:HO}
\end{equation}
For details, see \cite{Fuchs2014}.
The resulting potentials for the three interaction types, communication, trade and attack are shown in Fig. \ref{fig:V_r}. 
They follow a harmonic and a linear potential, where $\kappa_\beta$ is the respective ``force constant''.
%The corresponding equilibrium distance is at $r_m^\beta = \frac {b^\beta} {2\kappa^\beta}$.
%Potentials increase with distance without signs of saturation.  
For communication this result is consistent with real-world observations \cite{Cho2011}.
For trade and attacks, players need to reduce their distance to zero so that an interaction is possible.
We see that attractive forces are due to the exchange of messages and trade, whereas 
repulsive {\em and} attractive forces arise from hostile actions (``hit and run'' strategy). 
To confirm this finding in the real world, mobile phone data could be used to perform a similar study. 

%%%%%%%%%%%%%%%%%%%%%%%%%%%%%%%%%%%%%%%%%%%%%%%%%%%%%%%%%%%%
\section{How do people organize?}
\begin{figure}[t]
%\begin{center}
 \begin{minipage}[c]{0.4\textwidth}
\includegraphics[width=0.80\textwidth]{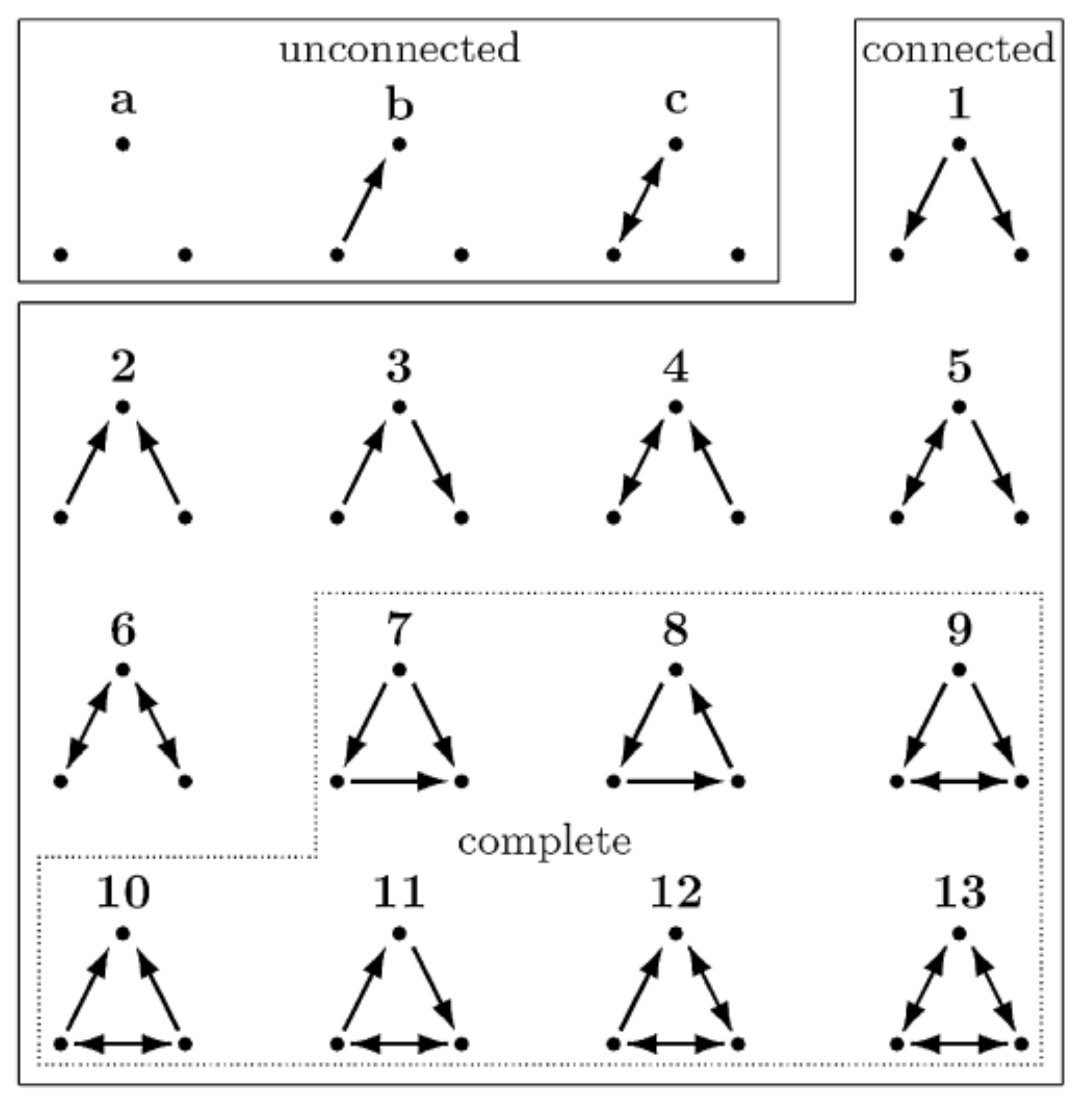}
  \end{minipage}\hfill
  \begin{minipage}[c]{0.6\textwidth}
%\end{center}
\caption{The 16 types of triads and their ids.
}
\label{triads}
  \end{minipage}\hfill
\end{figure}
%
%%%%%%%%%%%%%%%%%%%%%%%%%%%%%%%%%%%%%%%%%%%%%%%%%%%%%%%%%%%%%%%%%%%%%%%%%%
\subsection{Dynamics of the ``atoms of society'': triadic closure}

Some sociologists consider triangles (or triads), as the relation between three individuals,  
as the elementary building block of societies. 
The fractions of different types of triads within a society provide information about its structure, 
stability, and efficiency. Granovetter stated in 1973:  
``The triad which is most {\em unlikely} to occur, [\ldots] is that in which A and B are strongly linked, 
A has a strong tie to some friend C, but the tie between C and B is absent.'' 
Again he is  talking about {\em friendly} interactions. This statement means that  
one should expect over-representation of closed triads and a suppression of open ones. 
Does this mean that closed triangles should be under-represented in enmity networks? We will see. 
For directed networks, there exist 16 types of triads, see Fig. \ref{triads}. 
In dynamical terms this means that, over time, open triangles should tend to close. 
For example, in Fig. \ref{triadicclos} we would expect to see triad 6 close to form triad 13. 
Since we know networks over time, we can count transitions from open to closed triads.
The transition counts between all triad types are collected in Fig. \ref{triadicclos} (middle). 
 
%%%%%%%%%%%%%%%%%%%%%%%%%%%%%%%%%%%%%%%%%%%%%%%%%%%%%%%%%%%%%%%%%%%%%%%%%%
\subsubsection{Testing triadic closure---the triad-significance profile}

\begin{figure}[b]
\begin{center}
\includegraphics[width=0.30\textwidth]{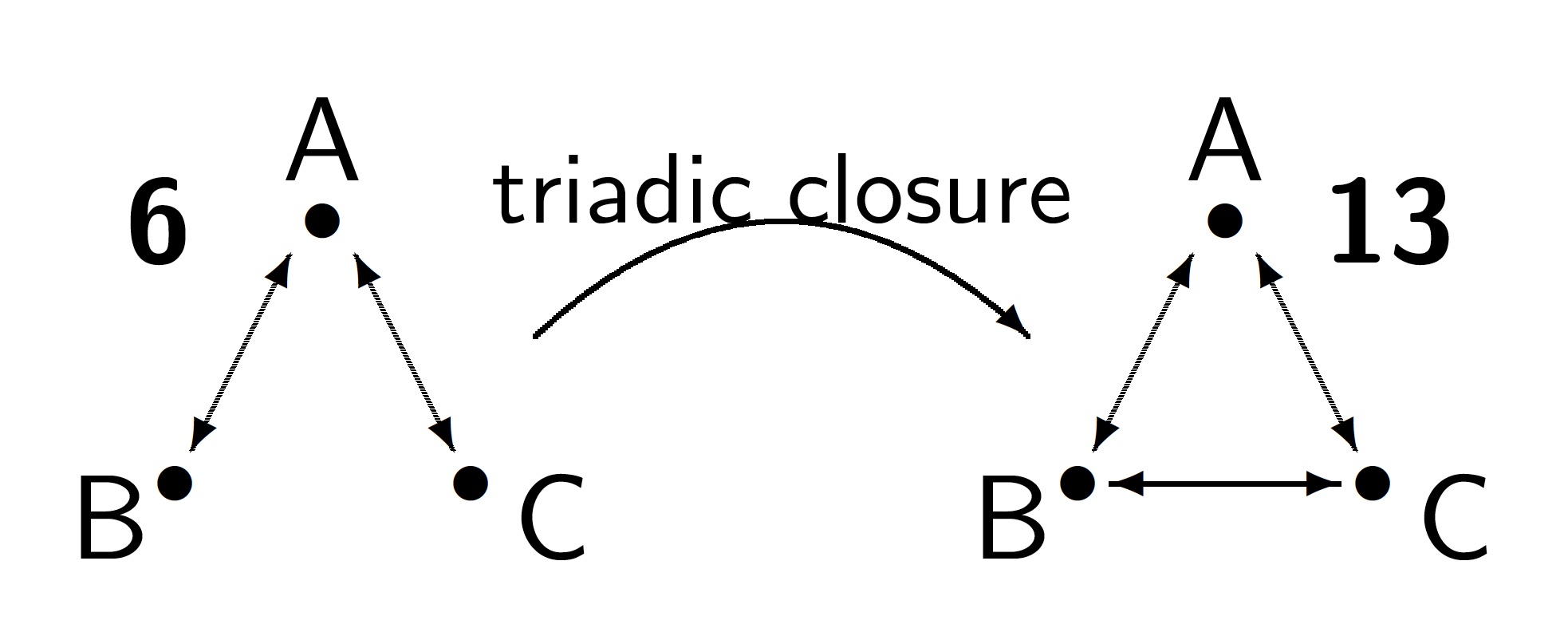}
\includegraphics[width=0.36\textwidth]{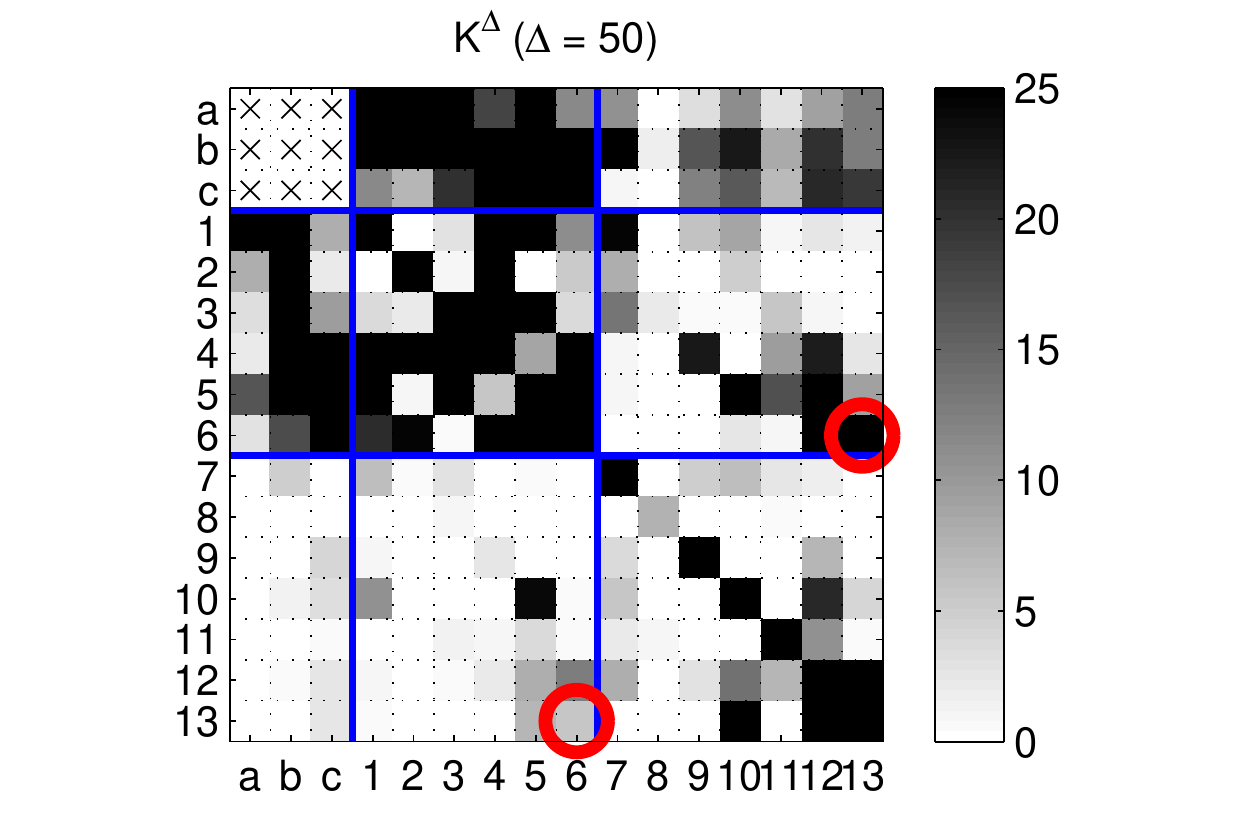}
\includegraphics[width=0.32\textwidth]{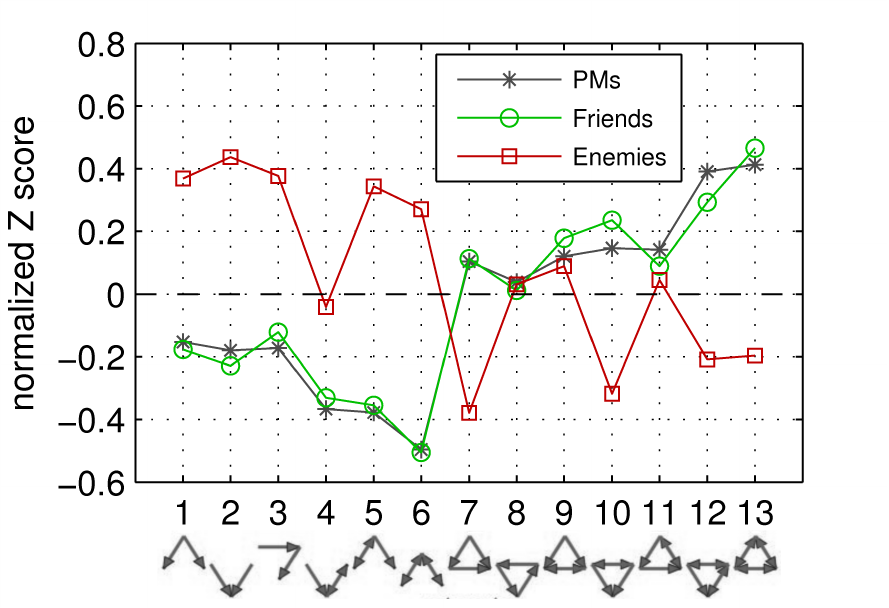}
\end{center}
\caption{(left) Illustration of triadic closure as the tendency to close open triads. 
(middle) Transition counts from one type of triad to another. 
Red circles mark the rates between $6 \to 13$, and from $13 \to 6$. 
Clearly, there is a tendency to close open triads.
(right) Triad significance profile for the three network types at day 445.
From \cite{Szell2010msd}.
}
\label{triadicclos}
\end{figure}

A way to visualize the over/under-representation of closed triads is to use the $Z$-score. 
It measures the over/under-representation of specific triads with respect to the number of 
triads that one would expect in a random graph with the same number of nodes and links. 
 The resulting numbers are collected in the  
{\em triad significance profile} shown in Fig. \ref{triadicclos} (right). 
For friendly interactions, i.e. friendship and 
communication networks, 
open triads have negative (normalized) $Z$ scores (under-represented), closed ones are positive (over-represented).
We see explicit evidence for triadic closure for friendship and communication networks in Fig. \ref{triadicclos}.
For negatively connoted ties, we find triad types 1--6  over-represented, and 7--13 under-represented in enemy networks.
Note the exceptions: triad id 4 is not clearly overrepresented, ids 9 and 11 are not clearly under-represented. 
If one wants to model social dynamics, triadic closure must be taken into account. 
%If it is ignored, wrong transitions are to be expected and any results that aim at modeling interactions in a predictive way will be severely wrong. 
We will see in the next section that triadic closure alone is sufficient to explain a number of  
statistical facts in networks associated with positive interactions. 
Triadic closure turns out to be a driving force for social network formation, well beyond Granovetter's initial ideas. 

%%%%%%%%%%%%%%%%%%%%%%%%%%%%%%%%%%%%%%%%%%%%%%%%%%%%%%%%%%%%%%%%
\subsection{Taking triadic closure seriously---understanding social multilayer network structure}

We now show that triadic closure is able to explain three important statistical facts in three 
positive networks (communication, trade, and friendship): 
the degree distribution is an approximate power law with exponent $q$, 
the node attachment kernel is an approximate power law with exponent $\gamma$, 
and the clustering coefficient as a function of node degree is an approximate power law with exponent $\beta$.
\begin{figure}[t]
%\begin{center}
\begin{minipage}[c]{0.4\textwidth}
\includegraphics[width=0.8\textwidth]{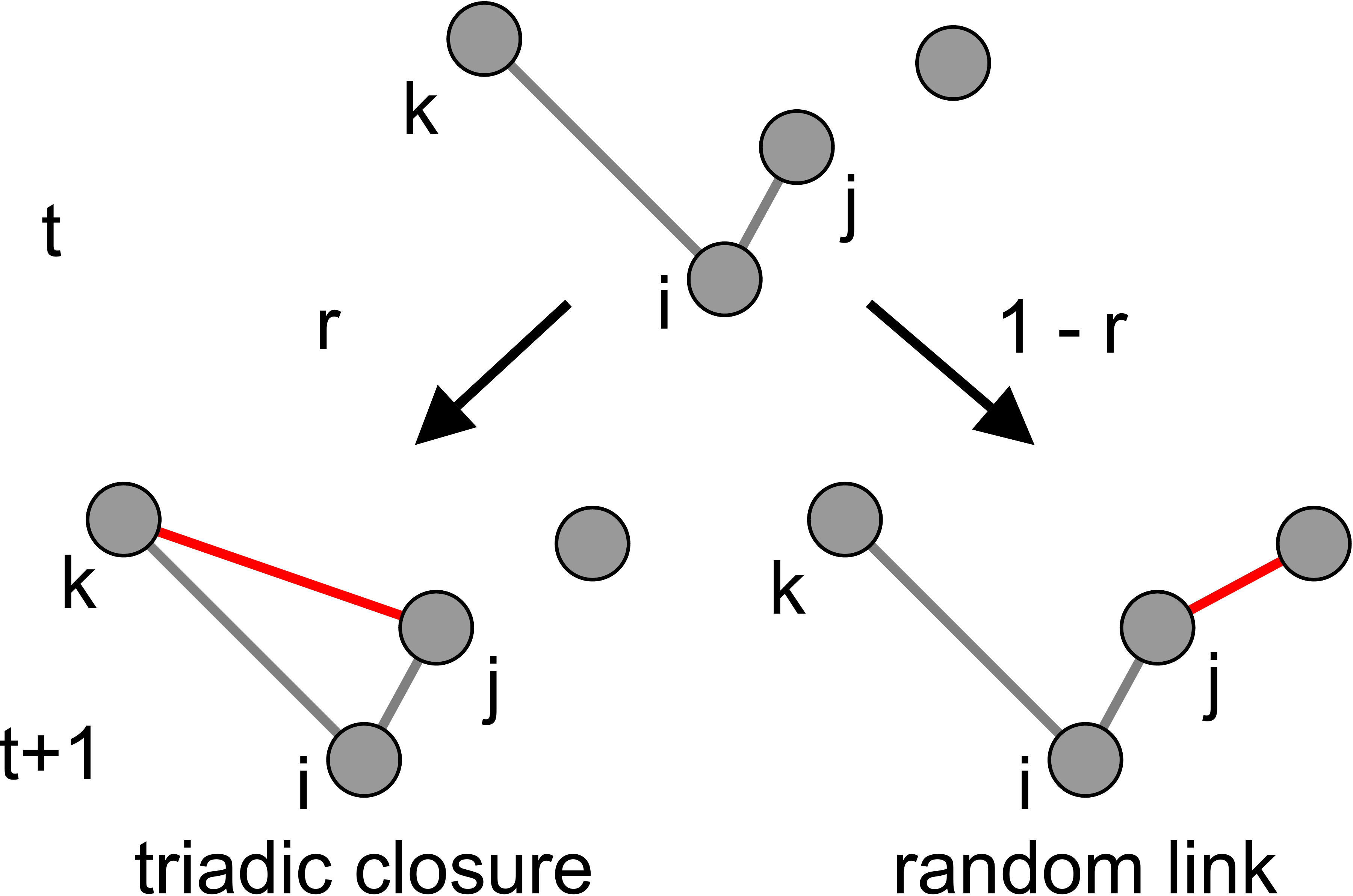}
  \end{minipage}\hfill
  \begin{minipage}[c]{0.6\textwidth}
%\end{center}
\caption{Simple triadic closure model. From time $t$ to $t+1$ a new link is created. 
With probability $r$ it closes a triangle, with $1-r$ it does not. From \cite{Klimek2013}.
}
\label{triadicmodel}
  \end{minipage}\hfill
\end{figure}
%
%The model  includes the addition and removal of nodes.

\begin{figure}[t]
 \begin{center}
 \includegraphics[width=0.95\textwidth]{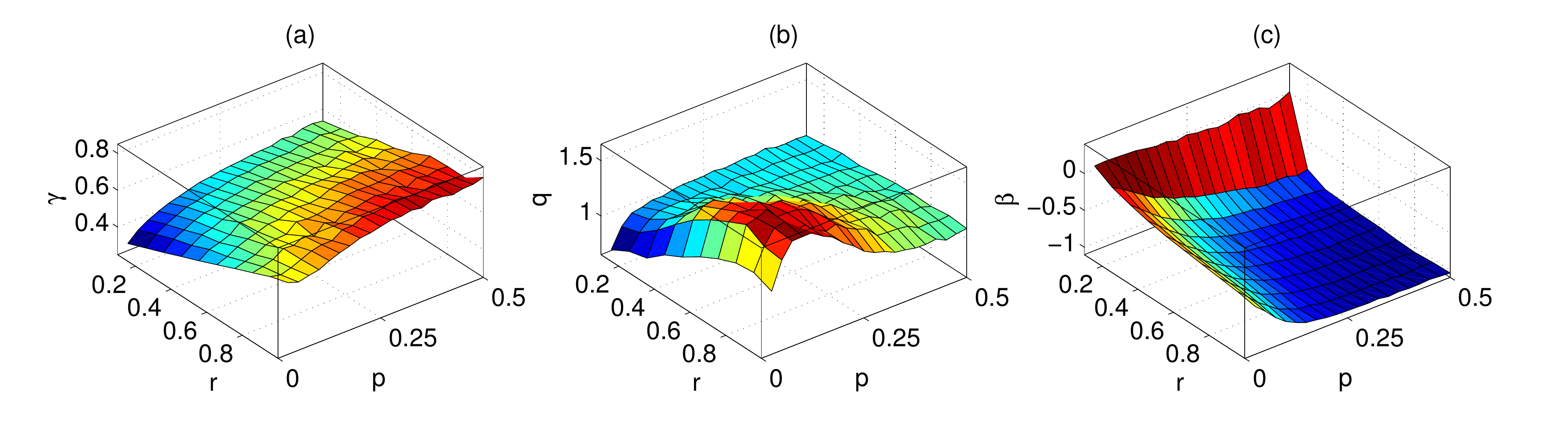}
  \end{center}
 \caption{Dependence of scaling exponents $\gamma$, $q$, and $\beta$ on model parameters $p$ and $r$. 
 (a) $\gamma$ increases in $p$ or $r$, and is confined to $0<\gamma<1$. 
 (b) $q$ is large for small $p$ and large $r$; it approaches 1 for large $p$. 
 (c) $\beta$ is close to zero for small $r$, and approaches $\beta = -1$, for large values of $p$ and $r$.
 $N=10^3$, $m=0$, averages over 500 realizations for each $(p,r)$.
 From \cite{Klimek2013}.
 }
  \label{slopes}
 \end{figure}

We use a simple model \cite{Klimek2013} that is shown in Fig. \ref{triadicmodel}. 
The network is initialized with $N$ nodes, each having one link to a randomly chosen node.
The dynamics is completely specified by the iteration of the following steps, 
\begin{enumerate}
\item Pick a node $i$ at random. If $i$ has less than two links, create a link between $i$ 
and any randomly chosen node, and continue with step (iii). 
If $i$ has two or more links, choose one of its neighbors at random, say, node $j$, and continue with step (ii).
\item With probability $r$ (triadic closure parameter), create a link between $j$ and another randomly 
chosen neighbor of $i$, say $k$. With probability $1-r$, create a link between $j$ and a node  randomly chosen 
from the entire network, see Fig. \ref{triadicmodel}.
\item With probability $p$ (node-turnover parameter) remove a randomly chosen node from the network along with all its links, and introduce a new node linking to $m$ randomly chosen nodes. Then continue with time-step $t+1$.
\end{enumerate}

For $p>0$ nodes have a finite lifetime, which implies that the network reaches a stationary state,  
where the total number of links $L(t)$, and the network measures  $\Pi(k)$, $P(k)$, and $c(k)$ 
fluctuate around stationary levels. 
The model is a variation of the model proposed in \cite{Davidsen2002}, which appears as the special case for $r=1$, 
in the above protocol. For similar models, see also \cite{Vazquez2003,Holme2002, Toivonen2006,Kumpula2007}.
The model is completely specified by four parameters, $N$, $r$, $p$, and $m$, all of which   
can be read off from the actual game data, i.e. $M^{\alpha}_{ij}(t)$. 
In this sense, the model does not have a single {\em free} parameter. 
All parameters, including the node- and link-generation rates
are obtainable from the game for the three interaction types. 
The model can now be simulated; the emerging networks are analyzed with respect to 
the degree distribution, the attachment kernel, and the clustering coefficients.

\subsubsection{Characteristic exponents}

\begin{figure}[b]
\begin{center}
\includegraphics[width=0.50\textwidth]{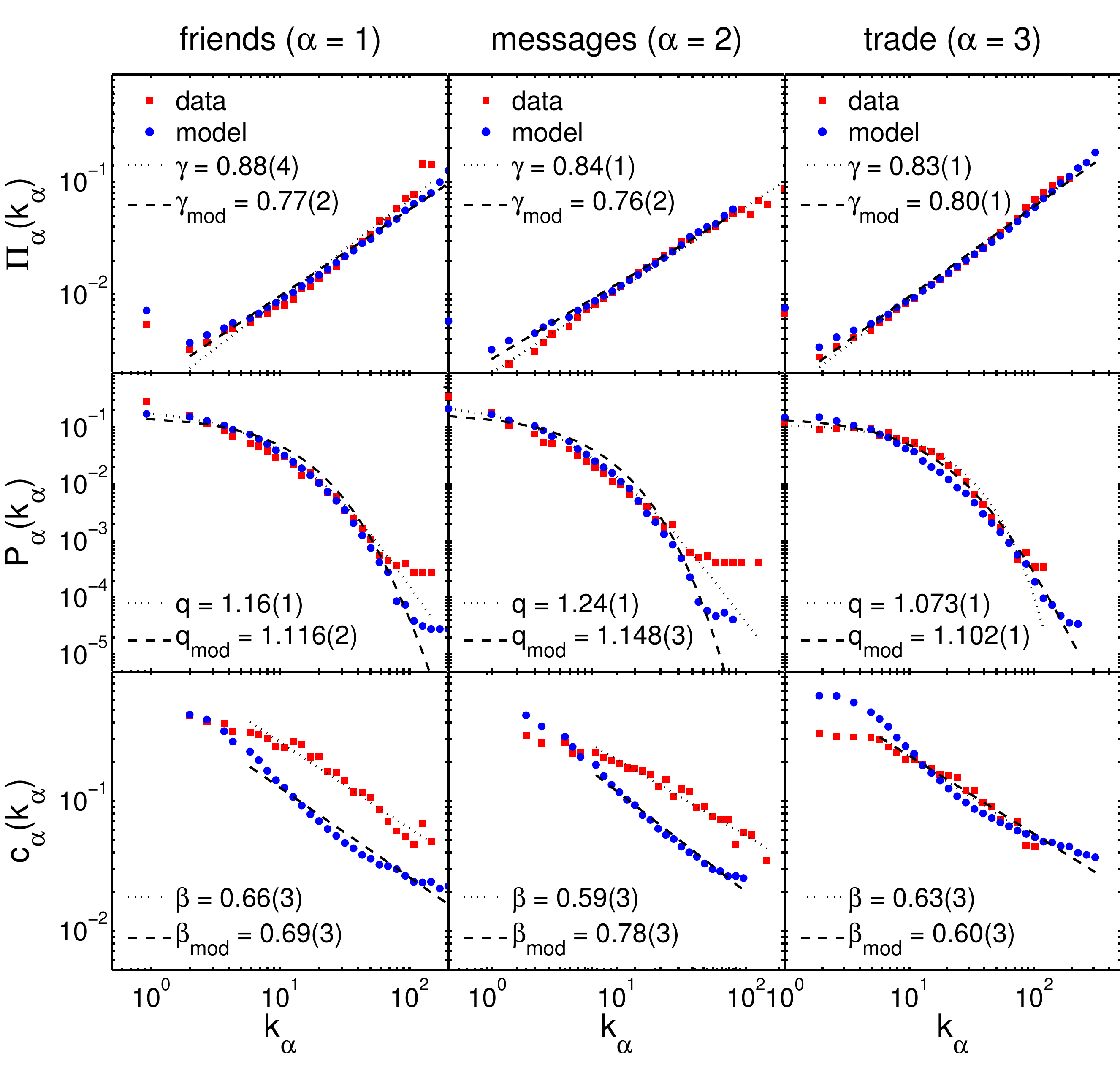}
\end{center}
\caption{Scaling exponents of the multilayer network as explained by the triadic closure model.
 Friendship ($\alpha=1$, left column), communication ($\alpha=2$, middle column), 
 and trade ($\alpha=3$, right column). 
(top row) Attachment kernels scale sub-linearly with the degree in each case. 
Data and model are barely distinguishable.
(middle row) Degree distributions for $\alpha=1,2,3$ and fits to a $q$-exponential. 
(bottom row) Clustering coefficients as a function of degree for data and model.
From \cite{Klimek2013}.
}
\label{data}
\end{figure}

Simulation results for the values of the characteristic exponents $\gamma, q$, and $\beta$ 
depend on the parameters $p$ and $r$, see Fig \ref{slopes}. 
Given $p$ and $r$ (as measured in the data), 
we can read off the corresponding values of the scaling exponents from Fig \ref{slopes}. 
These can be compared to the direct measurements from the data, i.e. the model can be validated.  
Figure \ref{data} shows the attachment kernel $\Pi_{\alpha}(k_{\alpha})$ (the probability for a new node to 
attach to an existing link with degree $k_{\alpha}$), 
the degree distribution $P_{\alpha}(k_{\alpha})$, and the clustering coefficients 
$c_{\alpha}(k_{\alpha})$, for the three sub-networks $M^{\alpha}$ in the empirical multilevel network data.
They are compared to the respective distributions of the calibrated model.
Data and model agree nicely.

These results suggest that triadic closure may play an even more 
fundamental role in social multilayer network formation than previously anticipated \cite{Rapoport1953, Granovetter1973}.
Given that {\em all} model parameters can be measured in the data, it is remarkable that the three important 
scaling laws are simultaneously explained by this radically simple model.
The exponents found in the model compare well to those of real-world networks. 
Sub-linear preferential attachment was reported in  scientific collaboration networks 
and the actor co-starring network, $\Pi(k) \sim k^{0.79}$ and $ \sim k^{0.81}$, respectively 
\cite{Jeong2003}. Degree distributions of many social networks often fall between 
exponential and power law distributions 
\cite{Barabasi1999,Newman2001,Onnela2007PNAS, Szell2010msd, Amaral2000}, 
and scaling of the average clustering coefficients as a function of  degree,  is observed in 
scientific collaboration and actor networks, $c(k)\sim k^{-0.77}$ and $\sim k^{-0.31}$, 
respectively (when the same fitting as in Fig. \ref{data} is applied). 
Mobile phone and communication networks give $\sim k^{-1}$ \cite{Onnela2007}.

%%%%%%%%%%%%%%%%%%%%%%%%%%%%%%%%%%%%%%%%%%%%%%%%%%%%%%%%%%%%%
\subsection{Degree distributions for negative ties are power laws---positive are not}

\begin{figure}[t]
%\begin{center}
\begin{minipage}[c]{0.4\textwidth}
\includegraphics[width=0.9\textwidth]{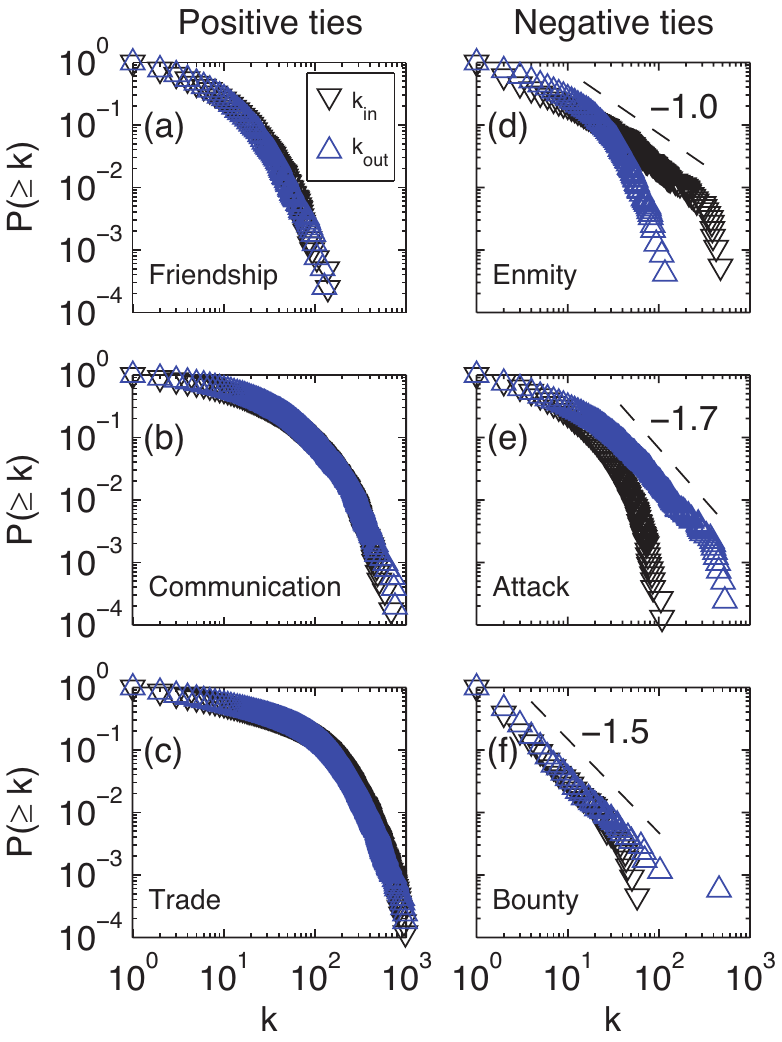}
  \end{minipage}\hfill
  \begin{minipage}[c]{0.6\textwidth}
%\end{center}
\caption{(left) Degree distributions for positive interactions (communication, trade, friendship) 
follow an approximate Poisson distribution, while negative interactions (enmity, attacks, bounties) 
show fat tailed distributions that could be power laws (right).
From \cite{Szell2010}.
}
\label{fig2PNAS}
  \end{minipage}\hfill
\end{figure}

For the (cumulative) in- and out-degree distributions, we find approximate power laws for 
aggressive behavior: attacking (out-degree for attacks), being declared an enemy (in-degree for enmity), 
and punishing/being punished (out- and in- degree for bounty). 
Power laws are absent for positive (friendship, communication, trade) and passive links (being attacked), 
see Fig. \ref{fig2PNAS}. 
This suggests different linking/rewiring processes for positive and negative ties. 
Moreover, we find that positive links are highly reciprocal (directed links go in both directions, $M_{ij}=M_{ji}$), 
while negative links are not \cite{Szell2010}. 
Low reciprocation in enemy networks may partially be explained by deliberate refusal of 
reciprocation to demonstrate aversion by total lack of response \cite{Szell2010msd}. 
For attack networks, it may originate from the asymmetry in the strength of the players 
(a strong player is more likely to attack someone weaker). 
We also find that positively connoted links show higher clustering coefficients than negatively connoted ones \cite{Szell2010}. 
High values of clustering are expected for positive interactions because they signal a cohesive structure 
and seems to benefit performance \cite{Coleman1988}. 
The significantly lower clustering for negative interaction types suggests that triadic closure 
\cite{Rapoport1953} is irrelevant for negative interactions. Their formation might be the result of 
a ``balance''  of signed motifs---which is at the core of social balance theory \cite{Heider1946}.

%%%%%%%%%%%%%%%%%%%%%%%%%%%%%%%%%%%%%%%%%%%%%%%%%%%
\subsection{Social Balance}

\begin{figure}[t]
%\begin{center}
\begin{minipage}[c]{0.55\textwidth}
\includegraphics[width=0.95\textwidth]{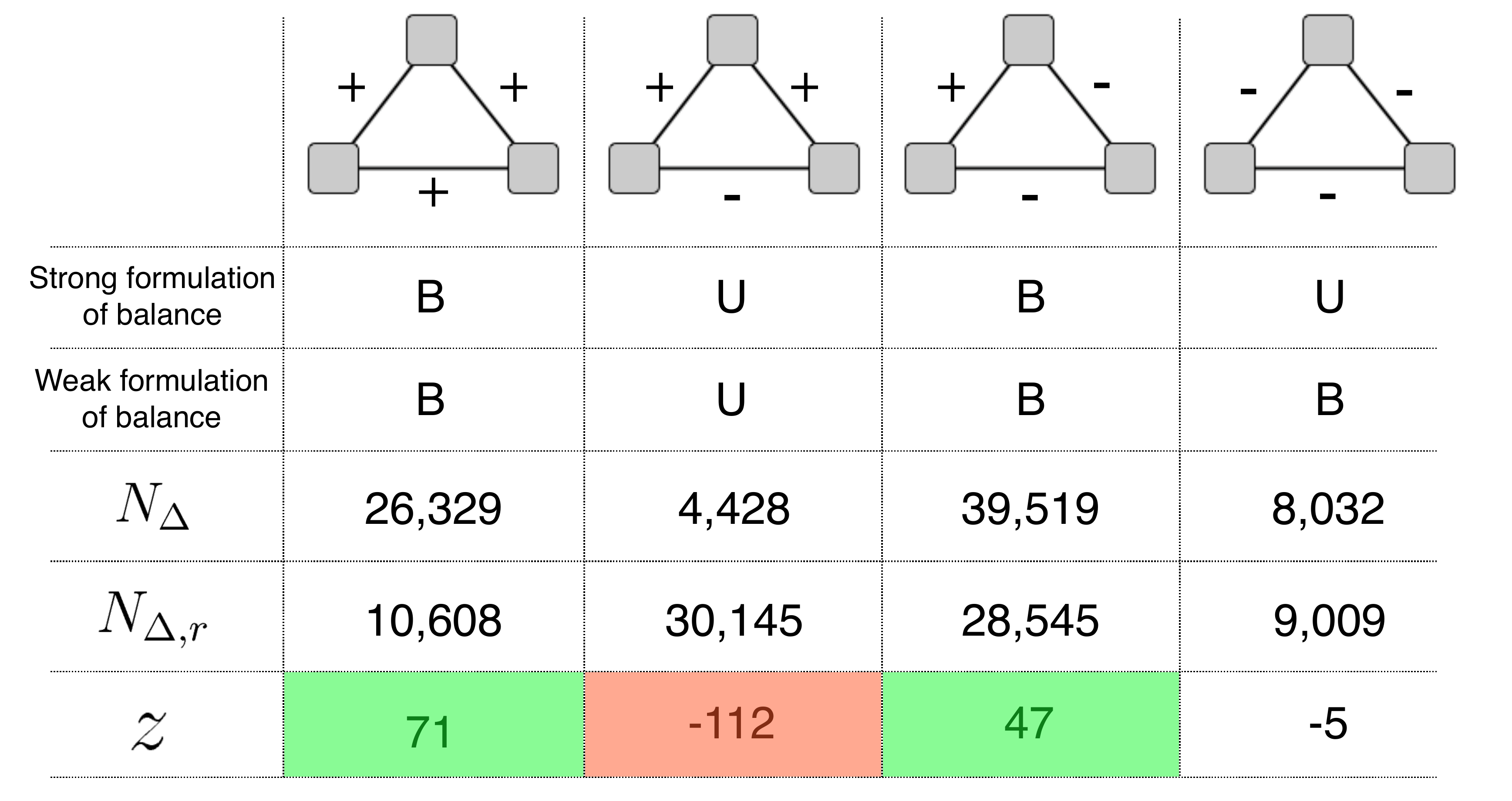}
  \end{minipage}\hfill
  \begin{minipage}[c]{0.45\textwidth}
%\end{center}
\caption{Signed triads, balanced (B) or unbalanced (U), according to the strong or 
weak formulation of structural balance. We see the number of each type of triad $N_{\Delta}$ in 
the friendship-enmity bilayer, the expected number of triads in a null model with 
sign-randomization, $N_{\Delta}^{\mathrm{rand}}$, 
and the corresponding Z-score (standard deviation from the null model). 
 $+++$ and $+--$ are over-represented, $++-$ are under-represented with extraordinary significance.
 From \cite{Szell2010}.
}
\label{fig4}
  \end{minipage}\hfill
\end{figure}

Social balance focuses on {\em signed} triads, where the sign is the product of the signs of its three links.
In the following we assign +(-)1 to a positive (negative) link, e.g., friendship links have +1, enemy links -1. 
Social balance theory, in its strong form \cite{Cartwright1956}, claims that positive triads are 
``balanced'' and negative triads are ``unbalanced'', see Fig. \ref{fig4}.  
Unbalanced triads are sources of social stress and tend to be avoided. They are therefore under-represented. 
There is a  ``weak'' formulation of structural balance  \cite{Davis1967} that assumes that triads 
with exactly two positive links are under-represented in real networks, 
while the three other triads should be abundant. In the weak formulation only situations like, 
``the friend of my friend is my enemy'' are unstable, whereas in the strong form of structural balance, 
``the enemy of my enemy is my enemy'' is also unstable, see Fig. \ref{fig4}. 
 
To test social balance, we focus on the bilayer network of friendship and enmity interactions. 
The number of the different signed triads is $N_{\Delta}$. 
They are compared to the expected number of such triads in a null model, $N_{\Delta}^{\rm rand}$, where we re-shuffle the 
signs of links. In Fig. \ref{fig4} the Z-score shows that $+++$ and $+--$ triads are heavily over-represented, 
while $++-$ triads are under-represented. Triads of type $---$ are under-represented to a lesser degree 
than the three other types, favoring the weak formulation of structural balance. 
 
 %%%%%%%%%%%%%%%%%%%%%%%%%%%%%%%%%%%%%%%%%%%%%%%%%%%%%%%%%%%%
\subsubsection{Origin of social balance}

\begin{figure}[t]
%\begin{center}
\begin{minipage}[c]{0.55\textwidth}
\includegraphics[width=0.95\textwidth]{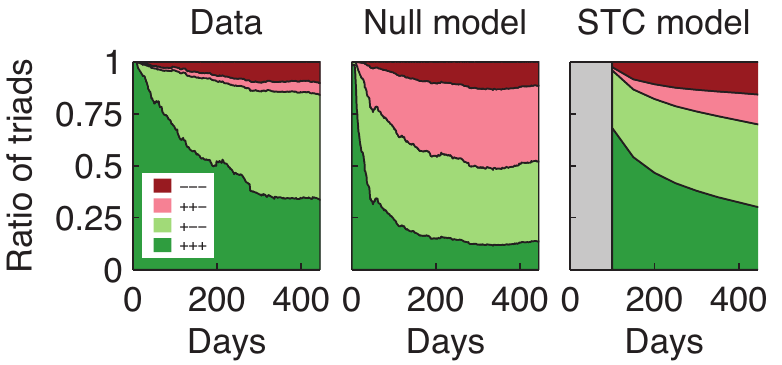}
  \end{minipage}\hfill
  \begin{minipage}[c]{0.45\textwidth}
%\end{center}
\caption{Ratio of signed triads over time as  
(left) seen in the data, (center) expected from the random null model, and  (right) simulation of signed triadic closure. 
%with a model for wedge transition rates. 
%Initial condition: Measured network of day 100. 
Measured ratios in the data deviate from those in the null model, except for $---$ triads. The 
model explains the observed ratios much better.
 From \cite{Szell2010}.
}
\label{fig5}
 \end{minipage}
\end{figure}

A dynamical analysis reveals that changes in the network are driven by the creation of new positive and negative links, 
and not by switching signs of existing links. To illustrate this, we define a {\em wedge} as a signed, 
open triad with two links and one link missing (hole). 
There are three possible wedge types: $++$, $+-$, $--$. We measure day-to-day transitions from 
wedges to other triadic structures. For almost all cases ($>99.9\%$), a wedge stays unchanged. 
In case of change, most often a hole is closed by either a positive or a negative link. 
Link removal is less frequent, and sign switches almost never occur. 
This result is in marked contrast with other dynamical models of structural balance \cite{Antal2006}, 
which assume that a given social network is fully connected from the start and that link-signs 
are the relevant dynamical parameters, which evolve to reduce stress in the system. 
In full agreement with the results in Figs. \ref{fig4} and \ref{fig5}, wedges of type $++$ close preferentially 
(about 7 times more likely) with a positive link, wedges of type $+-$ close preferentially 
(11 times more likely) with a negative link. 
%There is no clear sign preference in the closure of type $--$ wedges. 

We collect empirical transition rates in a transition matrix $A^{\mathrm{STC}}$, 
which we use in a simple dynamical model for {\em Signed Triadic Closure} (STC). 
$A^{\mathrm{STC}}$ is simply applied on the state vector that contains all signed wedges at time $t$. 
At time $t+1$, these wedges are either closed or left unchanged. 
This model reproduces the empirical observations reasonably well,  see Fig. \ref{fig5} (right).

%%%%%%%%%%%%%%%%%%%%%%%%%%%%%%%%%%%%%%%%%%%%%%%%%%%%
\subsection{Avatars organize in multiples of four}

Humans dominate their environment by the way they organize in groups.
Societies consist of hierarchically layered, nested groups of various quality, size, and structure, 
such as support cliques, sympathy groups, bands, cognitive groups, tribes, linguistic groups, 
and so on, \cite{Dunbar1995,Hill2003,Dunbar1993}.
%Anthropological data suggests that on average, each group at any given layer, consists of approximately three sub-groups.
Combining data on human group formation patterns, a discrete hierarchy of group sizes with a preferred scaling 
ratio close to $3$ was identified \cite{Zhou2005}. 
It was later confirmed for hunter-gatherer groups \cite{Hamilton2007} and mammal societies \cite{Hill2008}.
Do we see such a hierarchical organization in the \verb|Pardus| society? 
In particular, do we see the Dunbar numbers? 

\subsubsection{Dunbar numbers}

\begin{figure}[b]
%\begin{center}
\begin{minipage}[c]{0.55\textwidth}
\includegraphics[width=0.99\textwidth]{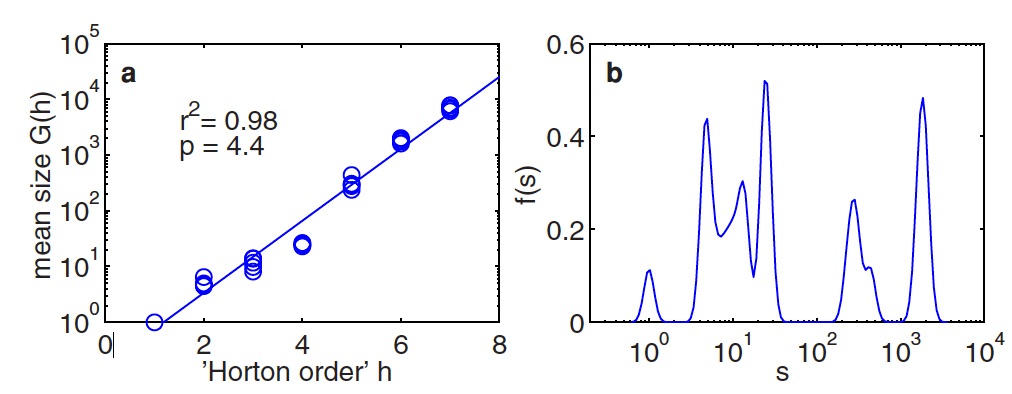}
  \end{minipage}\hfill
  \begin{minipage}[c]{0.45\textwidth}
%\end{center}
\caption{Group size scaling. (a) Horton plot: average size of groups per order.
 (b) Estimated probability density of group sizes $s$ in Pardus, obtained with a Gaussian kernel estimation
 ($\sigma=0.14$ acting on the logarithm of group sizes, $\ln(s)$).
  From \cite{Fuchs2014}.
}
\label{horton}
  \end{minipage}\hfill
\end{figure}

In a nutshell, the concept of the Dunbar numbers is that societies are approximately organized in 
multiples of three. 
In its simplest form, this  means that groups of size $3$, $3\times3$, $3\times3\times3$, ...,  $3^n$,  and so on, 
should be over-represented. 
In Fig. \ref{horton} we see two indications for that  hierarchical scaling. 
The so-called Horton plot shows the average size of groups per order. 
These orders have the following, somewhat subjective, meaning.
Horton order $h=1$, is the trivial group consisting of one person, the ``ego''.
Layer 2 ($h=2$) contains closest friends of the ego, defined by both a friendship 
marking and at least one communication event within the last 30 days.
Layer 3 ($h=3$) includes more casual relations, in particular all players that ego 
has marked as a friend, or by whom ego was marked as friend.
Layer 4 ($h=4$) contains the alliance  members of the ego. 
Layer 5 ($h=5$) is obtained by applying a community detection algorithm (Louvain algorithm) 
\cite{Blondel2008,Rubinov2010} to the communication network of the players.
We tested that layer 5 is an organisational layer in its own right, 
whose communities are predominantly subsets of the factions ($h=6$), and supersets of the alliances ($h=4$).
Layer 6 ($h=6$) contains the three factions (political parties), and layer 7 ($h=7$) is the entire society.
The size of the groups behind these orders scales like a power law with an exponent of roughly $4$, indicating the 
hierarchical scaling, see Fig. \ref{horton} (a).
The distribution of group sizes after a smoothing procedure (with a Gaussian kernel) 
is shown in Fig. \ref{horton} (b). It is immediately visible that groups of approximate sizes 
1, 4, 16, 30, 250, 1500 are over-represented.  This is in line with an hierarchical organization 
of about four,  $4^n$,  $1,          4,          16,          64,         256,        1024$. 
Clearly, this is not a perfect series, and the expected peak at 64 is clearly missing. 
Hierarchical organisation with a scaling ratio of 3.2 \cite{Zhou2005}, and 3.77 \cite{Hamilton2007}, 
was observed before in real-world societies.
The fact that we find these structural patterns in the virtual setting of \verb|Pardus|, 
where ``life'' is detached from most real-world constraints, suggests that this particular form of hierarchical 
organisation of societies is deeply rooted in human psychology.

%%%%%%%%%%%%%%%%%%%%%%%%%%%%%%%%%%%%%%%%%%%%%%%%%%%%
\subsection{The Behavioral Code}

\begin{figure}[t]
%\begin{center}
\begin{minipage}[c]{0.55\textwidth}
\includegraphics[width=0.95\textwidth]{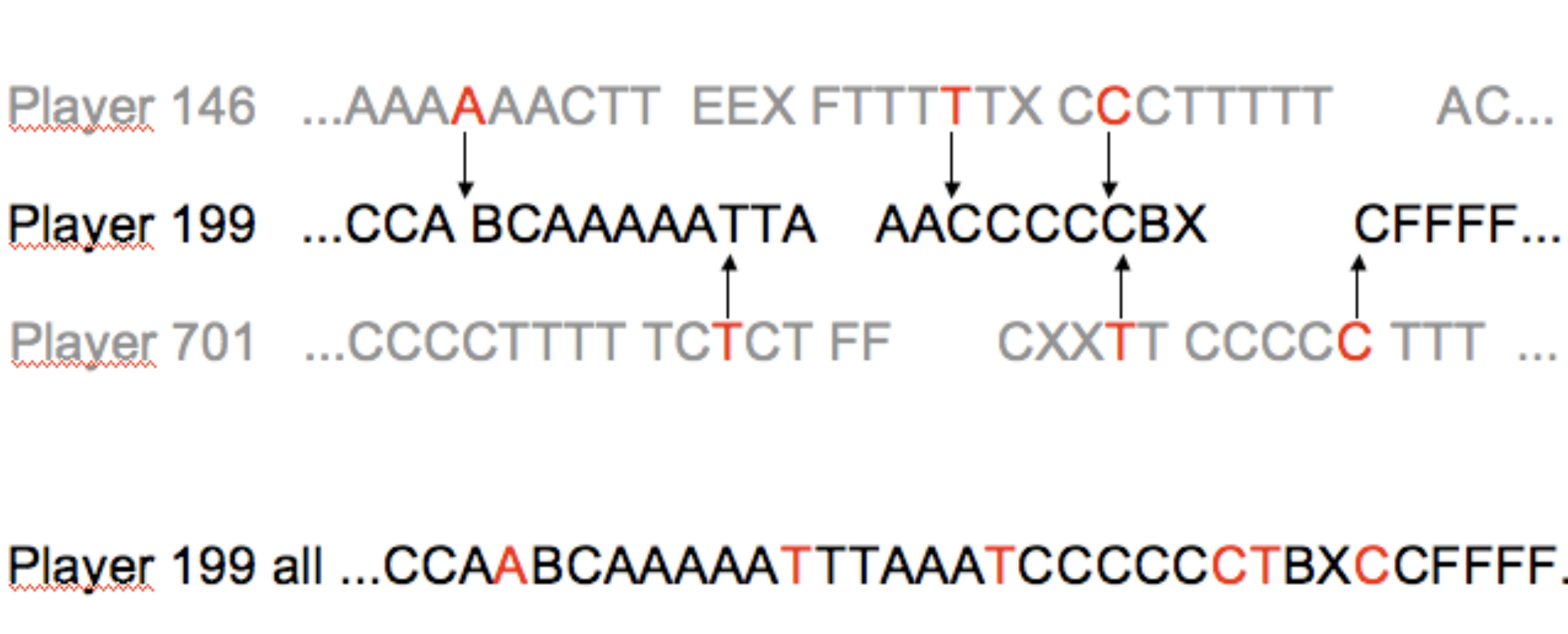}
  \end{minipage}\hfill
  \begin{minipage}[c]{0.45\textwidth}
%\end{center}
\caption{Segment of action sequences of three players. 
Some actions of players 146 and 701 are directed toward player 199. 
This results in a sequence of received-actions for 199, $R^{199}=\{ \cdots {\rm  ATTCT } \cdots   \}$. 
(bottom line) Combined sequence of actions (originated from--and directed to) player 199, $C^{199}$.
Red letters mark actions from others to player 199.   
}
\label{fig:seq}  
\end{minipage}\hfill
\end{figure}

To describe a human, one way of doing it is to list the temporal sequence of her actions.  
In real life there is a huge number of such actions, like cooking coffee, brushing teeth, 
washing cars, and so on. In the computer game there are much less actions that players 
can act out. But we have these action sequences ready for analysis. 
We limit ourselves to eight different actions that every player can execute at any time, and which are 
observable as changes in $M_{ij}^{\alpha} (t)$. 
These are  communication (C), trade (T), setting a friendship link  (F),  removing an enemy link (forgiving) (X), 
attack (A), placing a bounty on another player (punishment) (B),  removing a friendship link (D), 
and setting an enemy link (E).
While C, T, F and X are positive (good) actions,  A, B, D and E are hostile or negative (bad). 
We classify communication as positive because only a negligible fraction of communication takes 
place between enemies \cite{Szell2010msd}.
We ignore other possible actions like movement, production, working, sleeping, and so on. 
Segments of action sequences of three players are shown in Fig. \ref{fig:seq}. 

We consider three types of sequence. The first  is the (time-ordered) 
stream of $N$ consecutive actions $A^i=\{a_n | n=1, \cdots, N \}$,  
which player $i$ performs during his ``life'' in the game.
The second is the stream of actions that player $i$ receives from  all the other players, i.e. all the actions 
which are directed towards player $i$.  Received-action sequences we denote by $R^i=\{r_n | n=1, \cdots, L \}$.
The third sequence is the time-ordered combination of player $i$'s actions and received-actions, which is 
a chronological sequence of the elements of $A^i$ and $R^i$ in their order of occurrence. 
The combined sequence we denote by $C^i$; its length is $L+N$, see Fig. \ref{fig:seq}.  
The $n$th element of one of these  series is denoted by $A^i(n)$, $R^i(n)$, or $C^i(n)$.
We do not consider the actual time between two consecutive actions, which can range from seconds to weeks.  
We work in ``action-time''. 

%%%%%%%%%%%%%%%%%%%%%%%%%%%%%%%%%%%%%%%%%%%%%%%%%%
\subsubsection{Two ways of seeing the same data}

Since individual actions are directed, %(actions originate in one individual and are directed to someone), 
the information-content of both, the temporal multilayer data, $M_{ij}^{\alpha} (t)$, 
and the behavioral code ($A$, $R$, $C$) are identical, except that in the latter we use action time. 
The situation is similar to the Heisenberg- and Schr\"odinger picture in quantum mechanics. 
In the {\em multilayer picture} the focus is on the topological linking structure, 
temporal information is hard to visualize (``Heisenberg picture''). 
In the {\em Behavioral code picture} the temporal information is clear, 
linking structure  is harder to visualize (``Schr\"odinger picture''). 

%%%%%%%%%%%%%%%%%%%%%%%%%%%%%%%%%%%%%%%%%%%%%%%%%%%%
\subsubsection{Behavioral code and predicting behavior}

\begin{figure}[t]
%\begin{center}
\begin{minipage}[c]{0.55\textwidth}
\includegraphics[width=1.0\textwidth]{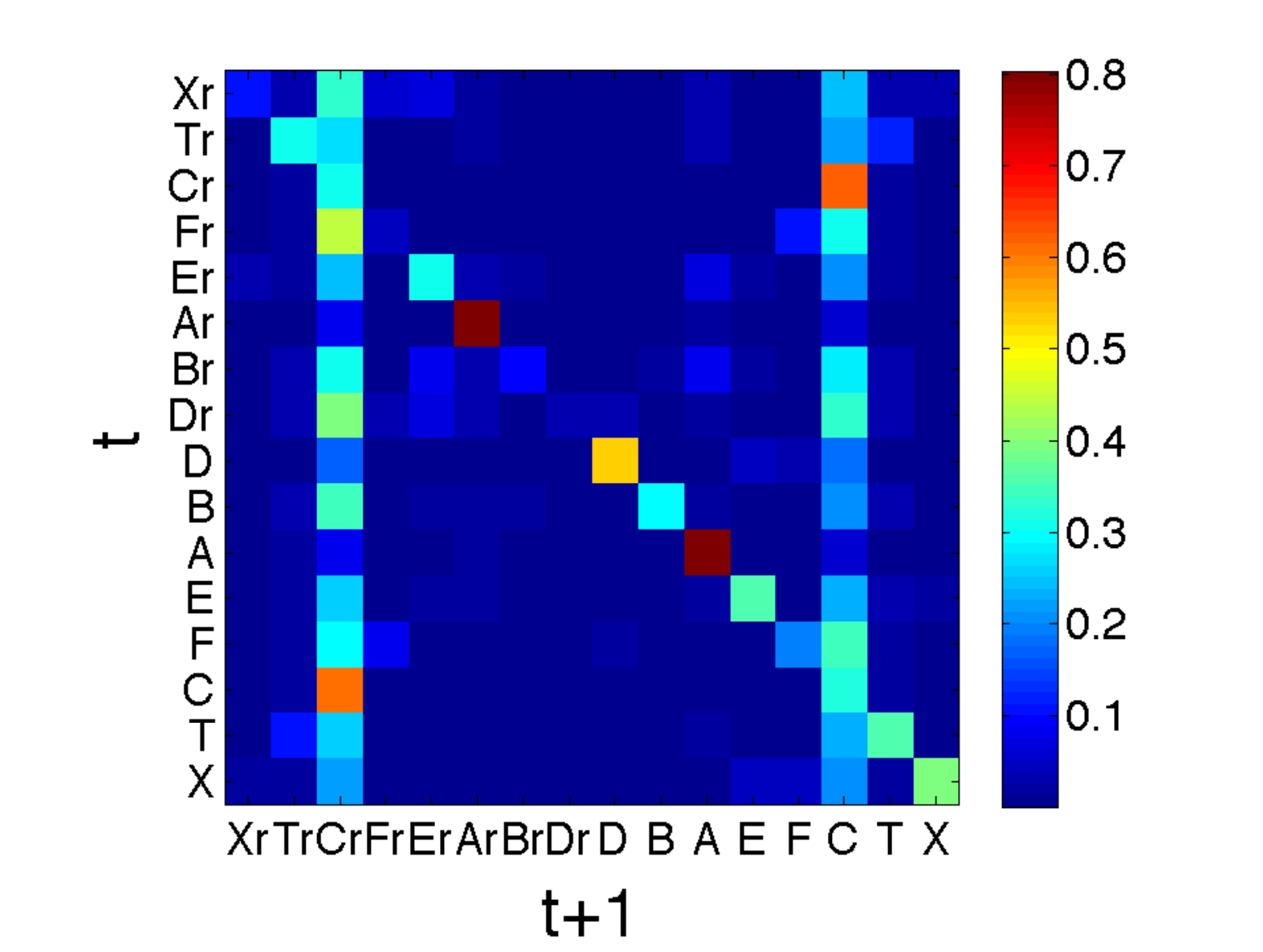}
  \end{minipage}\hfill
  \begin{minipage}[c]{0.45\textwidth}
%\end{center}
\caption{Transition probabilities $p\left( Y | Z \right)$ for actions (or received actions) $Y$ at time $t+1$, 
given that a specific action $Z$ was executed (or received) in the previous timestep $t$.
Received-actions are indicated by subscript $r$. Normalization is such that rows add up to one.
Large values in the diagonal show that human actions are often repetitive. 
Large values for $C \rightarrow C_r$ and $C_r \rightarrow C$ reveal that communication is an 
anti-persistent activity---it is more likely to receive a message after sending one,
and vice versa, than to send consecutive messages. 
 From \cite{Thurner2012}. 
  }
\label{fig:trans}
  \end{minipage}\hfill
\end{figure}

By $p( Y | Z )$ we denote the probability that an action of type $Y$ follows 
action of type $Z$ in the behavioral sequence of a player. 
$Y$ and $Z$  stand for any of the eight actions, executed or received (received is indicated by a subscript $r$). 
In Fig. \ref{fig:trans} the transition probability matrix, $p\left( Y|Z \right)$, is shown. 
The $y$ axis indicates the action (or received-action) happening at a time $t$, 
the probabilities for the actions (or received-actions) that immediately follow are given in the corresponding horizontal place. 
%This transition matrix specifies to what extent an action or a received action of a player is influenced by the action that was done or received at the previous time-step. In fact, if the behavioral sequences of players had no correlations, i.e.  the probability of an action, received or executed, is independent of the history of the player's actions, the transition probability $p\left( Y | Z \right)$  simply is  $p \left( Y \right)$, i.e. to the probability that an action or received action $Y$ occurs in the sequence is determined by its relative frequency only. Therefore, deviations of the ratio $\frac{p\left( Y | Z \right)}{p(Y)}$ from $1$ indicate correlations in sequences. In Fig.~\ref{fig:trans}~(b) we see the values of $\frac{p\left( Y | Z \right)}{p(Y)}$ for actions and received actions (received actions are indicated with the subscript $r$) classified only according to their positive (+) or negative (-) connotation. In brackets we report the {\em Z}-score with respect to the uncorrelated case. 
We find that the probability to perform good actions 
is significantly higher if in the previous time-step a positive action has been received. 
Similarly, it is more likely that a player is the target of a positive action, if at the previous timestep 
he executed a positive action. Conversely, it is highly unlikely that after a good action, 
executed or received, a player acts negatively, or is the target of a negative action. 
Instead, if a player acts negatively, it is very likely that he will perform another negative action in 
the following timestep. 
Finally, if a negative action is received, it is likely that another negative action 
will be received in the following timestep.
%, all other possible actions and received actions are under-represented. 
The high statistical significance of the cases $P(-|-)$ and $P(-_r|-_r)$ hints at a high persistence of negative actions 
in the players' behavior. For details see \cite{Thurner2012}.

An important finding is obtained by considering pairs of received actions, followed by performed actions. 
This approach allows us to quantify the influence of received actions on the performed actions. 
For these pairs we measure a conditional probability of $0.02$
of performing a negative action right after receiving a positive action. 
This value is significantly lower when compared to the probability of $0.10$, 
which is obtained from randomly reshuffled sequences. 
Similarly, we measure a probability of $0.27$ of performing a negative action, right after receiving a negative action.  
These results agree with a recent study, where the emotional content of posts in 
online fora was analyzed \cite{Holyst2011}.
The homo sapiens seems to become drastically more aggressive, almost by a factor of ten, 
immediately after being treated badly. 

%%%%%%%%%%%%%%%%%%%%%%%%%%%%%%%%%%
\subsubsection{Worldlines of players}

\begin{figure}[b]
%\begin{center}
\begin{minipage}[c]{0.55\textwidth}
\includegraphics[width=0.9\textwidth]{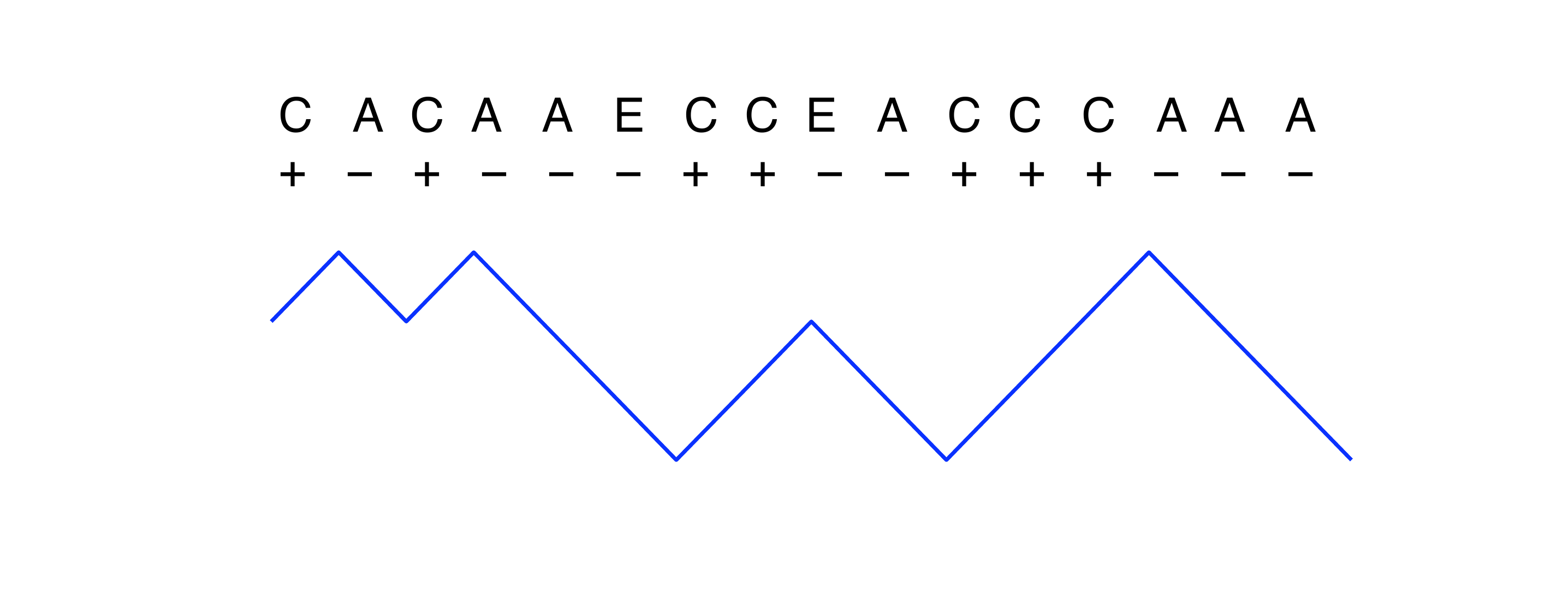}
  \end{minipage}\hfill
  \begin{minipage}[c]{0.45\textwidth}
%\end{center}
\caption{Illustration of a worldline $W_i^{\rm good-bad}$  as a binary random walk in ``good-bad'' action space. 
Positive actions (C, T, F or X) produce an upward move, negative ones (A, B, D and E) go downward. 
Good people have rising worldlines.
}
\label{fig:seqb}
  \end{minipage}
\end{figure}

\begin{figure}[t]
%\begin{center}
\begin{minipage}[c]{0.55\textwidth}
\includegraphics[width=0.7\textwidth]{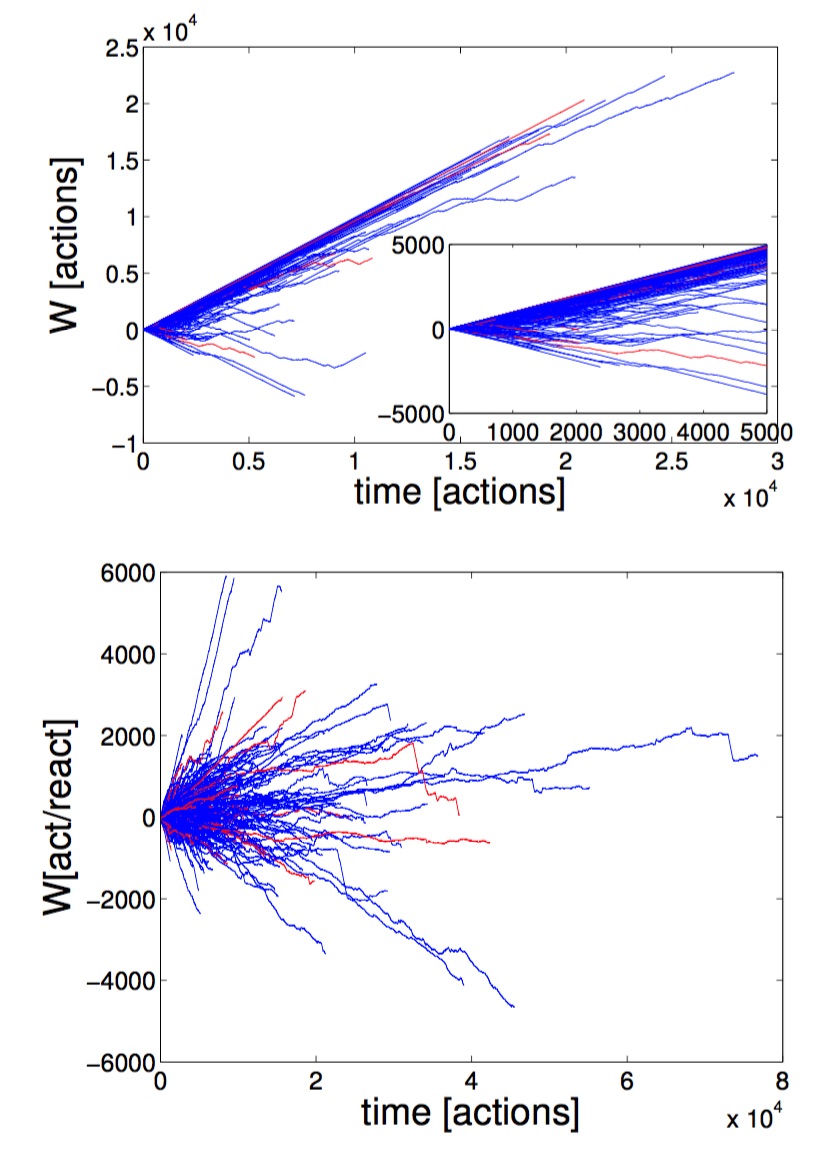}
  \end{minipage}\hfill
  \begin{minipage}[c]{0.45\textwidth}
%\end{center}
\caption{Worldlines of good-bad action random walks (top) of the 1,758 most active players.
(bottom) action-reaction worldlines of the same players. Red lines show female avatars.
}
\label{fig:world}
  \end{minipage}
\end{figure}

We can interpret action sequences as random walks. 
By assigning $+1$ to positive actions C, T, F or X, and $-1$ to negative ones A, B, D and E, 
we translate a sequence, $A_i$, into a binary sequence, $A_i^{\rm bin} $. 
The cumulative sum of the binary sequence is the ``worldline'' or a random walk for player $i$,  
$W_i^{\rm good-bad}(t) = \sum_{n=1}^{t} A_i^{\rm bin} (n)$, see Fig. \ref{fig:seqb}.  
Similarly, we define binary sequences from the combined sequence $C_i$, 
where we assign $+1$  to an executed action, and $-1$ to a received.
This sequence we call $C_i^{\rm bin}$; its cumulative sum, $W_i^{\rm act-rec}(t) = \sum_{n=1}^{t} C_i^{\rm bin} (n)$ 
is the ``action-receive'' worldline. 
Worldlines are shown in  Fig. \ref{fig:world} for good-bad action sequences (top), and  
action-reaction (bottom). 
Figure \ref{fig:world} (a) also  shows that the lifetime of players with many negative actions is often short. 
The average lifetime for players with a slope $A<0$  is $2,528  \pm  1,856$ actions, compared to 
players with a slope $A>0$ with $3,909 \pm  4,559$ actions. The average 
lifetime of the whole sample of (very active) players is $3,849 \pm 4,484$  actions. 

\begin{figure}[t]
\begin{center}
\includegraphics[width=0.45\textwidth]{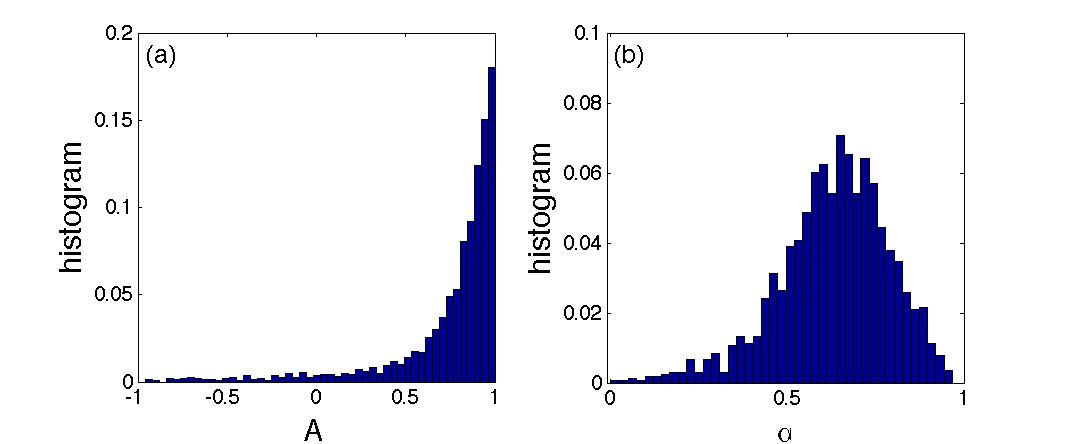}
\includegraphics[width=0.45\textwidth]{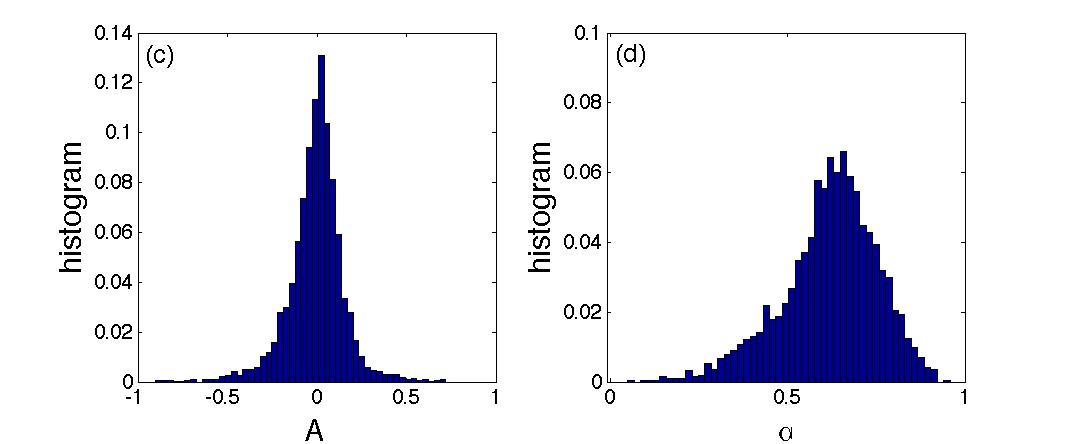}
\end{center}
\caption{
Distribution of worldline slopes, $A$, for good-bad action (a), and action-received action worldlines (c). 
Distribution of scaling exponents $\alpha$ for good-bad action (b), and action-received action worldlines (d).
}
\label{fig:worldb}
\end{figure}

To characterize worldlines we define the slope $A$ of the line connecting the 
origin of the worldline with its end point. It is  an approximate measure for ``altruism''. 
$A=1(-1)$ in the good-bad worldlines $W^{\rm good-bad}$ indicates that the player performed 
only positive (negative) actions. 
The histogram of the slopes is shown in Fig. \ref{fig:worldb} (a) for good-bad sequences.
%$\bar k^{\rm good} = 0.81 \pm 0.19$, $\bar k^{\rm bad} = -0.40 \pm 0.28$, and $\bar k^{\rm all} = 0.76 \pm 0.31$, respectively. Simulated random walks with the same probability $0.90$ of performing a positive action yield a much lower variation, $\bar k^{\rm sim} = 0.81 \pm 0.01$, pointing at  an inherent heterogeneity of human behavior.
For the action--received-action  worldline the slope is a measure of 
how well a person is integrated in her social environment. If $A=1$, the person only acts 
and receives no input, she is ``isolated'' but dominant. If  $A=-1$, the person is driven by the 
actions of others and never acts nor reacts. 
The histogram is shown in  Fig. \ref{fig:worldb} (c). 
%Most players are well within the $\pm45$ degree cone. Mean and standard deviation of slopes of good, bad, and all players are $\bar k^{\rm good} = 0.02 \pm 0.10$, $\bar k^{\rm bad} = 0.30 \pm 0.19$, and $\bar k^{\rm all} = 0.04 \pm 0.12$, respectively. Bad players are tendentially dominant, i.e. they perform significantly more actions than they receive. Simulated  random walks with equal probabilities for up and down moves for a sample of the same sequence lengths, we find again a much narrower distribution with slope $\bar k^{\rm sim} = 0.00 \pm 0.01$.

As a second measure we use the mean square displacement of  worldlines to quantify 
the persistence of action sequences 
\begin{equation}
S^2(\tau) = \langle  (\Delta W (\tau) -  \langle \Delta W (\tau) \rangle )^2 \rangle_t  \sim \tau^{2\alpha} \quad, 
\end{equation}
where $\Delta W(\tau) = W(t +\tau)- W(t)$, and $\langle . \rangle_t$ is the average over  $t$. 
The asymptotic exponent $\alpha$ quantifies the ``persistence'' of a worldline. 
$\alpha=1/2$  is the pure diffusion case,  
$\alpha > 1/2$ indicates  persistence, $\alpha < 1/2$, anti-persistence. 
Persistence means that the probability of making an up (down) move at time $t+1$ is larger (less) than $p=1/2$, 
given the move at time $t$ was up. 
%For calculating the exponents $\alpha$, we use a fit range of $\tau$ between 5 and 100. We checked from the mean square displacement of single world lines that this fit range is indeed reasonable.
The histogram of exponents  $\alpha$ for the good-bad random walk is shown in  Fig. \ref{fig:worldb} (b), for 
the action--received-action world line in (d). In both cases persistent behavior is obvious.
% in the second there is a slight tendency towards persistence. Mean and standard deviation for the good-bad world lines are $\alpha_{\rm good-bad}= 0.87\pm  0.06$,  for the action-received actions   $\alpha_{\rm act-rec}=0.59 \pm 0.10$. Simulated sequences of random walks have -- as expected by definition -- an exponent of $\alpha_{\rm rnd} = 0.5$, again with a very small standard deviation of about $0.02$.
%

%%%%%%%%%%%%%%%%%%%%%%%%%%%%%%%%%%%%%%%%%%%%%%%%%%%%%%%%%%%%
\subsubsection{Zipf's law in the human behavioral code}

The ensemble of sequences of all actions $A_i$ of all players $i$ allows us to analyse 
the frequencies of the occurring $n$-strings, see \cite{Zipf1949,Sinatra2010}. 
An $n$-string is a subsequence of $n$ adjacent actions in an action sequence. 
Given our 8-letter action alphabet, (A, B, C, D, E, F, T, X), there are $8^n$ different $n$-strings, or ``words'' $i$,  
that occur with probability $P_i^{(n)}$. 
\begin{figure}[b]
%\begin{center}
\begin{minipage}[c]{0.55\textwidth}
\includegraphics[width=0.99\textwidth]{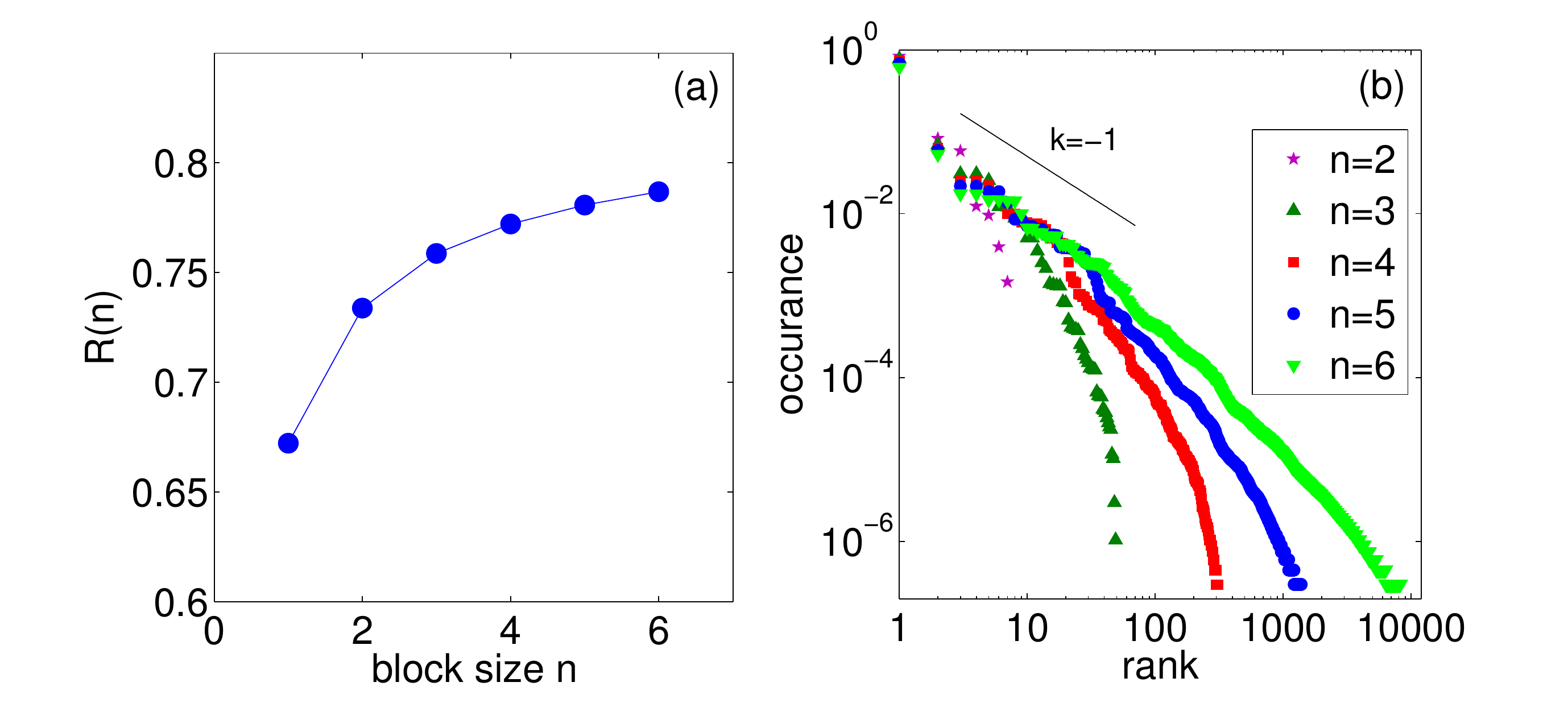}
  \end{minipage}\hfill
  \begin{minipage}[c]{0.45\textwidth}
%\end{center}
\caption{(b) Rank ordered probability distribution  of 1 to 6 letter ``words'' in the 8-letter action alphabet. 
The slope of $\kappa = -1$ (Zipf's law) is indicated. 
(a) Shannon $n$-tuple redundancy increases as a function of word length $n$, a sign for structures in the sequences.
 From \cite{Thurner2012}. 
}
\label{fig:entropy}
  \end{minipage}
\end{figure}
We partition the action sequences  into ``words'' of length $n$. 
Fig. \ref{fig:entropy} (b) shows the rank distribution of word occurrences for different lengths $n$. 
The distribution shows an approximate  Zipf law \cite{Zipf1949} (slope of $\kappa=-1$), for ranks up to about 100. 
For ranks between 100 and 25,000, $\kappa \sim -1.5$. 
The Shannon $n$-tuple redundancy, see e.g. \cite{dna,dna2},  
for sequences composed of 8 letters is 
\begin{equation}
 	R^{(n)} = 1+ \frac{1}{3n} \sum_{i=1}^{8^n} P_i^{(n)} \log_2 P_i^{(n)} \quad .
\end{equation}
%where $P_i^{(n)}$ is the probability of finding a paricular $n$-letter word $i$. 
For the equi-distribution, $P_i= 8^{-n}$, we have $R^{(n)} =0$, and 
in the other extreme of only one single letter in the sequence, $R^{(n)} =1$.
Figure \ref{fig:entropy} (a) shows $R^{(n)}$ as a function of $n$. 
$R^{(n)}$ increases with $n$, which indicates strong structure in the sequences, 
since Shannon entropy is not an extensive quantity for action sequences \cite{Hanel2011}.

%%%%%%%%%%%%%%%%%%%%%%%%%%%%%%%%%%%%%%%%%%%%%%%
\subsection{Network--network interactions}

Social interaction networks of  are not independent. 
How do they influence each other? An answer to this question would contribute much to a deeper understanding 
of how societies work. We try to take simple first steps in this direction 
by interpreting several measures that quantify inter-dependencies between pairs of networks.
We follow two approaches.

In the first, we focus on the link-overlap between networks and calculate the 
Jaccard coefficient, $J_{\alpha\beta}$, between two interaction layers $\alpha$ and $\beta$. 
It measures the tendency that links are simultaneously present in both layers, 
and is defined as the size of the intersection of the link sets, $\alpha$ and $\beta$, divided by the size of their union, 
$J_{\alpha\beta} = |\alpha \cap \beta|/|\alpha \cup \beta|$. 
The link overlap, $O_{\alpha \beta}=\frac{1}{2}\sum_{ij} M_{ij}^{\alpha} M_{ij}^{\beta}$, 
measures the overlap between networks  $\alpha$ and $\beta$. 
In the second approach, we compute Pearson correlation coefficients,  $\rho(k^{\alpha},k^{\beta}) = E\left[(k^{\alpha}-\bar k^{\alpha}) (k^{\beta}-\bar k^{\beta}) \right] /(\sigma_{k^{\alpha}} \sigma_{k^{\beta}} )$, 
between node degrees in the different layers. 
They measure to which extent degrees of avatars in one layer correlate with 
degrees of the same avatar in another. If $\rho(k_{\alpha},k_{\beta}) \sim 1$, players who have many 
(few) links in layer $\alpha$ have many (few) links in layer $\beta$. 
Note that correlation coefficients might be influenced by different network sizes or different average degrees. 
To account for this possibility, we additionally compute correlations 
$\rho(\mathrm{rk}(k^{\alpha}),\mathrm{rk}(k^ {\beta}))$ between {\em ranks} of node degrees. 
Overlap and correlation provide complementary views on the organization of social structures. 
All three measures are shown in Fig. \ref{nwnw} for all possible combinations of network layers. 
It suggests the following set of conclusions:

\begin{figure}[t]
%\begin{center}
\begin{minipage}[c]{0.55\textwidth}
\includegraphics[width=0.95\textwidth]{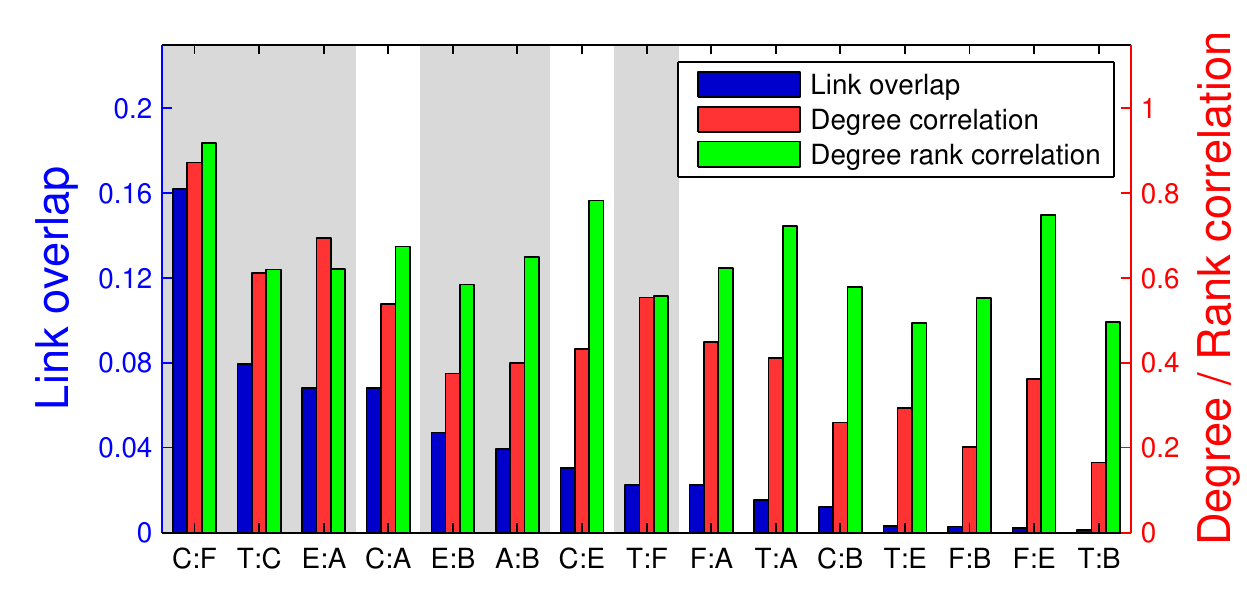}
  \end{minipage}\hfill
  \begin{minipage}[c]{0.45\textwidth}
%\end{center}
\caption{Measures to quantify network-network interactions. Node degree correlation (red), overlap (blue), 
and rank correlation (green). For an interpretation, see the text. From \cite{Szell2010}.
}
\label{nwnw}
  \end{minipage}
\end{figure}

{\em Communication--Friendship.} The pronounced overlap implies that friends tend to 
talk {\em with each other}, which is of course not unexpected. Strong correlation means that players 
who communicate with many (few) others tend to have many (few) friends, see also \cite{Szell2010msd}. 
%where a high fraction of communication partners was shown to be friends.

 {\em Trade--Communication.} The high overlap shows that trade partners have a tendency to communicate 
 with each other, while high correlations indicate a tendency of communicators also being traders. 

{\em Enmity--Attack.} The high overlap shows that enemies tend to attack each other, 
or that attacks are likely to lead to enemy markings. The high correlations imply that aggressors (or victims) 
tend to be involved in many enemy relations.  

{\em Communication--Attack.} The relatively high overlap shows that there is a tendency for 
communication taking place between those players that attack each other. 
A relatively high correlation implies that players who communicate with many (few) others 
tend to attack or be attacked by many (few) players.  Aggression is not anonymous, but is mostly 
accompanied by communication.

 {\em Enmity--Bounty and Attack--Bounty.} The situation is similar to the Enmity--Attack case.

{\em Communication--Enmity.} The situation is similar to the Communication--Attack case. 

{\em Trade--Friendship.} Similar to Trade--Communication, however, with a smaller overlap. 
It is more difficult for traders to become friends than to just communicate.

{\em Friendship--Attack.} The low overlap shows that attacks tend to {\em not} take place between friends, 
or that fighting players do {\em not} tend to become friends. The relatively high correlations mean 
that players with many (few) friends do attack or are attacked by many (few) others.

{\em Trade--Attack.} Is similar to the Friendship--Attack case.

{\em Communication--Bounty.} Similar to Communication--Attack and Communication--Enmity, however, 
with much smaller overlap and degree correlations.

{\em Trade--Enmity.} For this and all other interactions, overlap vanishes. 
Players who trade with each other almost never become enemies and vice versa. 

{\em Friendship--Bounty.} Is similar to the Com\-mu\-ni\-ca\-tion--Bounty case.

{\em Friendship--Enmity.} The degree (rank) correlation is substantial, suggesting that 
players who are socially active tend to establish both, positive and negative links. 
However, vanishing overlap indicates the absence of ambivalent relations. Friends can not be enemies.

{\em Trade--Bounty.} This interaction shows the smallest values for all measures. 
%This could be due to substantial differences in network sizes. 
The relatively small correlation suggests that players who are experienced in trade 
have a tendency to {\em not} act out negative sentiments by spending money on bounties.
\\
 
The values of the two correlation measures must be interpreted with some care \cite{Szell2010}. 
Low values of $\rho(k_{\alpha},k_{\beta})$ indicate that hubs in one network are not necessarily 
hubs in another (see e.g. the Trade--Enmity case). This suggests that avatars play very different 
roles in different network layers. For example they can be central for flows of information 
but peripheral for flows of goods \cite{Haythornthwaite2001}.

%%%%%%%%%%%%%%%%%%%%%%%%%%%%%%%%%%%%%%%%%%%%%%%%
\section{Gender differences}

\begin{figure}[t]
%\begin{center}
\begin{minipage}[c]{0.55\textwidth}
\includegraphics[width=0.6\textwidth]{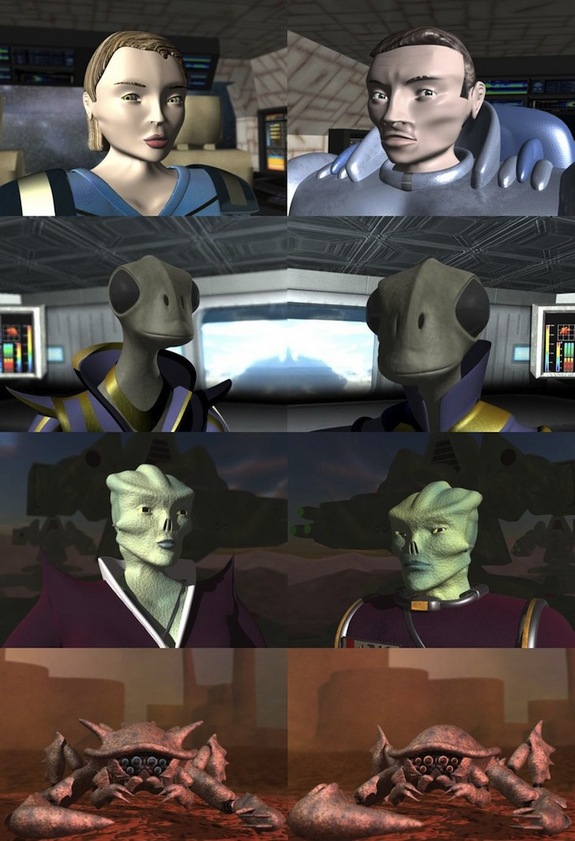}
  \end{minipage}\hfill
  \begin{minipage}[c]{0.45\textwidth}
%\end{center}
\caption{Players choose a male or female gender when joining the game. 
Some avatars that players can choose from. 
All possible avatars come in two genders.
}
\label{fig:multiplexitygender}
  \end{minipage}
\end{figure}

When signing up for the first time, players chose to be a male or female avatar, Fig. \ref{fig:multiplexitygender}. 
We have no information about the biological sex of players.  
Selecting a gender different from the biological is called {\em gender swapping}, 
which is common in online games \cite{Grosman2010}. A survey on 8,694 players 
in {\em Everquest} found 15.5\% gender-swappers, 17\% of the males and 10\% of the  females 
\cite{Griffiths2003}. Similar values are reported for {\em Second life} (10\% swapping of all, 
16\% of males, 2.7\% of females) \cite{De2006}, or for {\em World of Warcraft} (23\% of males, 3\% of 
females) \cite{Yee2007}.

We observe that on average females are less risk-taking but wealthier \cite{Szell2012C}.
Females accumulate significantly more wealth (4 sigma level) than males. 
At the same time, male players experience significantly more deaths (2 sigma level), 
due to more risk-taking and/or aggressive behavior. This points to a much larger 
engagement of females in economic, rather  than destructive activities. 
Concerning overall activity, experience points, kills and collected bounties, 
female and male players perform comparably  ($H_0$ can not be rejected).

Females show homophily, males are heterophiles \cite{Szell2012C}.
Homophily is the tendency of individuals to associate and link with similar others \cite{Mcpherson2001}.
A straightforward way to measure 
homophily is to compare the numbers of directed links between all gender-combinations  (MM, MF, FM, FF)
in all network layers to the corresponding numbers from surrogate data, where the gender of 
nodes is randomized (re-shuffled) but the topology of the network is left intact. 
To measure statistical significance of differences between the various combinations of genders, we compare 
each real network to 1,000 reshuffled surrogate networks.
Female-to-female trading  and communication are the most significantly over-represented link types, 
with a Z-score of approximately 4 sigmas \cite{Szell2012C}. 
Male-to-female trades ($Z=2.7$) and communication ($Z=2.7$) 
are also strongly over-represented, whereas  the opposite, female-to-male trades and communication 
is much less substantial ($Z=1.4$ and $1.6$, respectively). 
Male-to-male trades and communication are under-represented ($Z= -2.3$ for both cases). 
Negative link types show no  significant homophily in either direction.

%%%%%%%%%%%%%%%%%%%%%%%%%%%%%%%%%%%%%%%%%%%%%%%%
\subsection{Gender differences in networking}

How do male and female players create and manage their local networks? Are they different? 

%%%%%%%%%%%%%%%%%%%%%%%%%%%%%%%%%%%%%%%%%%%%%%%%
\subsubsection{Gender differences in network topology}
 
\begin{figure}[t]
\begin{center}
\includegraphics[width=0.35\textwidth]{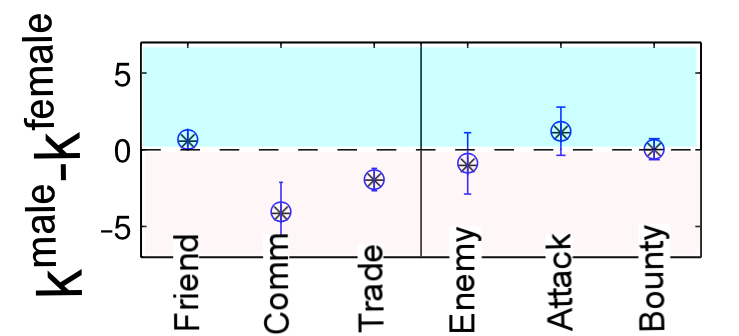}
\includegraphics[width=0.35\textwidth]{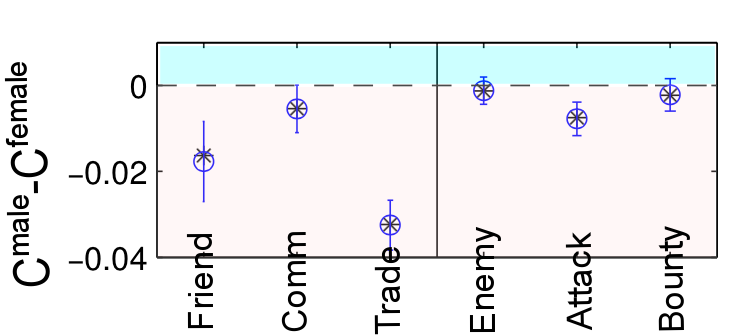}\\
\includegraphics[width=0.35\textwidth]{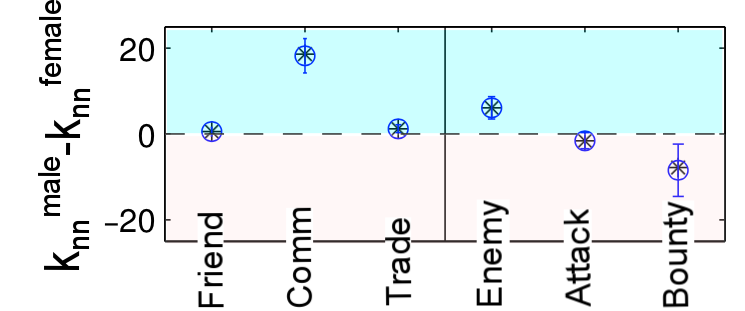}
\includegraphics[width=0.35\textwidth]{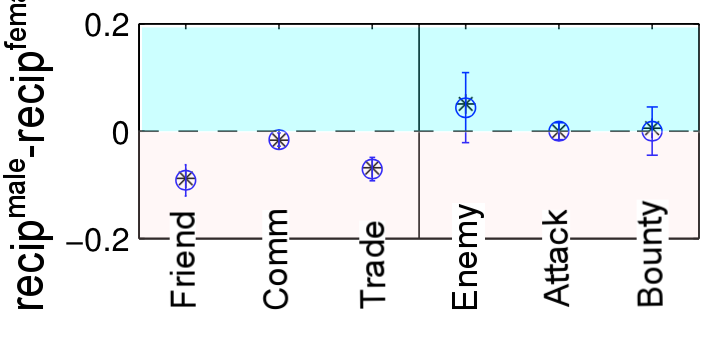}
\end{center}
\caption{Network differences for male and female players on day 856  (male  minus female). 
(a) average degree ${k}$, (b) clustering coefficient $C$, (c) average neighbour degree ${k}^{nn}$. 
Females have higher average degrees in communication and trade networks, as well as a 
higher clustering coefficient in trades, but considerably lower ${k}^{nn}$ for communication. 
(d) reciprocity between male-male and female-female links. 
Females are much more reciprocal in friendships and trades. 
Errorbars are standard deviations of the network measures obtained from 8 male control groups, 
each of the same size as the female sub-group. 
%A $t$-test supports these results by significant rejection of the equal means hypothesis. 
}
\label{fig:propertydifferences}
\end{figure}

\begin{figure}[t]
\begin{center}
\includegraphics[width=0.27\textwidth]{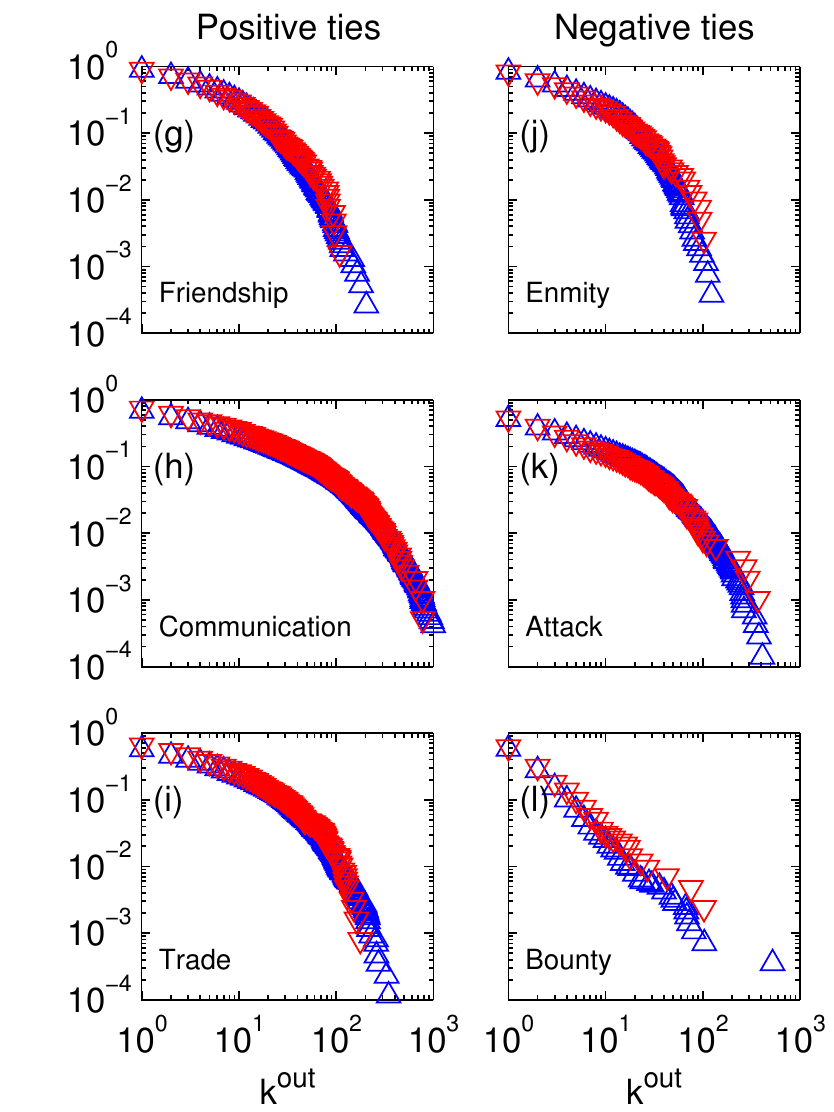}
\includegraphics[width=0.27\textwidth]{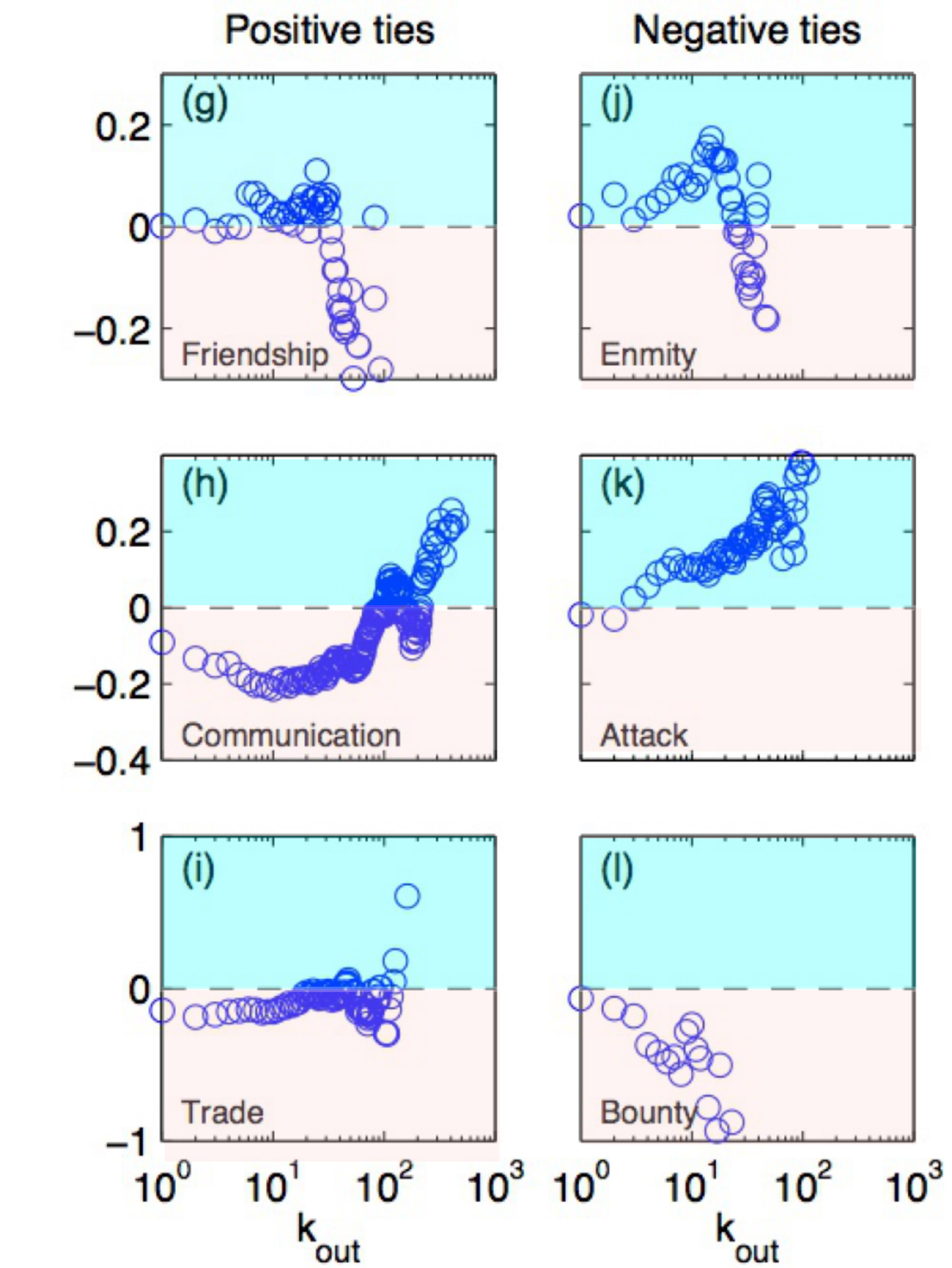}
\end{center}
\caption{(left panel) degree distributions for six social interaction layers for males (red) and females (blue). 
(right panel) differences (male minus female) out-degree, plotted against the degree. 
Gender differences become  apparent. 
Female players appear to be super-likers, and super-haters. 
Females have more communication partners for almost all degrees. 
For aggression-management, males prefer attacks, females 
prefer bounties---at all degrees. From \cite{Szell2012C}.
}
\label{fig:propertydifferencesb}
\end{figure}

Figure \ref{fig:propertydifferences} shows gender differences in four network properties as observed on day  856, 
illustrating that males and females structure their local network layers in very different ways. 
We compute the average degree ${k}$, the clustering coefficient $C$, and the average nearest neighbour 
degree $k^{\mathrm{nn}}$. We then subtract the value for the female group from the average of 8 male control groups 
of equal size. This difference is shown in the panels. Errorbars are the standard deviations of the means of the control groups. 
We find the following differences: 

{\em Females have more communication partners.}
As seen in Fig. \ref{fig:propertydifferences} (a), on average females have about 
5 more communication and 2-3 more trading partners than males.

{\em Females organize in clusters.}
Female trading networks show a clustering coefficient that is much (about 25\%) higher than the one of males,  
Fig. \ref{fig:propertydifferences} (b). This means that females tend to trade with people who trade among themselves. 
Also the clustering of female friendship networks is significantly higher than those of males, showing a preference for 
stability in local networks \cite{Granovetter1973}. Surprisingly, also for attacks females are more likely than men to 
attack people who are already in conflict with each other. 

{\em Males prefer well-connected communication partners.}
From Fig. \ref{fig:propertydifferences} (c) we learn that the communication partners of males 
have more communication partners than the communication partners of females. The same tendency  
is seen for male enmity networks, meaning that the typical enemy of a male player has more enemies than the typical 
enemy of a female. In relative terms, both effects are in the 10\% range \cite{Szell2012C}.

{\em Females reciprocate friendships.}
Females invest more effort in reciprocating positive links. 
Figure \ref{fig:propertydifferences} (d) shows that females reciprocate more friendship and 
trading links than males. For most negative links, there are 
no substantial gender differences. 
\\

On average, females have 5 more communication partners,  
females build more triangles, males link to well-connected communicators, and 
females reciprocate more. This has practical consequences for social life: 
while females focus on stable structures (high clustering), males seem to optimize 
communication speed by linking to well-connected communication partners. 
Male networks are less stable, but information spreads faster. 

In  the left panel of Fig. \ref{fig:propertydifferencesb} we compare degree distributions of the different interaction layers,   
for females (red) and male (blue). In the right panel we see the degree-differences as a function of 
the degree. These plots illustrate that women are super-communicators 
(dominating, whenever many friends are involved), 
and females are the super-enemies (whenever the degree in enmity networks becomes very large). 
Males play out aggression though attacks, while females prefer bounties to manage their negative 
feelings towards others---at all degrees.

%%%%%%%%%%%%%%%%%%%%%%%%%%%%%%%%%%%%%%%%%%%%%%
\subsubsection{Gender differences in temporal behavior}

\begin{figure}[t]
%\begin{center}
\begin{minipage}[c]{0.55\textwidth}
\includegraphics[width=0.9\textwidth]{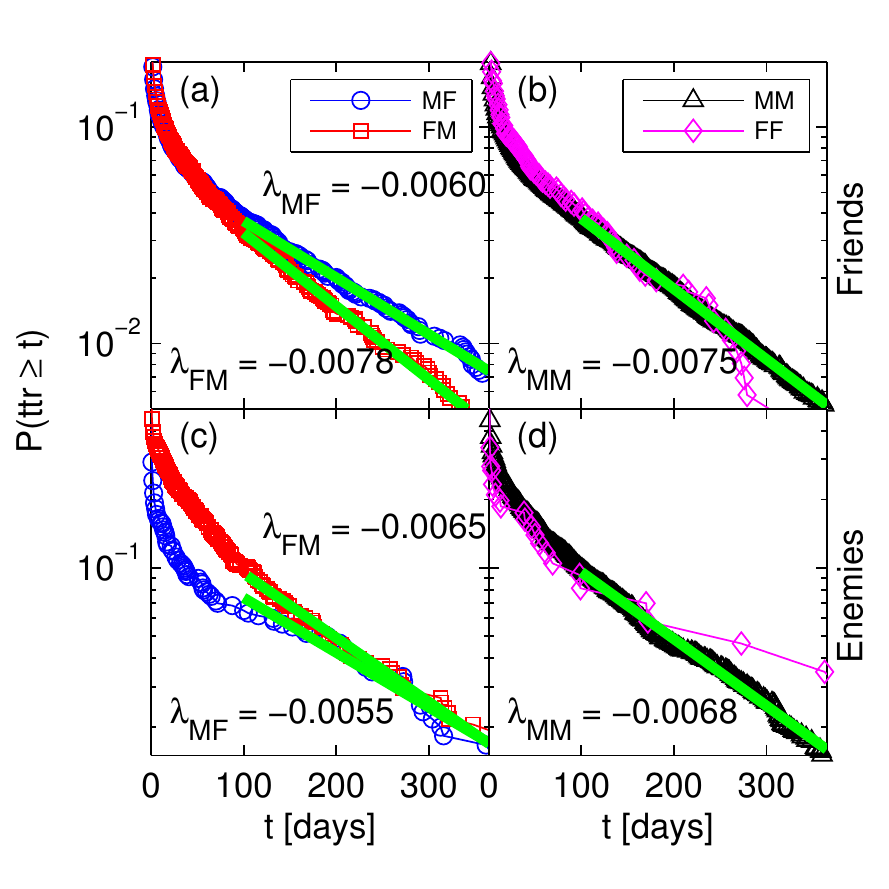}
  \end{minipage}\hfill
  \begin{minipage}[c]{0.45\textwidth}
%\end{center}
\caption{Time-to-respond distributions for different gender combinations. 
	(a)  Time for females to reciprocate a friendship link from a male initiator (MF), and vice versa (FM). 
	%Probabilities fall sub-exponentially in the first 30 days, later  exponentially, $P(\mathrm{ttr}\geq\!t) \sim e^{-\lambda t}$ with long-time decay rates $\lambda$ depending on gender pairs: 
	On long time scales ($>30$ days), males are much faster to reciprocate female friendship initiatives 
	than the other way round. % ($\lambda_{\mathrm{MF}} = 	-0.0060$). 
	(b) Equal sex reciprocation MM, and FF. Decay times are somewhere between the MF and FM case.
	(c) Time-to-respond for enemy links. Males are considerably slower to reciprocate within the first 180 days 
	if the initiator was a female than the other way. 
	(d) Equal sex reciprocation for enemy links. 
	%Fit ranges are 100 to 365 days, fits of FF curves were not feasible.
}
\label{fig:timetoreciprocate}
  \end{minipage}
\end{figure}

Imagine one player marks another as a friend. How long does it take to respond to this action? 
We can measure the time-to-respond to any action that is directed from one player to another. 
Males appear to respond fast (slow) to female friendship (enmity) initiatives.
We measure the time-to-reciprocate (ttr) it takes for individuals to reciprocate actions of a given type. 
In Fig. \ref{fig:timetoreciprocate} (a) and (b) we show the cumulative distributions for the time-to-reciprocate 
for friendship and enmity links, for the four possible gender permutations:  MM, FF, MF, FM. The  
first letter denotes the gender of the initiator, the second of the reciprocator. 
%Reciprocation time probabilities follow a sub-exponential decrease for about the first 30 days after a link was initiated, beyond which distributions become approximately exponential, $P(\mathrm{ttr}\geq\!t) \sim e^{-\lambda t}$. 
The decay rate $\lambda$ for MF friendship reciprocation is $\lambda_{\mathrm{MF}} \sim -0.0060$. 
For female initiation and male reciprocation it is $\lambda_{\mathrm{FM}} \sim -0.0078$. 
The MM rate is similar to the FM case, $\lambda_{\mathrm{MM}} \sim -0.0075$ (b). 
Correspondingly, the half-life for MF reciprocation is about 116, for FM it is only 89 days. 
The situation changes for enemy links. 
Here a difference is found within the first 150 days of an enemy marking: males 
reciprocate much faster than females. 
The approximate rates are $\lambda_{\mathrm{FM}} \sim -0.0065$, 
and  $\lambda_{\mathrm{MF}} \sim -0.0055$, respectively. 

In summary, we see that males respond {\em fast} to female friendship initiatives---females  {\em slow} to male ones.
Females respond {\em fast} to female friendship initiatives---males {\em slow} to males ones.
And finally, males respond {\em fast} to enemy activities from males---but respond {\em slow} to female aggression

%%%%%%%%%%%%%%%%%%%%%%%%%%%%%%%%%%%%%%%%%%%%%%%%%%%%%%%%%%%
\section{Mobility---how avatars move in their universe}

\verb|Pardus| is a world with a well-defined space. Players can move on a 2-dimensional surface, see Fig. 
\ref{fig:Illustration} (c). Since we know the location of all avatars at any point in time, 
we can study mobility patterns, and compare them to those of humans 
on the 2-dimensional surface of Earth. 
We locate players in one of the $N=400$ nodes, the so-called sectors (cities), that are linked by $K=1160$ wormholes 
(roads).
Sectors are arranged into 20 different {\em clusters}, which are perceived by the players as different
political or socio-economic regions, similar to countries. 
Each cluster is shown with a different background color in Fig. \ref{fig:Illustration} (c). 
Players usually have a ``home cluster'', where they focus their
socio-economic activities over long time periods. 
Occasionally, they move to sectors belonging to other clusters to explore the
universe, to relocate their home (migrate), or in response to extreme events, such as wars.

%%%%%%%%%%%%%%%%%%%%%%%%%%%%%%%%%%%%%%%%%%%%%%%%
\subsection{Jump- and waiting time distributions}

\begin{figure}[t]               
%    \begin{center}
    \begin{minipage}[c]{0.55\textwidth}
        \includegraphics[width=0.99\textwidth]{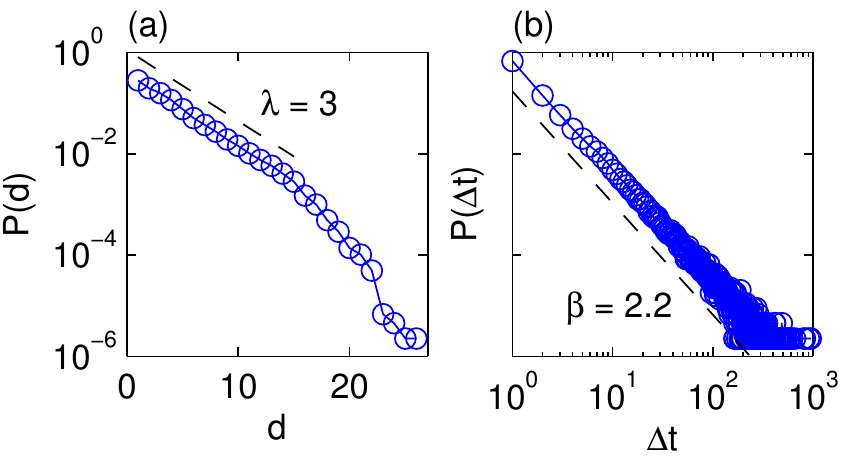}   
  \end{minipage}\hfill
  \begin{minipage}[c]{0.45\textwidth}
%    \end{center}
\caption{Distribution of jump distances $d$ (a), and waiting times $\Delta t$ (b).  
A {jump} occurs whenever the sector position changes from one day to the following.  
The distribution of jump distances has a characteristic length of $\lambda \sim 3$.  
The waiting time $\Delta t$ is the number of consecutive days a player 
spends in the same sector. The distribution is an approximate power law with an 
exponent of $\beta \sim 2.2$.
From \cite{Szell2012}. 
      }
\label{fig:distributions}
  \end{minipage}
\end{figure}

In Fig. \ref{fig:distributions} we show the distributions of jump (travel) distance and the waiting times 
between movements (jumps) from players' trajectories over 1,000 days. 
In Fig. \ref{fig:distributions} we show the distributions of jump (travel) distance and the waiting times 
between movements (jumps) from players' trajectories over 1,000 days. 
The length $d$  (integer) of a jump is measured in terms of network distance, and ranges from 1
to $d_{\mathrm{max}} = 27$, the diameter of the network. The distribution of jump distances for all players
over the observation period, is seen in
Fig. \ref{fig:distributions} (a).  For $d \leq 15$, the distribution is approximately exponential 
\begin{equation}
P(d) \sim e^{-\frac{d}{\lambda}} \quad ,
\label{eq:distancedistribution}
\end{equation}
with a characteristic jump length of $\lambda \sim 3$. 
The existence of a typical travel distance was found in real-world mobility data \cite{Roth2010,Bazzani2010}. 
%In some cases, players stay in the same sector for a number of consecutive days. On average, a player does not change sector in approximately $75\%$ of the days. 
The distribution of waiting times, $\Delta t$ (in days), 
between all consecutive jumps is seen in Fig. \ref{fig:distributions} (b). 
It follows an approximate power law 
\begin{equation}
P(\Delta t) \sim \Delta t^{-\beta} \quad , 
\label{eq:waitingtimedistribution}
\end{equation}
with $\beta \sim 2.2$, in agreement with recent measurements on human mobility \cite{Barabasi2005}. 
We find that mobility patterns are strongly influenced by the presence of clusters, the
socio-economic regions \cite{Szell2012}. 

\subsection{Long-term memory and mobility}

\begin{figure}[t]
%\begin{center}
\begin{minipage}[c]{0.55\textwidth}
   \includegraphics[width=0.7\textwidth]{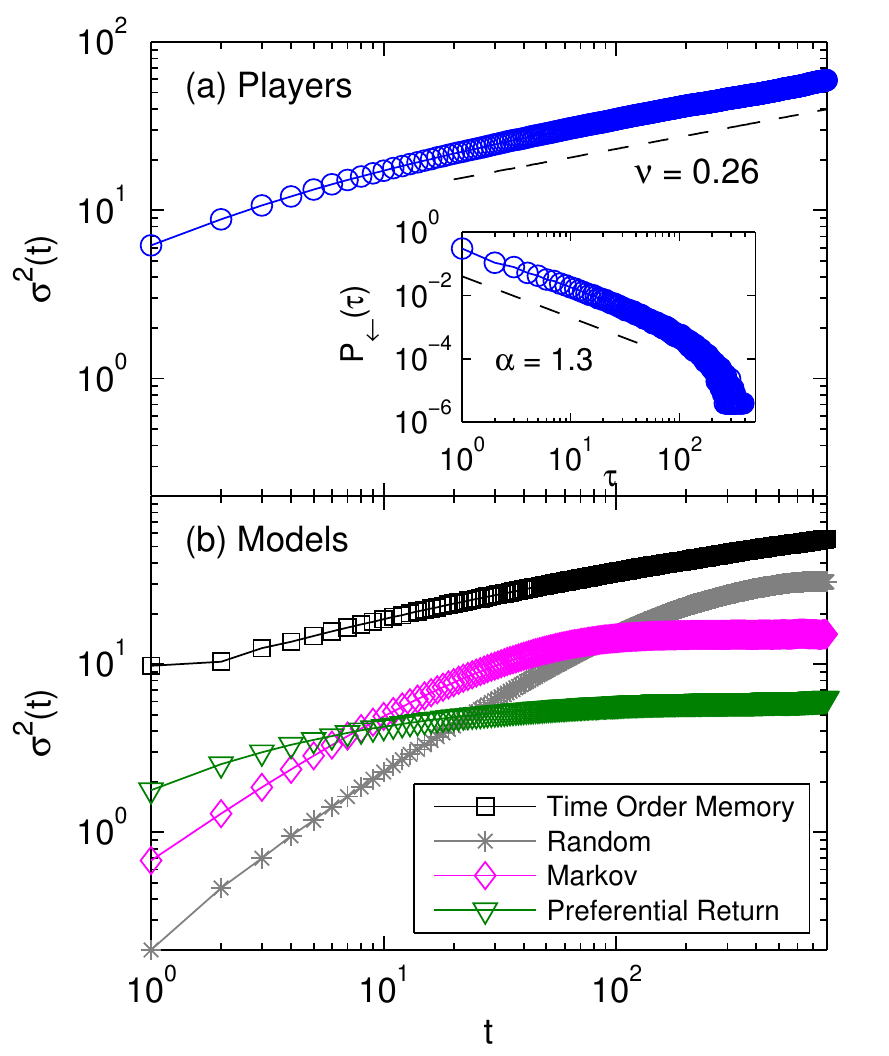}
  \end{minipage}\hfill
  \begin{minipage}[c]{0.45\textwidth}
%\end{center}
\caption{(a) Mean square displacement (MSD) of players' trajectories follows a power law,  
$\sigma^2(t) \sim t^{\nu}$, with sub-diffusive exponent $\nu \sim 0.26$. 
The inset shows the probability, $P_{\hookleftarrow}(\tau)$, for a player to return to a previously visited 
sector after $\tau$ jumps.  
%The curve follows a power law $P_{\hookleftarrow}(\tau) \sim \tau^{-\alpha}$ with an exponent of $\alpha \approx 1.3$ and an exponential cutoff.  
(b) MSD for various models to explain the observed scaling exponent $\nu$. 
Random walkers, a Markov model, and a preferential return model can not explain $\nu$.
A model with long-time memory (Time Order Memory) reproduces the exponent almost perfectly.
Curves are shifted vertically for clarity.
From \cite{Szell2012}. 
     }
\label{fig:msdptau}
  \end{minipage}
\end{figure}

To understand the diffusion of avatars over the transport network, we show the mean square displacement 
(MSD) of their positions in Fig. \ref{fig:msdptau} (a),  $\sigma^2(t) \sim t^{\nu}$, with $\nu \sim 0.26$.
This indicates anomalous, sub-diffusive behaviour.
This is not an effect from the specific topology of the \verb|Pardus| universe. 
To see this, in Fig. \ref{fig:msdptau} (b), we show the situation for random walkers on the wormhole network (gray stars).
It produces the expected MSD result that is expected from standard diffusion, i.e. with an exponent 
$\nu \sim 1$, for up to $t \sim 100$ days, at which the fine size of the universe begins to play, which causes 
the saturation.
We tested several potential models to explain the anomalous diffusion behaviour, including a simple 
Markov model that is based on the observed node--node transition probabilities, 
and a preferential-return model that takes higher-order memory effects into account \cite{Sinatra2010nms,Song2010}.
Both are not able to capture the correct scaling pattern of the MSD, 
in particular $\nu \sim 0.26$, as seen in Fig. \ref{fig:msdptau} (b).

The reason for failure is that the probability to move to a certain location does not depend on the 
current location, nor on the order of previously visited locations.
Instead, we observe that individuals tend to return to sectors they have visited recently.
%Consequently a sector that has been visited many times but with the most recent visit dating back one year has a lower probability to be visited again than a sector that has been visited just a few times but with the last visit dating back only one week.
To model this mechanism we measure the return time distribution
in the jump timeseries for returning (for the first time) to
the currently occupied sector after $\tau$ jumps, and find $P_{\hookleftarrow}(\tau)\sim \tau^{-\alpha}$, with 
$\alpha \sim 1.3$, see Fig. \ref{fig:msdptau} (a).
We use this information to design a ``Time Order Memory'' (TOM) model that 
incorporates a power law distribution of first return times, power law distributed waiting 
times, and exponentially distributed jump distances. These ingredients are sufficient to reproduce 
the sub-diffusive behaviour in Fig. \ref{fig:msdptau} (a). 
The model works as follows: an individual rests in a given sector for a
number of days drawn from the waiting time distribution. Then, she jumps. 
There are two possibilities: (i) with probability $v$ she returns to an already
visited sector, (ii) with probability $1-v$ she jumps to a sector
she never visited before. For (i), one of the previously visited sectors is chosen 
with $P_{\hookleftarrow}(\tau)$. In case (ii), she draws a distance $d$ 
from the distance distribution, and jumps to a randomly selected sector at that distance. 
The model parameters, $\lambda$, $\beta$, and $\alpha$, are all fixed by the data. 
By averaging over all jumps and players, the probability of returning to an already visited location is found as $v \sim 0.83$. 
We get $\nu_{\mathrm{TOM}} = 0.23 \pm 0.02$, in agreement with the observed exponent.
Black squares in Fig. \ref{fig:msdptau} (b) indicate that the model works indeed.

In summary, we find that mobility in the \verb|Pardus| world is not all that different 
from mobility on Earth. Locations are visited in specific temporal patterns, leading to strong 
memory effects that are essential to understand the statistics of observed mobility trajectories. 
Neglecting either spatial or temporal factors make it hard to understand the statistics  of human mobility. 
This might be true in the real world too.
Interestingly, a thorough understanding of human mobility is still outstanding, 
since some results appear to be contradictory  \cite{Barthelemy2010}. 
Some report  fat-tailed distributions of trip lengths \cite{Brockmann2006, Song2010}, 
others   exponential or binomial distributions \cite{Bazzani2010,Roth2010,Barthelemy2010}.

%%%%%%%%%%%%%%%%%%%%%%%%%%%%%%%%%%%%%%%%%%%%%%%%%%%%%%%%%%%%%
\section{The wealth of virtual nations}

 Almost universally, wealth is not distributed uniformly within societies. 
 Even though wealth data have been collected in various forms for centuries, 
 the origins behind wealth-, and hence, social inequality are not yet fully understood. 
 This is not different in \verb|Pardus|. However, there we can figure out what it needs to be 
 wealthy in terms of your position in the social multilayer network.
Can we finally understand the origin of wealth inequality?

%%%%%%%%%%%%%%%%%%%%%%%%%%%%%%%%%%%%%%%%%%%%%%%%%%
\subsection{More on the Pardus economy}

From interaction data, $M_{ij}^{\alpha}(t)$, and player states, 
$\sigma_i^{\rm X}(t)$, we can reconstruct all economic activities in the \verb|Pardus| society. 
In particular the input-output production matrix of the economy and the variety of goods are pre-defined.
Goods are of uniform quality (homogeneous); consumables and equipment can be 
partially substituted by other types of consumable and equipment.
Intermediate goods are needed for production in exact proportions.
There are 5 commodities (natural resources), 19 intermediate goods, and 5 end-products, i.e. consumables.
Although capital requirements to establish production facilities are low, there are barriers to entering production. 
Incumbents may threaten or harm potential new entrepreneurs. Game rules set a maximum number of 
production facilities per player. %Many operate the maximum number of factories.
Production facilities %are fixed assets with infinite durability and 
can {\em not} be sold. 
Investments in production facilities therefore motivate players to stay in the sector.
No labor is needed for production itself, but transport of raw materials and intermediate 
goods requires effort and resources. 
Because of transport costs, facilities effectively only compete with similar facilities nearby.
This leads to local oligopolies.
%A special kind of goods are various forms of equipment, i.e. items like a space ship or weapons. 
%Equipment can only be produced in special facilities which also act as warehouses and selling points. 
%Equipment is durable, but has a finite lifetime. Maintenance applied to equipment 
%can increase the lifetime. 
%When a player sells equipment after usage, it is scrapped.
%%\footnote{Players may own only a limited amount of equipment, resulting in an incentive to sell from time to time.}.
Owners of production facilities are completely free to set the price at which they %buy raw materials and 
sell their products.
There exist non-player facilities (belonging to the game) whose prices depend on local supply and demand.
The monetary currency is called {\em credits}. There is no credit or banking system, 
all transactions are payed and cleared immediately. 
There is no inflation.

%%%%%%%%%%%%%%%%%%%%%%%%%%%%%%%%%%%%%%%%%%%%%%%%%%%%%%%%%%%%%
\subsection{Wealth}

\begin{figure}[t]
\begin{center}
   \includegraphics[width=0.6\textwidth]{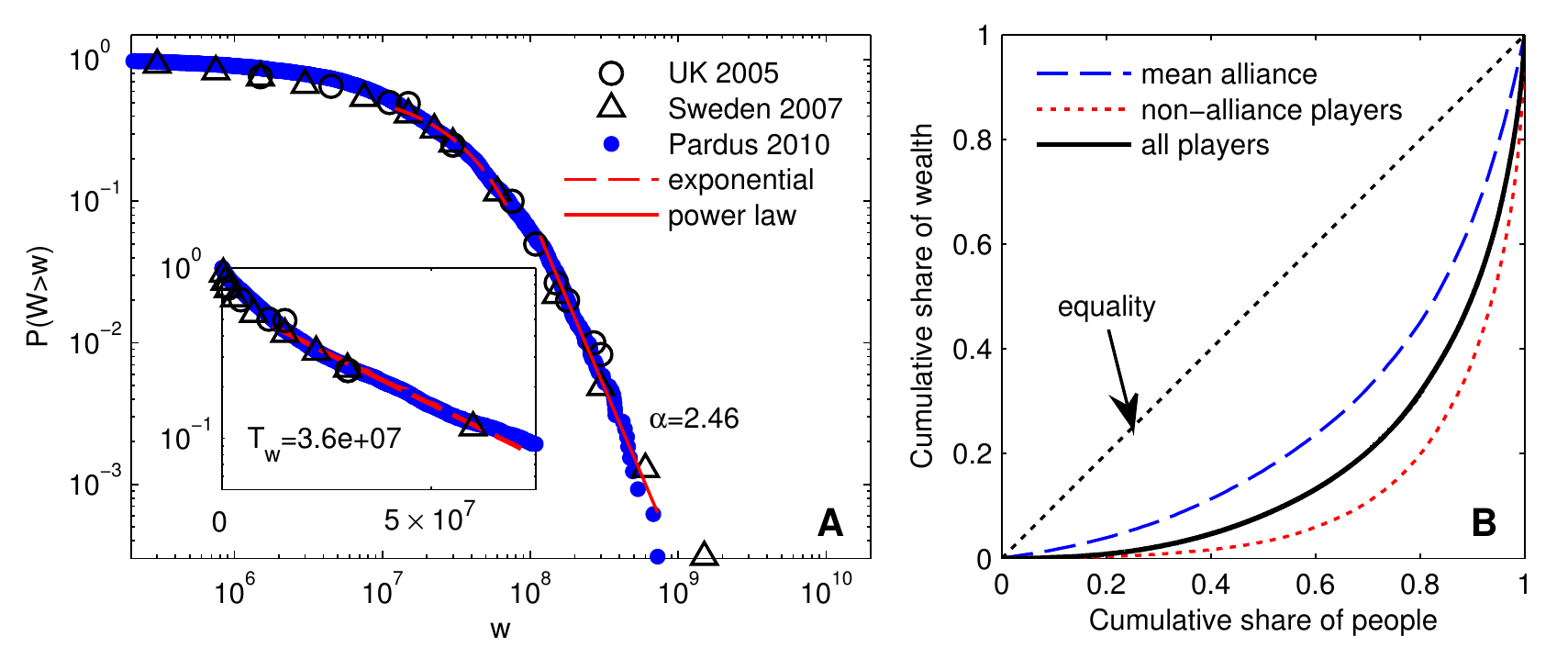}
   \includegraphics[width=0.34\textwidth]{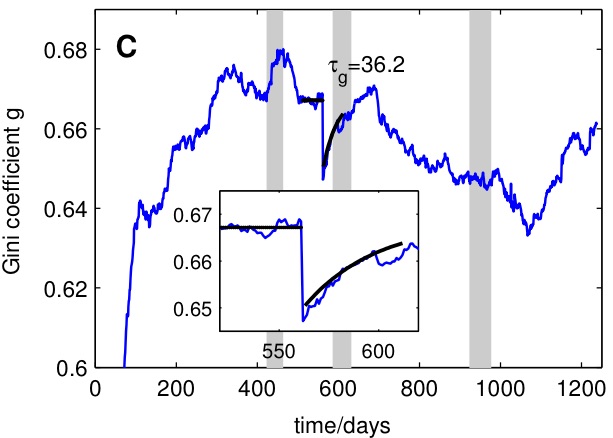}
\end{center}
\caption{(a) Cumulative wealth distributions for Sweden, the UK, 
and for the Pardus society on day 1200. Note the similarity.
(b) Corresponding Lorenz curve of wealth in Pardus. 
For every alliance, a separate Lorenz curve is calculated; the dashed blue line is their average.
(c) Gini index over time, $g(t)$. A Christmas charity event on day 562 leds to a re-distribution 
from the wealthy to the poor, resulting in a downward jump of the Gini index.
The inset shows the exponential recovery to previous levels.
Gray areas indicate periods of war. 
From \cite{Fuchs2014B}.
}
\label{fig:Pardus_wealth_cum}
\end{figure}

There are various ways to accumulate wealth: 
 trading,
 collecting and selling natural resources,
 producing goods,
 working for hire (common jobs are courier, hunter, or bounty hunter),
 receiving donations or other payments,
 increase of the alliance funds,
 robbing, and stealing. 
The wealth of player $i$ is the sum of the value of his assets,  
cash $v_{l,i}$,  
 equipment $v_{e,i}$,   
 share of alliance funds $v_{af,i}$,  
 and inventory $v_{inv, i}$; details in \cite{Fuchs2014B}, 
 \begin{equation}
 	\sigma_i^{\rm wealth} (t) = v_{l,i}(t) + v_{e,i}(t) + v_{af,i} + v_{inv, i} \quad .
\end{equation}
Ways to reduce wealth include
 consumption,
 paying for maintenance, %(because of degradation or of damage from fights),
 investing in production facilities or equipment,
 discarding goods,
 becoming victim of theft or robbery,
 giving to fellow players, or paying into alliance funds,
 a decrease of the alliance funds,
 or making adverse trades.

The cumulative wealth distribution of \verb|Pardus| players, in comparison to the UK and Sweden, 
is shown in Fig. \ref{fig:Pardus_wealth_cum} (a) New and inactive players were excluded.
The bulk of the distribution is compatible with an exponential
with decay $T_w=3.6\times10^7$ credits. The tail can be seen as an approximate power law 
with exponent $\alpha \sim 2.5$.
The situation is compatible with real world data \cite{Dragulescu2001}. 

%%%%%%%%%%%%%%%%%%%%%%%%%%%%%%%%%%%%%%%%%%%%%%%%%%%%%%%%%%%%%
\subsection{Inequality}

Figure \ref{fig:Pardus_wealth_cum} (b) shows the Lorenz curve for the Pardus society (black line). 
It is the share of the total wealth as a function of the fraction of the people holding that share. 
The closer the Lorenz curve is to the diagonal (black dotted line) the more egalitarian is the wealth distribution. 
Uniform wealth distribution corresponds to the diagonal.
Associated to the Lorenz curve is the Gini index, $g=1-2A$, with $A$ being the area (integral) 
under the curve \cite{Gini1912}. We find $g=0.65$. 
We show the Lorenz curves for all players and for those that are not organized in any alliance (red dotted line). 
These players  generally operate individually, and show a much more pronounced wealth 
inequality than the entire society, the respective Gini index being $0.70$. 
In contrast, the Lorenz curve for the various alliances (dashed blue line) indicates 
that people within the alliances tend to be much more equal in wealth, when compared to the entire society.
The Gini index for the alliances is  $0.50$.
The main reason for this higher equality is the smaller fraction of poor players in alliances: 
while $79\%$ of the total population, and $92\%$ of the richest $10\%$, are alliance members, 
only $28\%$ of the poorest $10\%$ are.

Figure \ref{fig:Pardus_wealth_cum} (c) shows the time evolution of the Gini index. 
After an initial steep rise in the first 150 days, the Gini index $g$ fluctuates between 
$0.68$ and $0.63$,  similar to many  Western countries. 
A sharp drop of $g$ from $0.67$ to $0.65$ occurred on  Christmas day 2008.
On this day, a charity event took place, where thousands of players donated cash for the less wealthy.
The gained level of equality is lost exponentially fast, within a few days previous Gini index levels are reached, see inset.
This indicates a remarkable stability of the shape of the wealth distribution.
It is  hard to decrease inequality by re-distributing wealth. Is the origin of wealth inequality 
based in social behavior?

%{\bf Evolution of the wealth distribution over time.}
%The average wealth in the Pardus society $\langle w(t) \rangle$ grows over time. Brackets indicate the average over all players present at time $t$. The daily average change we denote by $\Delta w(t) = \langle w(t) \rangle-\langle w(t-1) \rangle$, and is presented in Fig. \ref{fig:ginietc_evolutionb} for different cohorts. Cohorts are defined by the time when the players joined. Cohort 1 contains all players who joined on the first day, cohort 2 joined between day 2 and day 200, cohort 3 between day 201 and 400, etc. For each cohort we computed its average wealth from the individual wealth timeseries of its members. Clearly, those players that joined early on have a faster wealth growth (slope), especially in the early years of the Pardus universe. It is visible that average wealth increases less during war periods (gray shaded areas). 

%\begin{figure}[t]
%\begin{center}
%\includegraphics[width=0.4\textwidth]{plots_chapter1/wealthTS_latecomers_February-eps-converted-to.pdf}
%\end{center}
%\caption{Wealth increase over time for different cohorts of players. Cohort 1 is the group of players that joined the game 
%from the very beginning. The wealth increase (slope) of the first cohort is significantly larger than for the latecomers.
%}
%\label{fig:ginietc_evolutionb}
%\end{figure}

%%%%%%%%%%%%%%%%%%%%%%%%%%%%%%%%%%%%%%%%%%%%%%%%%%%%%%%%%%%%%
\subsection{Behavioral factors for wealth} 

%%%%%%%%%%%%%%%%%%%%%%%%%%%%%%%%%%%%%%%%%%%%%%%%%%%%%%%%%%%%%
\subsubsection{Influence of activity on wealth} 

We see a trivial strong linear relation between the average wealth of a player and her total activity.
The corresponding Pearson correlation coefficient is $\rho=0.535$ ($p$-value $<10^{-6}$). 
We conduct a partial correlation analysis and find several significant behavioral factors that explain wealth \cite{Fuchs2014B}. 
We find that the more a player trades compared to his other actions, the higher is his wealth-gain.
This is not surprising since trade is the main source of income in the game.
We also find that the more of a player's actions are attacks, the lower is his wealth-gain.
This suggests that revenue from attacks through robbery and bounty hunting does hardly bear  
the associated costs, e.g. for repairing damage caused by fights.
There might be secondary damaging effects of aggressive behavior, 
such as reduced willingness of others to socialize (and trade) in the future.
Another explanation might be  that attacks are sometimes carried out 
without economic interests, but just for creating terror.

%%%%%%%%%%%%%%%%%%%%%%%%%%%%%%%%%%%%%%%%%%%%%%%%%%%%%%%%%%%%%
\subsubsection{Influence of achievement-factors on wealth}

Wealth and other achievement factors, such as skills, XPs, and faction rank, 
are strongly correlated with total activity.
For this reason, we define the wealth-gain of player $i$ as $ \eta_i(t) = w_i(t)/a_i(t)$,
where $a_i(t)$ is the cumulative activity of a player approximated by the total amount of APs he has ``spent''. 
$\eta_i(t)$ can be seen as the efficiency of gaining wealth.
To further exclude these spurious correlations, partial correlation coefficients are calculated.
Age and faction rank are a significant factors ($p$-value  below 1\%).
%Faction rank is a significant positive factor for wealth with a significance level below 0.01\% for all days. 
% High faction rank means `political' influence in the game.
Players that are not in any faction, i.e. on average less social, 
have the smallest possible value of faction rank, and are generally poorer.
There is a significant fraction of rich people with low combat skill.
Otherwise, we find no correlation between combat skill and wealth-gain.
Farming skill has a consistently positive and mostly significant correlation with wealth. 
%Farming skill is associated with the collection of resources, which generates income.

\begin{figure}[t]
%\begin{center}
\begin{minipage}[c]{0.65\textwidth}
\includegraphics[width=0.95\textwidth]{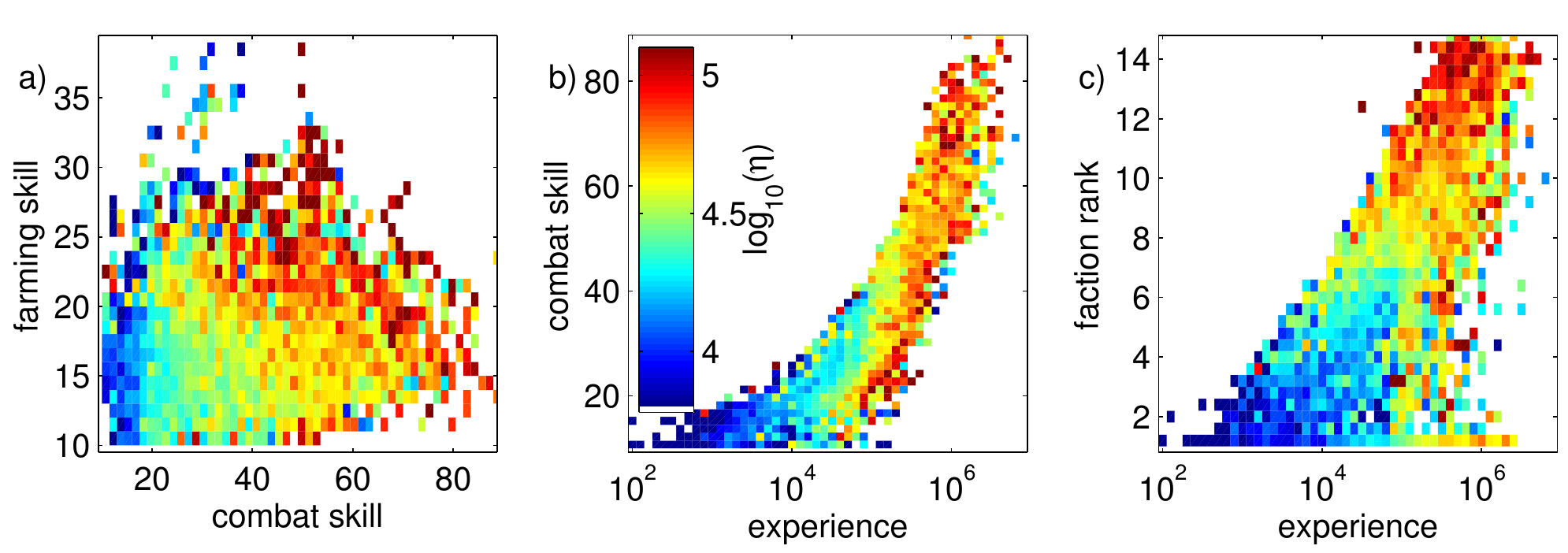}
  \end{minipage}\hfill
  \begin{minipage}[c]{0.35\textwidth}
%\end{center}
\caption{Two-dimensional binned averages of the wealth-gain as a function of achievement-factors. 
Colors represent the logarithm of the average wealth-gain, $\log_{10}(\eta)$, over all players that fall into that bin.
Blue corresponds to low, red to high values, empty bins are white.
A XP and faction rank, B XP and combat skill, C combat skill and farming skill.
From \cite{Fuchs2014B}.
}
\label{fig:OtherProperties2D_w}
  \end{minipage}
\end{figure}

In Fig. \ref{fig:OtherProperties2D_w} we show the wealth-gain as a function of a combination 
of different performance factors. 
High farming skills and intermediary combat skills correlate with high wealth-gains that are shown as by colors
(high wealth-gain is red). 
Below a certain level of experience points, no high wealth-gain seems possible, and 
a high faction rank together with high experience point scores are a good predictor for wealth. 
For details, see \cite{Fuchs2014B}.

\subsubsection{Wealth depends on how social you are}

%Alliances can locally coordinate production capacities to build up entire production chains. For an optimal production chain, it is sometimes necessary to increase the production capacity of a certain intermediate good. This is often done by luring a new member into the corresponding business and by paying her for the construction of an additional production facility.

\begin{figure}[b]
%\begin{center}
\begin{minipage}[c]{0.65\textwidth}
\includegraphics[width=0.97\textwidth]{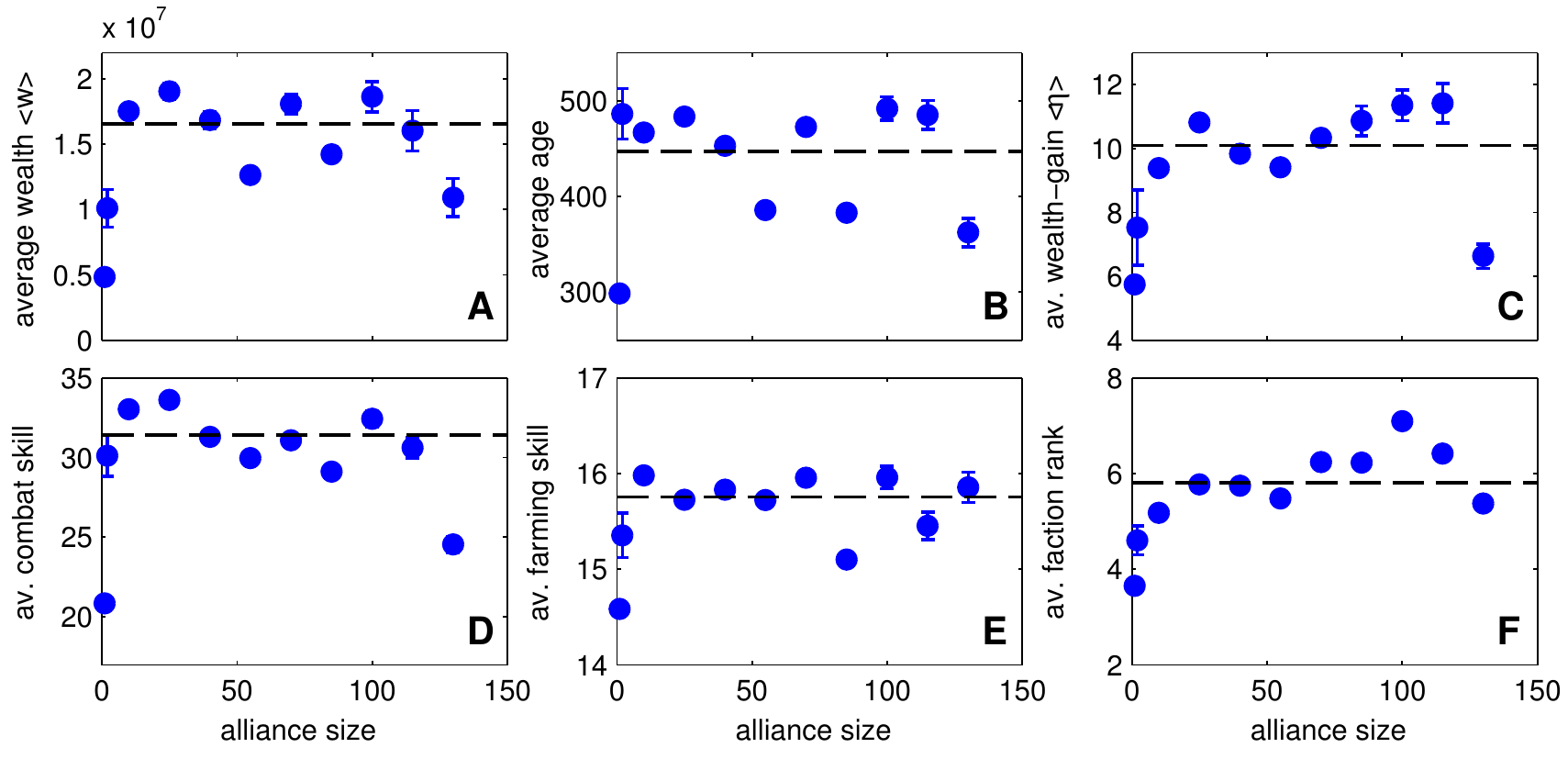}
  \end{minipage}\hfill
  \begin{minipage}[c]{0.35\textwidth}
%\end{center}
\caption{(a) Wealth, (b) age, (c) wealth-gain, (d) combat skill, (e) farming skill, (f) faction rank 
as a function of alliance size.
The first bin contains players that are in no alliance, the second bin has players in alliances of size two.
Members of the smallest alliances show low wealth and achievement scores.
Also the largest groups show lower levels.
%Error bars denote the standard errors. 
The line is the  average over all players in alliances with at least three members. 
From \cite{Fuchs2014B}.
}
\label{fig:alliSize}
  \end{minipage}
\end{figure}

Alliance members on average are somewhat richer than non-alliance members, 
both in absolute terms and in wealth-gain.
Members also have better skills and a higher faction rank. 
As seen in Fig. \ref{fig:alliSize} the size of an alliance has little influence on wealth and other factors,
except for players that are in no alliances or in alliances with only two members.
These players are consistently poorer than players in groups with three and more members.
Members of the largest alliances show performance indicators below average (dashed line).
It does not matter if you are member of big  or small groups;  it matters if you are a member of at least one group.  

%%%%%%%%%%%%%%%%%%%%%%%%%%%%%%%%%%%%%%%%%%%%%%%%%%%%%%%%
\subsection{Wealth and  position in the multilayer network}

We use the trade, communication, friendship, and enemy interactions layers, $M_{ij}^{\alpha} (t)$, 
to determine the in- and out-degree, $k^{\mathrm{in}}_i$ and  $k^{\mathrm{out}}_i$, the nearest-neighbor degree, 
 $k^{\mathrm{nn}}_i$, and the clustering coefficient, $c_i$, for every player $i$.
To show the dependence of wealth-gain on various combinations of these network parameters, 
we plot two-dimensional binned averages of 
wealth-gain versus pairs of network measures in Fig. \ref{fig:NWProperties2D_w}.
The color is the logarithm of wealth-gain, $\log_{10}(\eta)$, from blue (lowest) to red (highest). 
We find the following results:

\begin{figure}[t]
%\begin{center}
\begin{minipage}[c]{0.65\textwidth}
\includegraphics[width=0.97\textwidth]{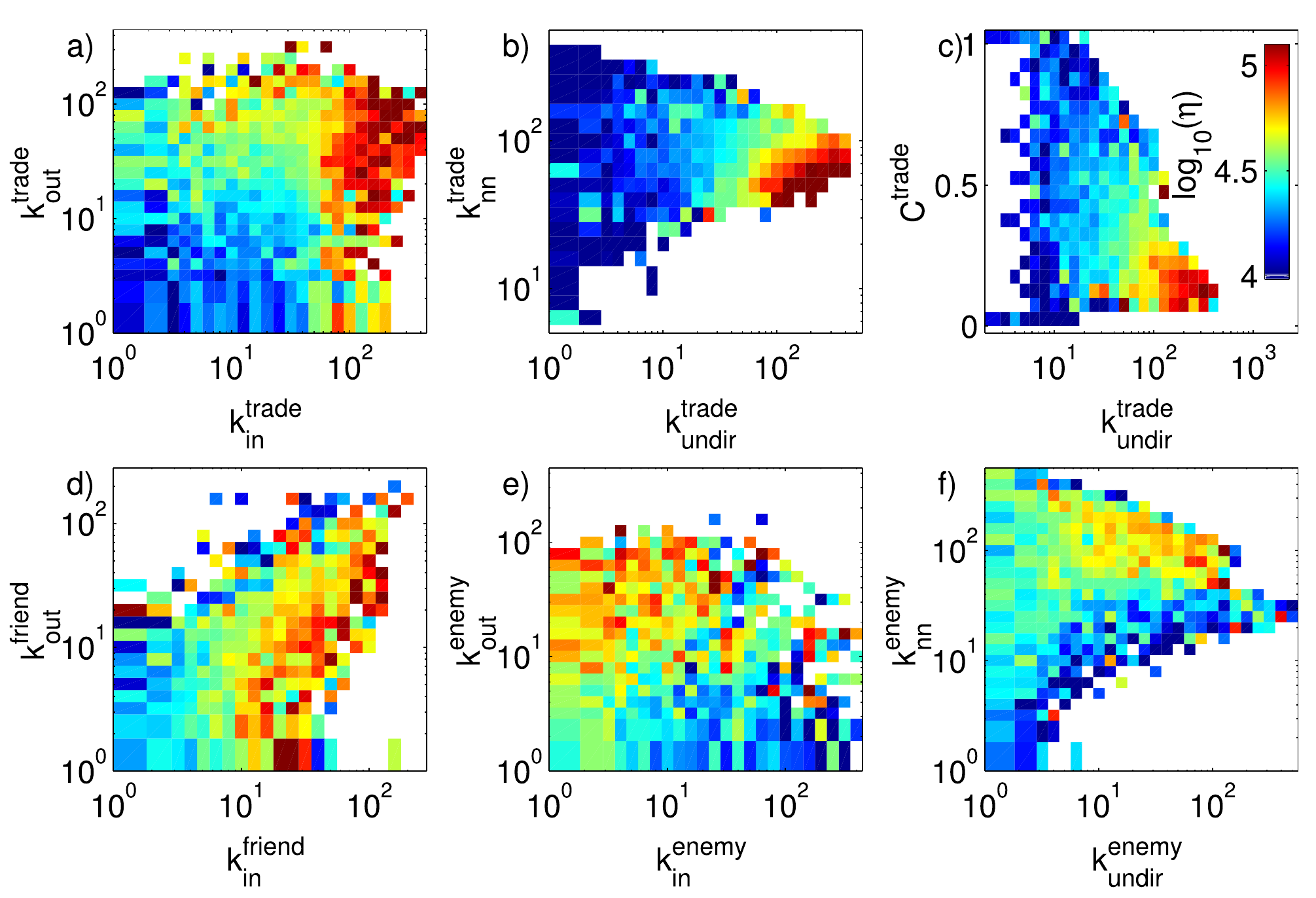}
  \end{minipage}\hfill
  \begin{minipage}[c]{0.35\textwidth}
%\end{center}
\caption{Wealth-gain ($\log_{10}(\eta)$) as a function of network properties,  
blue (lowest) to red (highest), empty bins are white. 
(a) trade in- and out-degree, (b) trade undirected degree and nearest-neighbor degree, 
(c) trade undirected degree and clustering coefficient,
(d) friend in- and out-degree, (e) enmity in- and out-degree, (f) enmity undirected degree, 
and nearest-neighbor degree.
%Data are taken every 240 days. 
From \cite{Fuchs2014B}.
}
\label{fig:NWProperties2D_w}
\end{minipage}\hfill
\end{figure}

{\em Trade}. The trade layer has the strongest impact on wealth. 
 Trade in-degree has a significant, positive partial correlation with wealth.
The in-degree is defined as trade with a player's production facilities and is therefore a proxy for his production.
Figure \ref{fig:NWProperties2D_w} (a) confirms the positive connection between trade in-degree and wealth, 
while showing much less influence from trade out-degree.
Concerning the undirected degree of the trade network versus the nearest-neighbor degree (b), we see that 
the richest are found to have an intermediate nearest-neighbor degree of about 
$k^\mathrm{trade}_\mathrm{nn} \sim 35-70$, well below their undirected degree.   
This means that they are selling to people that are less connected in the trade network than they are themselves. 
From Fig. \ref{fig:NWProperties2D_w} (c) we gather that high wealth-gain is made with a combination of high degree and 
a relatively low clustering coefficient, $C^\mathrm{trade} \sim 0.1$.  
This means that rich players avoid cyclical structures in their trading 
networks, which allows them to act as ``brokers'' between players that do not directly trade with each other.
The partial correlation between wealth and the trade clustering coefficient is negative.

{\em Communication}. Communication in-degree has a significantly positive partial correlation.
High communication in-degree means good access to information, which is expected to be profitable.
%The Communication out-degree shows positive partial correlation on most days.
%A player's communication out-degree is the number of players she {\em tries} to influence.
Since communication links are reciprocal, and in- and out-degree are highly correlated, 
there might be a spurious effect of the communication in-degree.
The communication nearest-neighbor degree has a negative and mostly significant partial correlation.
This might indicate that it is advantageous to converse with players who are less informed than oneself.

{\em Friendship}. In Fig. \ref{fig:NWProperties2D_w}  (d) the situation for the in- 
and out-degrees for the friendship layer is shown. 
Players with high wealth-gain are those that are liked by more players than they like themselves, 
$k_\mathrm{in}^\mathrm{friend} > k_\mathrm{out}^\mathrm{friend}$.  
Poor players have marked others as friends more often on average 
than they are marked. The role of these asymmetries might be a fruitful direction for future study. 

{\em Enmity}. We see that people with above-average wealth-gain 
are rarely marked as an enemy by others, but do mark others as enemies, (e). 
Players who have been marked as enemy by many others are generally poor.
In agreement with this finding, the enmity in-degree has a significant negative partial correlation with wealth, 
while the enmity out-degree has a weak significant positive correlation with wealth \cite{Fuchs2014B}.
This suggests that players with high wealth-gain actively invest in a good reputation. 
Finally, players with above average wealth-gain have a high nearest-neighbor degree (f).
Players with high enmity (in)degree are ``public enemies'' \cite{Szell2010msd}.
A high $k_\mathrm{nn}^\mathrm{enemy}$ indicates that one is the enemy of public enemies 
and that one has few private enemies.
\\

We learn that being wealthy in \verb|Pardus| is not mere luck---it is highly structural. 
It depends on the actions that you do and toward whom you direct them, or in different words, 
in which sectors of the social multilayer network you are located. 

%%%%%%%%%%%%%%%%%%%%%%%%%%%%%%%%%%%%%%%%%%%%%%%%%%%%%%%%
\section{Towards a new social science?}

What have we learned about the homo sapiens?
Methodically we created a unique lab situation. 
For the first time, we are confronted with almost complete 
information about an entire human society with little-to-no interference with their observers. 
However virtual \verb|Pardus| may be, it is a society where decisions are made by humans, 
not by bacteria, rats, or algorithms. 
Sociological predictions and hypotheses, sometimes century old, and  can now be tested on big data, without 
the danger of privacy violations, and sometimes, questions  can be decided: 
%In this situation the question arises: What do you ask if all information is available? 
in summary, we find
\begin{itemize}
\item The strength of human interactions is measurable. Local tie strength is related 
(maybe not causally) to the betweenness of links, which is a global property. 
We confirm the weak ties hypothesis \cite{Szell2010msd}; 
but we can go much further by measuring the strength of ties
as a function of the interaction density between humans \cite{Thurner2014}. 
  
\item Humans are triangle-closers. We confirm triadic closure \cite{Szell2010msd}. 
The new discovery is that the mechanism 
of triadic closure is so dominant, that it might be sufficient to describe the basic statistics of 
human interactions \cite{Klimek2013} and group formation \cite{Klimek2016}. 
  
\item Humans organize in stable signed triangles. We confirm previously conjectured 
social balance to unprecedented precision levels. 
We find evidence for the weak form of social balance \cite{Szell2010,Sadilek2018} .

\item We find that humans tend to organize in group sizes that are (roughly) multiples of four \cite{Fuchs2014}, 
a recently conjectured social organizational principle that might bear significance 
for human group formation \cite{Zhou2005}. 

\item  Males and females organize their local social networks in very different ways:  
females tend to focus on higher clustering and thus stable networks, males focus on more
``fuzzy'' networks that allow for fast communication across large parts of the society, but which are less stable.  

\item Good and bad actions lead to different organization of interaction networks.
Positively connoted interactions show Poissonian degree distributions and high reciprocity, 
negative ones show fat tails, and are hardly reciprocal \cite{Szell2010}. 

\item Females and males handle aggression differently. 
Males tend to act out aggression directly through 
attacks, females tend to delegate aggressive acts to others by placing bounties on others' heads \cite{Szell2012C}. 
  
\item Humans become vastly more aggressive when confronted with hostile behavior towards them.
The likelihood of negative responses after receiving unfriendly treatment increases about 10-fold \cite{Thurner2012}. 

\item Short-term efforts for wealth redistribution are short-lived.  
Wealth is found to be a function of your position in the social multilayer network. 
Wealth is a consequence of your behavior that determines that position.  In that sense 
wealth seems to be a structural phenomenon \cite{Fuchs2014B}.

\item We discover new social laws in temporal behavior, such as the 
characteristic exponential decay in response times for reciprocating actions \cite{Mryglod2015}. 
\end{itemize}

To what extent can we trust these findings?
Is \verb|Pardus| a good ``model'' for real societies? 
We don't know yet, but we provided evidence that there are striking similarities in 
the structures of communication and friendship networks \cite{Szell2010msd}, 
mobility patterns \cite{Szell2012},
social balance \cite{Szell2010},
gender specific patterns \cite{Szell2012C},
and wealth distributions \cite{Fuchs2014B}, 
when compared to their real-world analogs.

Would we finally agree that social dynamics will become a quantitative-predictive experimental 
science in the near future---that sociology eventually becomes a sub-discipline of physics, as Comte 
envisioned?
I think we demonstrated in this lectures, that this is a possibility. 
The limit to understanding societies will neither be data size nor computational power---we are almost there. 
It is not hard to imagine that the data, as we analyzed it here, can be replaced 
by real data, including mobility, friendship interactions, 
financial transactions, medical data, trading, shopping, surfing, and so on.
 I believe the limits to understanding societies, 
as soon as we aim for an understanding at a deeper level, 
is the co-evolutionary complexity of social systems 
that might impose bounds to predictions;  and of course, 
as always in science, the ultimate limit is the quality of our questions.

Works summarized here were supported by EC FP7 INSITE,  COST MP0801, 
and the Austrian Science Fund FWF under P23378-G16. 

%\bibliography{varennabib.bib}
\bibliography{references}
\bibliographystyle{varenna}

%\begin{appendix}
%\section{Appen \label{apA}}
%\end{appendix}

\end{document}